\documentclass[a4paper,11pt]{article}

\usepackage{jcappub} 

\usepackage{graphicx}
\usepackage{subcaption}
\usepackage{rotating}
\usepackage[T1]{fontenc}
\usepackage{multirow}
\usepackage[]{algorithm2e}
\usepackage[title]{appendix}
\usepackage{lineno}
\usepackage{bm}
\usepackage[normalem]{ulem}

\newcommand{\gem}{{$\gamma$EM}}
\newcommand{\gems}{{$\gamma$EMs}}

\usepackage{comment}

\subheader{LAPTH-068/22}

\title{Mind the gap: The discrepancy between simulation and reality drives interpretations of the Galactic Center Excess}

\author[a,b]{Sascha Caron}
\emailAdd{scaron@nikhef.nl}
\author[c,d]{, Christopher Eckner}
\emailAdd{eckner@lapth.cnrs.fr}
\author[a,b]{, Luc Hendriks}
\emailAdd{luc@luchendriks.com}
\author[e,f]{, Gu{\dh}laugur J{\'o}hannesson}
\emailAdd{gudlaugu@hi.is}
\author[g]{, Roberto Ruiz de Austri}
\emailAdd{rruiz@ific.uv.es}
\author[h]{, Gabrijela Zaharijas}
\emailAdd{gabrijela.zaharijas@ung.si}

\affiliation[a]{Theoretical High Energy Physics, Radboud University
Nijmegen, Heyendaalseweg 135, 6525 AJ Nijmegen, the Netherlands}
\affiliation[b]{Nikhef, Science Park 105, 1098 XG Amsterdam, the Netherlands}
\affiliation[c]{LAPTh, CNRS, USMB, F-74940 Annecy, France}
\affiliation[d]{LAPP, CNRS, USMB, F-74940 Annecy, France}
\affiliation[e]{Science Institute, University of Iceland, IS-107 Reykjavik, Iceland}
\affiliation[f]{Nordita, KTH Royal Institute of Technology and Stockholm University, Roslagstullsbacken 23, SE-106 91 Stockholm, Sweden}
\affiliation[g]{Instituto de F\'isica Corpuscular, IFIC-UV/CSIC, Valencia, Spain}
\affiliation[h]{Center for Astrophysics and Cosmology, University of Nova Gorica, Vipavska 13, SI-5000 Nova Gorica, Slovenia}

\abstract{
The Galactic Center Excess (GCE) in GeV gamma rays has been debated for over a decade, with the possibility that it might be due to dark matter annihilation or undetected point sources such as millisecond pulsars (MSPs). This study investigates how the gamma-ray emission model ($\gamma$EM) used in Galactic center analyses affects the interpretation of the GCE's nature.
To address this issue, we construct an ultra-fast and powerful inference pipeline based on convolutional Deep Ensemble Networks. We explore the two main competing hypotheses for the GCE using a set of $\gamma$EMs with increasing parametric freedom. We calculate the fractional contribution ($f_{\mathrm{src}}$) of a dim population of MSPs to the total luminosity of the GCE and analyze its dependence on the complexity of the $\gamma$EM. For the simplest $\gamma$EM, we obtain $f_{\mathrm{src}} = 0.10 \pm 0.07$, while the most complex model yields $f_{\mathrm{src}} = 0.79 \pm 0.24.$
In conclusion, we find that the statement about the nature of the GCE (dark matter or not) strongly depends on the assumed $\gamma$EM.
The quoted results for $f_{\mathrm{src}}$ do not account for the additional uncertainty arising from the fact that the observed gamma-ray sky is out-of-distribution concerning the investigated $\gamma$EM iterations. We quantify the reality gap between our $\gamma$EMs using deep-learning-based One-Class Deep Support Vector Data Description networks, revealing that all employed $\gamma$EMs have gaps to reality.
Our study casts doubt on the validity of previous conclusions regarding the GCE and dark matter, and underscores the urgent need to account for the reality gap and consider previously overlooked ``out of domain'' uncertainties in future interpretations.}

\begin{document}

\maketitle

\section{Introduction}
\label{sec:intro}

The all-sky gamma-ray emission in the GeV-TeV range contains a treasure trove of information on the Galactic interstellar medium, the Galactic cosmic ray (CR) population and CR accelerators. It primarily originates in the interstellar emission (IE), produced by interactions between CRs and the interstellar medium, in addition to the contribution of a large number of individual sources. This emission was mapped in detail in the GeV energy range by the Large Area Telescope (LAT) onboard the {\em Fermi Gamma-Ray Space Telescope} \cite{Ackermann:2012pya}, operating over the past decade, and at TeV energies, by the High-Altitude Water Cherenkov Gamma-Ray Observatory (HAWC) \cite{Zhou:2017lgv} and High Energy Stereoscopic System (H.E.S.S.) \cite{Aharonian:2006au, Abramowski:2014vox, Abdalla:2017xja}
telescopes. Some of the major issues, limiting the amount of information that was so far extracted from the data, are the relatively poor angular resolution of the detectors, limited statistics of the observations, as well as degeneracy between various emission components that is challenging to break with traditional analysis methods. 

A prime example of a sky region that is particularly prone to these challenges is the $\mathcal{O}$($10^\circ$) region surrounding the Galactic center (GC). The Galaxy is mostly transparent to photons in this energy range except at the highest energies \citep{2006ApJ...640L.155M, 2018PhRvD..98d1302P}, and the emission is therefore integrated over the entire diameter of the Galaxy, likely making it the most complex region of the sky.  This includes the GC which houses the supermassive black hole SgrA$^{\ast}$ \cite{Genzel:2010zy} and has the highest star formation rate (SFR) 
in the Galaxy \cite{Barnes_2017}, both of which are expected to impact the gamma-ray emission. In particular, the population of CRs could differ from that in the rest of the Galaxy (due to, e.g., freshly injected CRs \cite{Petrovic:2014uda}); their diffusion could be position-dependent \cite{Gaggero:2017jts} and strong Galactic winds could be present in  this region \cite{Carlson:2016iis}. Due to the high SFR a large number of individually unresolved sources is also expected to be  present in the region, with their cumulative emission overlapping and possibly mimicking other emission components. The base of the Fermi bubbles (FB), a large structure centered at the GC \citep{2010ApJ...717..825D, TheFermi-LAT:2017vmf}, is also expected to contribute to the measured emission, though the bubbles are challenging to detect close to the Galactic plane. In addition, the GC is expected to be the brightest place of a signal of dark matter (DM) particle pair-annihilation, at the verge of the sensitivity of current telescopes for WIMP DM models. 

\subsection{Understanding the GeV gamma-ray emission from the GC -- a brief review}
\label{sec:gce_short_review}

Excess emission over standard astrophysical backgrounds was detected in this region and dubbed the Galactic center excess (GCE).  First discovered in 2009 \citep{2009arXiv0910.2998G}, this excess peaks at $1-3$ GeV with a spatial extension of $\sim10^{\circ}$ (see also \cite{Vitale:2009hr, Hooper:2010mq,Abazajian:2010zy,Hooper:2011ti,Abazajian:2012pn,Hooper:2013nhl,Gordon:2013vta,Macias:2013vya,Daylan:2014rsa,Abazajian:2014fta,Zhou:2014lva,Huang:2015rlu,Fermi-LAT:2015sau}), is robust against variations in the IE models (IEMs) \cite{Calore:2014nla}, and is nearly symmetric around the GC \cite{Calore:2014xka,TheFermi-LAT:2017vmf,DiMauro:2021raz,Cholis:2021rpp, McDermott:2022zmq}.
Possible explanations for the GCE include a dim population of point sources, in particular, millisecond pulsars (MSPs) \cite{Abazajian:2010zy, Abazajian:2012pn, Macias:2013vya, Bartels:2015aea, Lee:2015fea, Ploeg:2017vai, Eckner:2017oul, Fragione:2017rsp, Fragione:2018jxd, Gonthier:2018ymi, Ploeg:2020jeh}, enhanced gamma-ray emission due to cosmic-ray interactions in dense molecular clouds \cite{deBoer:2017sxb, Carlson:2014cwa} or Galactic winds/past transient event(s) \cite{Petrovic:2014uda, Gaggero:2015nsa}. Intriguingly, the GCE is as well as consistent with the expectations for thermal dark matter pair-annihilating into, e.g., quark final states and a mass around $40-60$ GeV while the emission's spatial profile is consistent with an adiabatically contracted Navarro-Frenk-White (NFW) profile \cite{Goodenough:2009gk, Abazajian:2012pn, Macias:2013vya, Daylan:2014rsa, Calore:2014nla}. While there is consensus in the community about the presence of the GCE, its properties are less firmly established. Early works on the subject (e.g., \cite{Gordon:2013vta, Calore:2014xka}) already pointed out that large uncertainties in the diffuse background modeling impact the reconstruction of the GCE's spatial and spectral properties. As illustrated by the summary figure 1 in \cite{Dinsmore:2021nip} that compares GCE spectra obtained in a large number of works, it is not clear whether the GCE's spectrum extends to energies beyond $\sim10$ GeV. There is a recent debate about the preferred spatial morphology of the GCE based on the results of template fitting. The authors of \cite{DiMauro:2021raz, Cholis:2021rpp, McDermott:2022zmq} reconstruct the GCE as entirely consistent with spherical symmetry while -- at least in the latter two analyses -- accounting for the uncertainty of the IE by varying the assumptions on cosmic-ray propagation parameters. In contrast, \cite{Pohl:2022nnd} demonstrates that re-simulating the gas distribution in the GC region leads to a strong preference for a GCE following the asymmetric stellar density of the Galactic bulge, without the need for DM as claimed in \cite{DiMauro:2021raz}. The preference for the stellar bulge morphology has been pointed out many times in the literature by independent groups and employing diverse inference methods \cite{Macias:2016nev, Bartels:2017vsx, Macias:2019omb, Abazajian:2020tww, Calore:2021jvg}.

The preference for a stellar bulge component can be interpreted as corroborative evidence for the point-source nature of the GCE. The stellar mass of the Galactic bulge of about $10^{10}\;M_{\odot}$ \cite{Cao:2013dwa, Portail:2016vei} is dominated by old stellar populations (> 5 Gyr). The innermost $\sim 200$ pc of the MW feature an additional very-high-density region of stars, the nuclear stellar cluster \cite{Launhardt:2002tx, Portail:2016vei}. Due to their age or stellar density, these components of the MW are ideal places for the formation and presence of MSPs \cite{1998MNRAS.301...15D, Hui:2010vt, Mirabal:2013rba, Eckner:2017oul}. However, the question of the point-source nature of the GCE has commonly been treated fully decoupled from the question of the preferred spatial profile of the GCE. Early works that leveraged the discriminating power of small-scale photon clustering in the residual gamma-ray emission of the excess employed wavelet transforms \cite{Bartels:2015aea,Bartels:2016uxz} or, so-called, non-Poissonian template fitting (NPTF) \cite{Lee:2015fea}. Both of these seminal papers found evidence for the point-source nature of the GCE. Yet, the results of \cite{Bartels:2015aea} were later questioned by an independent study utilizing a dataset with a larger exposure and an updated gamma-ray source catalog \cite{Zhong:2019ycb}. The initial results of the NPTF method were followed by further analyses using the same approach, which cast doubt on the robustness of the technique. In \cite{Leane:2019xiy} an independent group published its results that the NPTF approach always reconstructs an injected DM-like signal -- with an overall luminosity below a certain threshold -- as point-source-like emission. These claims were rebutted in \cite{Chang:2019ars, Buschmann:2020adf} showing explicitly that diffuse background mis-modeling can induce the observed behavior of the NPTF method. Employing instead an improved  diffuse background model leads again to the conclusion that the GCE is, at least partially, made out of point sources. The response to these rebuttals was given in form of two publications \cite{Leane:2020nmi, Leane:2020pfc} showing that an unmodeled asymmetry of the GCE emission can induce spurious evidence for the point-source nature of the excess. 

These recent developments in exploring the limitations of frequently employed inference techniques, in particular template fitting and NPTF, revealed that background mis-modeling plays a crucial role in examining the nature of the GCE. To mitigate these uncertainties, several schemes have been proposed and applied in the literature: Introducing nuisance parameters on small- and large angular scales via spherical harmonic marginalization \cite{Buschmann:2020adf}, varying diffuse templates via Gaussian processes \cite{Mishra-Sharma:2021oxe} or adaptive template-fitting with \texttt{SkyFACT} \cite{Storm:2017arh, Bartels:2017vsx} that combines template fitting with image reconstructions techniques. The latter method is utilizing a large number of nuisance parameters $\mathcal{O}(>10^5)$ and associated regularization terms to prevent over-fitting and to optimize the compiled background model in a data-driven way. The authors of \cite{Calore:2021jvg} employed a \texttt{SkyFACT}-optimized IEM together with the 1-point probability distribution function (1pPDF) technique \cite{Zechlin:2015wdz} to discern the contribution of sub-threshold sources to the gamma-ray emission of the GC. The analysis showed that the 1pPDF's sensitivity to dim sources largely depends on the chosen IEM. However, the \texttt{SkyFACT}-optimized model is consistent with 4FGL catalog sources and it exhibits the highest sensitivity to point sources below the \textit{Fermi}-LAT detection threshold. 


Another means to extract the small-scale photon clustering in images of the GC is via the help of deep learning and Convolutional Neural Networks (CNNs).
A range of new analysis methods was proposed to address this question \cite{Caron:2017udl,List:2020mzd,Mishra-Sharma:2020kjb,List:2021aer}, with mixed results. Our previous work \cite{Caron:2017udl} pioneered the application of CNN to this region of the sky and provided a proof of principle that such methods could indeed successfully discern between the genuinely  diffuse and point source origin of the emission (albeit in a rather limited  setting). Building on that effort, we take a significantly more  ambitious approach in this work, improving both the network architecture and the modeling of astrophysical emission, as detailed below. 

\subsection{The issue of discrepancies between model and reality}

As discussed earlier, one of the biggest general issues when dealing with the GCE is that its properties strongly depend on our knowledge (or lack thereof) of the bright fore-/background emission originating from the rest of the Galaxy. While we have a good broad description of the IE, detailed knowledge of the distribution of CRs and their targets (gas and radiation) near the GC is imperfect. This could be dealt with by either improving our models via observational data or advanced inference techniques (see, e.g., \cite{2021A&A...655A..64M, Mertsch:2022oee, Karwin:2022xgn, Shmakov:2022vuc} or by adding flexibility to the IEM to parameterize away our imperfect knowledge. Yet, there is always a balance between allowing for too much freedom, which may explain any data, and not allowing for enough, which results in spurious deficits and excesses that require additional, and possibly erroneous, model components.  This problem is inherent to both the more traditional maximum likelihood and other methods (see e.g. \cite{Leane:2020nmi,Buschmann:2020adf}) as well as in our application of CNNs which rely on a gamma-ray emission model (\gem) and simulations of reality from it. 

The difficulty of harmonizing our incomplete knowledge of the full complexity of observational data with the real world is often called the ``reality gap'' and, as a result, a DM or other components might be needed to describe the data only due to a wrong or missing component in the \gem. 

\begin{figure}[t]
\includegraphics[width=\textwidth]{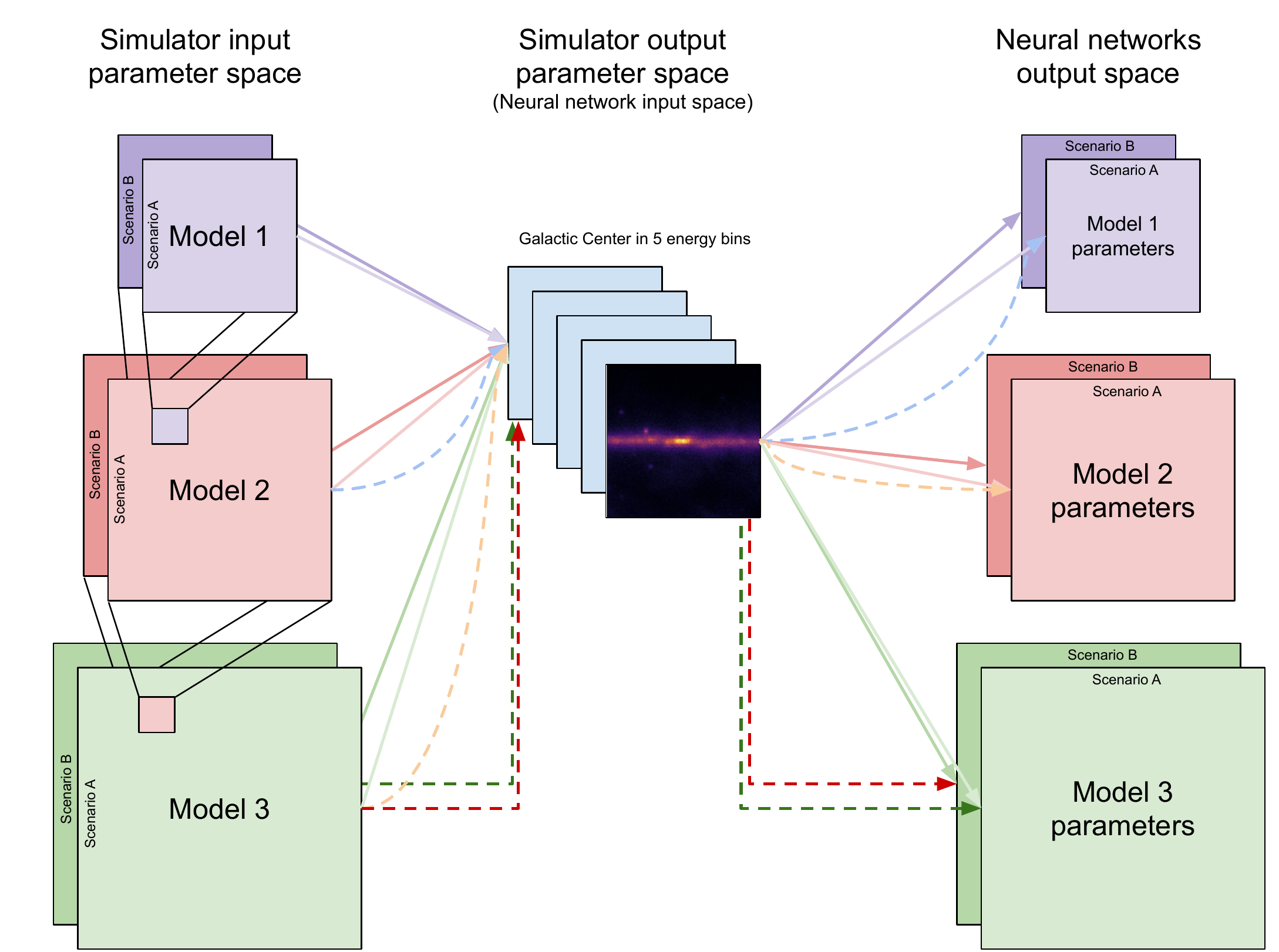}
\centering
\caption{Visualisation of domain adaptation. The background model to generate the GC images is expanded in dimensionality to make more complex (and hopefully more accurate models). Model 1 is embedded into Model 2, which is embedded in Model 3. Additionally, we define two scenarios with separate parameter spaces but following the same domain adaptation procedures. This results in six different background models. The output space of all background models is always the same: a GC image with 5 energy bins. In each respective case, this output is the input of the neural networks that predict the input parameters of the models. For every background model, we trained a neural network. Hence, there is a ``Model 1, scenario A network'' which is trained on images generated from background Model 1, Scenario A, and also for Model 2 and 3 and Scenario B. Additionally, it is possible to apply a network trained on simulations from Model 1 on simulations from Model 2 for example. The arrows represent all experiments we have done: the six straight arrows represent networks that are evaluated and trained on the same background model. The curved dashed arrows represent experiments where the neural network of a trained model is used to predict parameters from simulations of a more complex model (i.e.~a network trained on Model 1 which is applied on Model 2, and the same for Model 2 and 3). The elbow-connected dashed lines represent experiments where a neural network trained on Scenario A is applied to Scenario B and the other way around.}\label{fig:domainadaptation}
\end{figure}

In this work, we aim at answering the question of the nature of the GCE by leveraging the small-scale photon clustering in gamma-ray data of the GC. At the same time, we investigate how much the background model complexity influences the results of our study. To this end, we compile three different \gem~that are supposed to simulate the GC region. We dub these instances Models 1, 2 and 3, 
which have 6, 17 and 24 free parameters respectively. {Note that Model 1A -- the nomenclature becomes clear in the next paragraph -- is chosen to be similar to the \gem~adopted in Ref.~\cite{Caron:2017udl}, allowing us to make a direct comparison with our previous work.}

The main idea that we follow is to start with \gem s exhibiting a minimal number of parameters and then gradually increase the complexity of the model through human intervention motivated by inspecting the residuals of the resulting best fit, thus also testing the robustness of our determination of the GCE properties when varying the background model. 
In machine learning (ML) language such approaches are known as a change of the ``domain'' (or ``domain shift''). Methods known as ``domain adaptation'' aim to minimize the difference between two domains,
so that models trained in a source domain can generalize (and make correct predictions) to a target domain.
A typical example is that an ML inference model for road signs trained from US images (e.g. a particular GCE simulation) may not be applicable to European domains (the real data) and may yield incorrect results that uncertainties may not entirely reflect.
Here both data sets (simulated data from a compiled \gem~and real data) share the same feature space: diffuse emission following the distribution of gas and the structure of radiation fields in the Galaxy, bright and dim gamma-ray point sources, the Fermi Bubbles and the excess emission in the GC.
In particular, we try to test in what way such a gap between simulations and reality is causing different interpretations of the GCE.
Testing domain changes uses here (training) data in Simulation Model 1 to solve the task of interpreting the GCE in a different domain (e.g. Simulation Model 2). In our case Model 1 is nested in Model 2 etc., see Fig. \ref{fig:domainadaptation}.  Therefore, to adapt to the different domains, we try to learn from the more general \gem~a generalized distribution that is also applicable to the sub-model (and hopefully to the real data).

In addition, we consider two independent (non-nested) \gem-{\it sets}, dubbed Scenario A and B (see Fig. \ref{fig:domainadaptation} where scenario B is presented as `parallel' to Scenario A ). The two scenarios are motivated by uncertainties in the modeling of the dim point source components and are therefore not nested. Such an approach, of slowly adding parameters until a \gem~is found that leads to featureless residuals, is complementary to techniques like the approach of \texttt{SkyFACT}.
 
\subsection{Contribution to the field}

Our work presented here improves, extends and complements our previous proof-of-principle study \cite{Caron:2017udl}. In addition, we advance the current state-of-the-art of GCE analyses in the following aspects:
\begin{itemize}
    \item We create a series of increasingly complex template-based $\gamma$EMs for the GC region, adjusting the flexibility of the IEM parameters while modeling the GCE with a combination of DM and MSP templates. We generate Monte Carlo MSP populations using observation-based priors \cite{Bartels:2018xom}.
    \item We develop a high-speed inference pipeline using Deep Ensemble Networks (DENs) that reconstructs best-fitting model parameters and associated uncertainties for Monte Carlo data or real \textit{Fermi}-LAT gamma-ray sky observations. We demonstrate that the network can effectively separate gamma-ray data into emission components, accurately determining the spatial shape and total flux of the GCE.
    \item We discover that the fractional contribution ($f_{\mathrm{src}}$) of a dim MSP population to the GCE's total luminosity depends on the complexity of the $\gamma$EM used, indicating a reality gap between our models and real data. We employ One-Class Deep Support Vector Data Description (Deep SVDD) networks to quantify this reality gap, providing a powerful way to determine if the \textit{Fermi}-LAT gamma-ray sky is out-of-distribution for a specific $\gamma$EM. We conclude that a reality gap persists across all $\gamma$EMs explored, leading to an additional out-of-distribution uncertainty on $f_{\mathrm{src}}$ predictions not captured by our networks. This uncertainty is likely to be present in most previous GCE studies.

\end{itemize}


\subsection{Outline}

This paper is organized as follows:  In Sec.~\ref{trainingdata} we elaborate on the LAT data selection and we describe the building blocks of our \gem s, detailing how the simulations are performed and how the first model iteration is constructed. In Sec.~\ref{sec:NN} we describe the inference techniques utilized in this work, i.e.~DENs and a likelihood-based approach. Sec.~\ref{sec:results_model1} states our results from the first step of the simulation with the simplest \gem~that we introduced in Sec.~\ref{trainingdata}. The following Sec.~\ref{sec:discusion} is dedicated to a discussion of the approach leading to an extension of our initial \gem. We describe the construction of the more complex \gem~iterations Model 2 and 3 and we present the results obtained with respect to each of them. In Sec.~\ref{sec:robustness} we assess the robustness of the results presented in the previous sections. We initiate Sec.~\ref{subsec:likelihood} by describing how the likelihood-based analysis compares to the ML approach. Subsequently, in Sec.~\ref{sec:crossdomain-maintext}, we engage in a more detailed exploration of the reality gap. The paper is concluded in Sec.~\ref{sec:conclusion}. In addition, we give details about some of the analysis aspects in appendices: In Appendix \ref{app:param_comparison_rest} we summarize the predictions of all six neural networks regarding the astrophysical emission components. Appendix \ref{app:msp_threshold} offers a more detailed discussion of the physical consequences of MSP scenario B in terms of sources below and above the \textit{Fermi}-LAT detection threshold. We provide in Appendix \ref{app:residuals} the full list of spatial and spectral residuals according to the DNN results. Lastly, Appendix \ref{sec:diffuse_bkg_check} supplements Sec.~\ref{sec:crossdomain-maintext} by stating the results obtained for those \gem s and scenarios not discussed in the main text.

\section{Data and gamma-ray emission models} \label{trainingdata}

\subsection{Fermi data selection}

We use Pass 8 (R3) data \cite{2013ApJ...774...76A,2018arXiv181011394B} taken from the time interval from the 4th of August 2008 to the 2nd of April 2018 ($\sim$10 years). To reduce the residual CR background and increase the resolution, we select events in the ULTRACLEANVETO event class of the FRONT type.  We also apply the standard zenith angle cut ($<100^\circ$) as well as the time cut filters DATA$\_$QUAL==1 $\&\&$ LAT$\_$CONFIG==1. All \emph{Fermi}-LAT data selection, preparation, simulation and manipulation is conducted via the \emph{Fermi Science Tools}\footnote{\url{https://github.com/fermi-lat/Fermitools-conda}} (version 2.0.8).

Our Region Of Interest (ROI) is a $30^\circ \times 30^\circ$ square centered on the Galactic center and we bin it using square spatial bins of size $0.25^{\circ}\times0.25^{\circ}$ 
using the plate carr\'ee projection\footnote{The size of the ROI we chose is constrained  by the requirement of having a good angular resolution while keeping the size of training images feasible for our analysis (see Section \ref{sec:NN}). From a physics perspective,  a larger region could also potentially over-constrain some of the components, if their spatial morphology is incorrectly modeled closer to the GC.}. We consider the energy range 500 MeV - 500 GeV, which we split into 5 energy bins (0.5-1 GeV, 1-2 GeV, 2-7 GeV, 7-20 GeV, 20-500 GeV). We do not use data below 500 MeV because at lower energies the angular resolution, as determined by the 68\% containment of the point spread function (PSF), becomes worse than $1^\circ$. As the GCE emission peaks around 2 GeV we chose finer binning around that energy and we combine energy bins above 20 GeV in a single bin, due to the lower statistics at higher energies. 

The energy bins outlined above have been obtained by resuming a finer native energy binning of 30 logarithmically spaced bins. This resummation prescription is also applied to all simulated data in order to properly sample the LAT's instrument response functions. Moreover, we use energy dispersion information for the simulation of our templates by using the flag \texttt{edisp\_bins=-1}. This flag means that the energy dispersion correction only operates on the input spectra by adding an extra energy bin (of a width consistent with the initial specification) to the lower and upper end of the originally defined energy range.

\subsection{Emission components and Model 1 set up}

\subsubsection{Conventional astrophysical gamma-ray emission components}
\label{sec:astro_models}

Modeling the diffuse emission in the region  of the Galactic center is a complex task \cite{TheFermi-LAT:2017vmf,Calore:2014xka,Buschmann:2020adf}. Here we adopt one of the IEMs from Fermi LAT’s 1st SNR catalog \cite{Acero:2015prw} that was used to estimate the systematic uncertainty of the IEMs in that work.  The models are all based on GALPROP calculations that are detailed in \cite{Ackermann:2012pya} but have been adjusted to match the LAT data. 
The chosen IEM is based on a distribution  of the CR sources which is assumed to be traced by the distribution of pulsars (Lorimer \cite{2006MNRAS.372..777L}) with the height of the CR propagation halo $z = 10$ kpc and spin temperature of atomic hydrogen gas $T_S = 150$ K.  In this model, the gas-related gamma-ray emission (CO and HI components) is split into four Galactocentric rings (0–4 kpc, 4–8 kpc, 8–10 kpc and 10–30 kpc), allowing for additional freedom in the IEM by scaling them independently. {In addition, these IEMs include an inverse Compton (IC) gamma-ray component induced by leptonic cosmic rays colliding with the cosmic microwave background and interstellar radiation fields.}

The IEM is supplemented by an isotropic gamma-ray background model (chosen w.r.t.~to the \textit{Fermi}-LAT IRF file (P8R3$\_$ULTRACLEANVETO$\_$V2) and conversion type (FRONT)\footnote{iso$\_$P8R3$\_$ULTRACLEANVETO$\_$V2$\_$FRONT$\_$v1.txt}. 

\begin{figure}[t]
\centering
\includegraphics[width=0.6\linewidth]{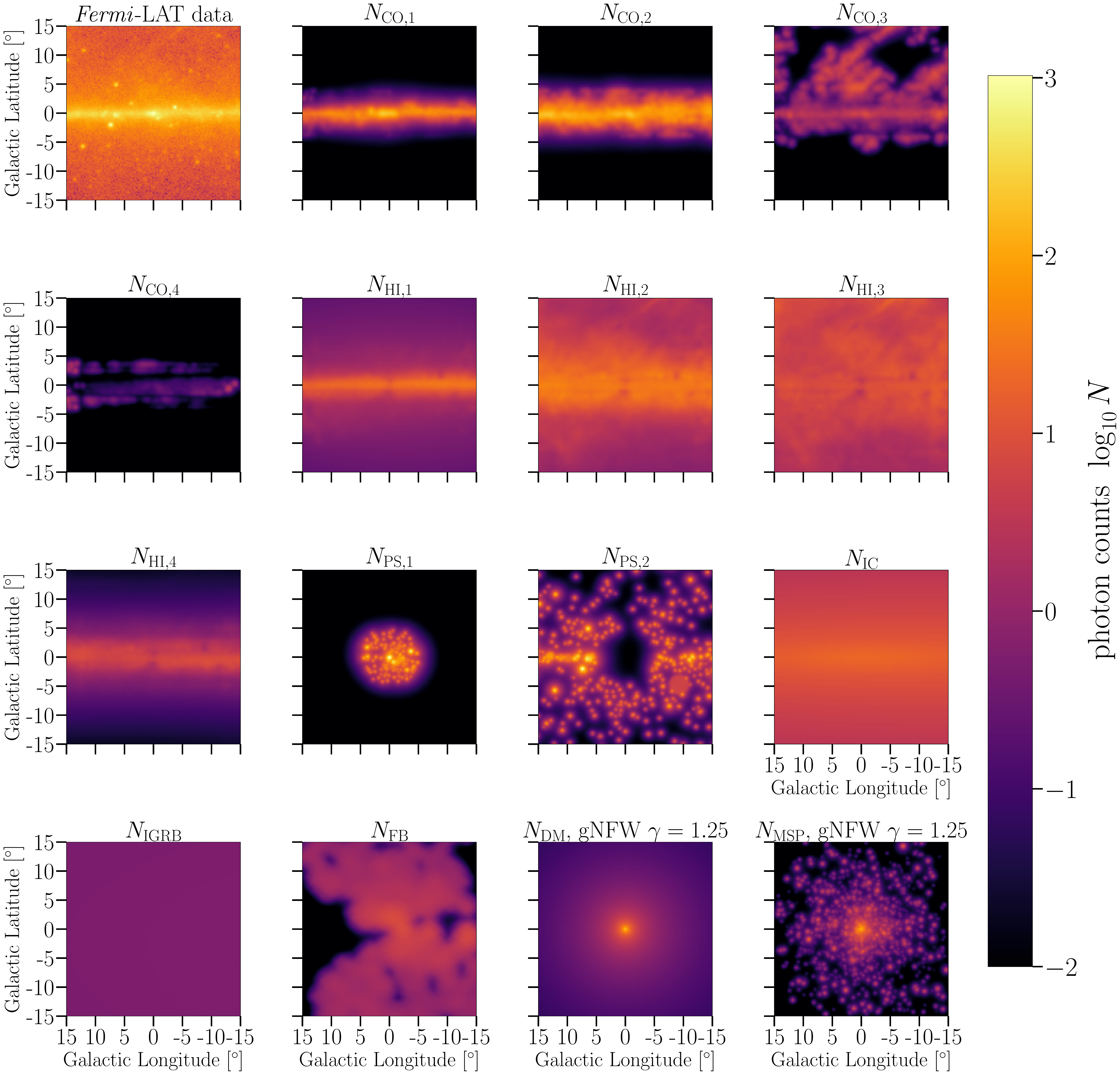}
\includegraphics[width=0.55\linewidth]{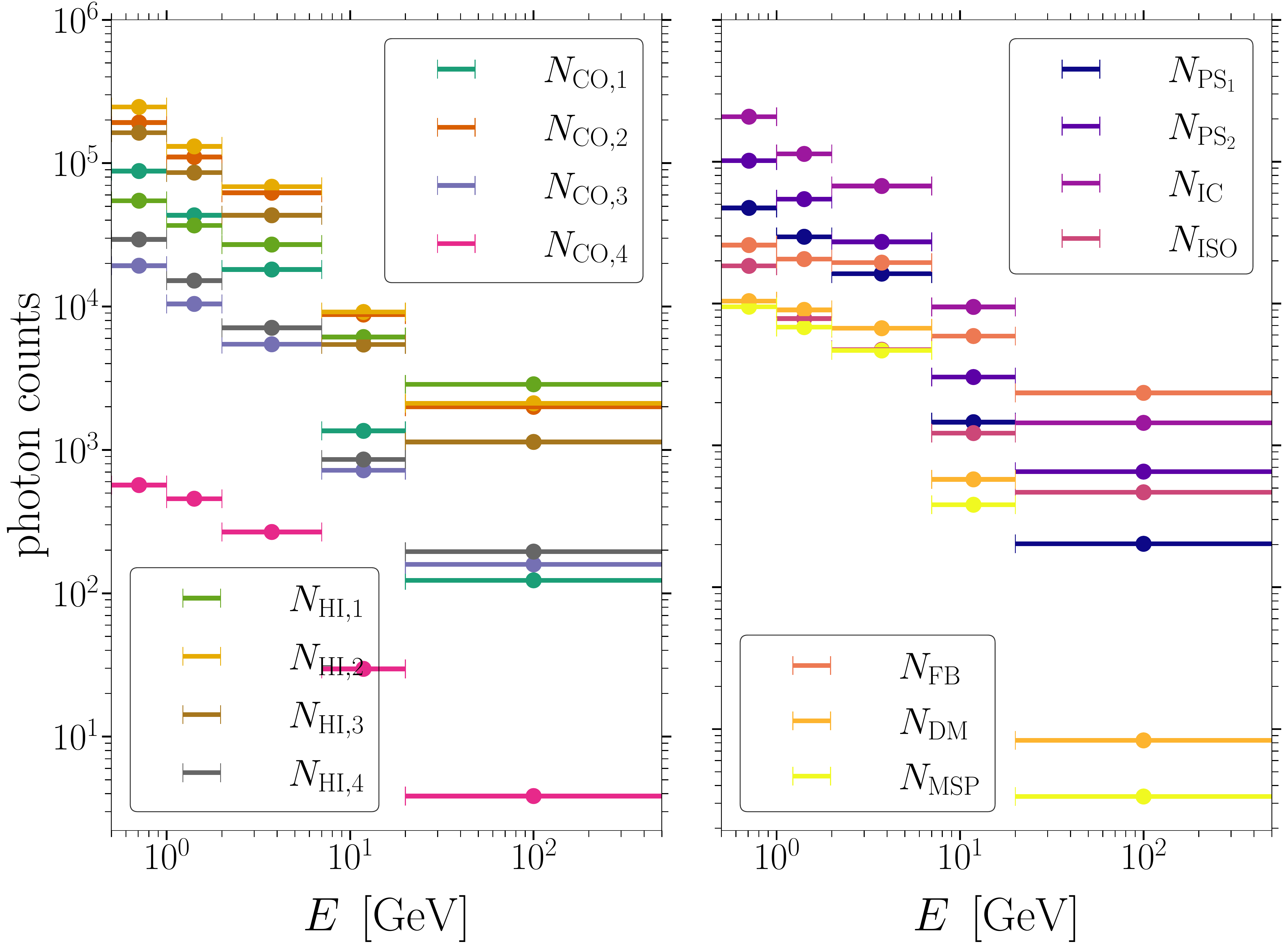}
\caption{Compilation of the spatial (\emph{upper panel}) and spectral (\emph{lower panel}) morphology of all gamma-ray templates used in our baseline setup to model the gamma-ray emission in the GC region. The upper panel's first image displays the \emph{Fermi}-LAT data in our ROI between 1 GeV and 2 GeV, which is the same energy bin chosen for the remaining templates. The templates are the output of the Fermi Science Tools routine \texttt{gtmodel} and hence display the expected events from the respective flux model for the given \textit{Fermi}-LAT observation time in the infinite statistics limit. The color indicates the base-10 logarithm of the number of expected gamma-ray events per spatial pixel. The spectral properties of the DM and MSP templates follow the best-fit results for Model 2A as stated in the text (c.f.~Fig.~\ref{fig:summaryplot}, which fixes $\gamma=1.25$ in Eq.~\ref{eq:gNFW_profile}. The adjacent MSP template is based on the same spatial profile whereas the spectral parameters read $\sigma_L = 0.76$ and $F_{\mathrm{MSP}}= 4.1\times10
^{-7}\;$ph/cm$^2$/s.} \label{fig:templates}
\end{figure}

In addition to the IE and the isotropic background, we include emission from the base of the Fermi bubbles \cite{Herold:2019pei}. The template of the non-uniformly spatially structured, data-driven characterization of the bubbles is shown among other components in Fig.~\ref{fig:templates}. This emission is very uncertain close to the Galactic plane and at the same time {can} {have} a significant impact on the results, as already argued in \cite{TheFermi-LAT:2017vmf,List:2020mzd}. We adopt the same \texttt{LogParabola} spectrum \cite{Herold:2019pei}  

\begin{equation}
\label{eq:log_parabola}
\frac{\mathrm{d}N}{\mathrm{d}E} = F_{0}\left(\frac{E}{E_0}\right)^{-\alpha-\beta\ln\!{\left(E/E_0\right)}}
\end{equation} 
for the base of the FBs, with parameters $F_0 = 5\times10^{-10}\;\mathrm{ph}\,\mathrm{cm}^{-2}\,\mathrm{s}^{-1}\,\mathrm{MeV
}^{-1}$, $\alpha$ = 1.6, $\beta$ = 0.09, $E_0$ = 1 GeV.

We also explicitly include all sources listed in the fourth \textit{Fermi}-LAT source catalog DR2 (4FGL-DR2) \cite{Ballet:2020hze}, which is  built using the same data period we adopt here. We model sources within a radius of 20$^\circ$ around the GC via two templates; one covering sources within a radius of 5 degrees from the GC and another incorporating the rest of the sources. We use the best-fit 4FGL-DR2 source spectra as given in the catalog as our baseline. Note that we do not supplement the 4FGL-DR2 sources with contributions from the unresolved part of the respective population, which would add to the diffuse background modeling. Such a contribution was studied in \cite{Fermi-LAT:2015bhf}, where it was shown that  it could be sizeable, though its  properties are highly uncertain. To answer our initial science question, we however include the contribution from unresolved MSPs in the GC region (as described below).

\begin{figure}[t]
\centering
\includegraphics[width=0.49\linewidth]{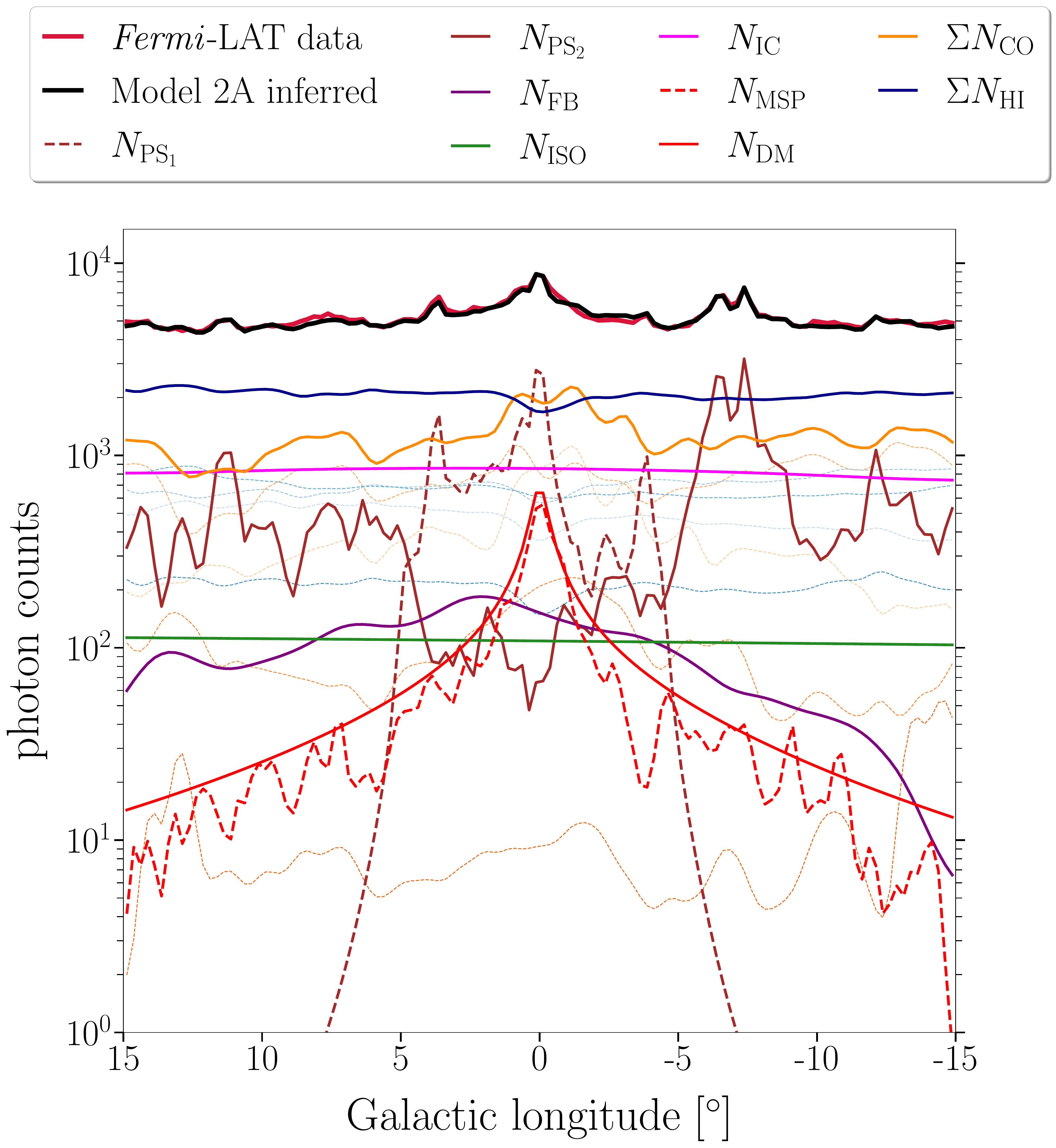}
\hfill
\includegraphics[width=0.49\linewidth]{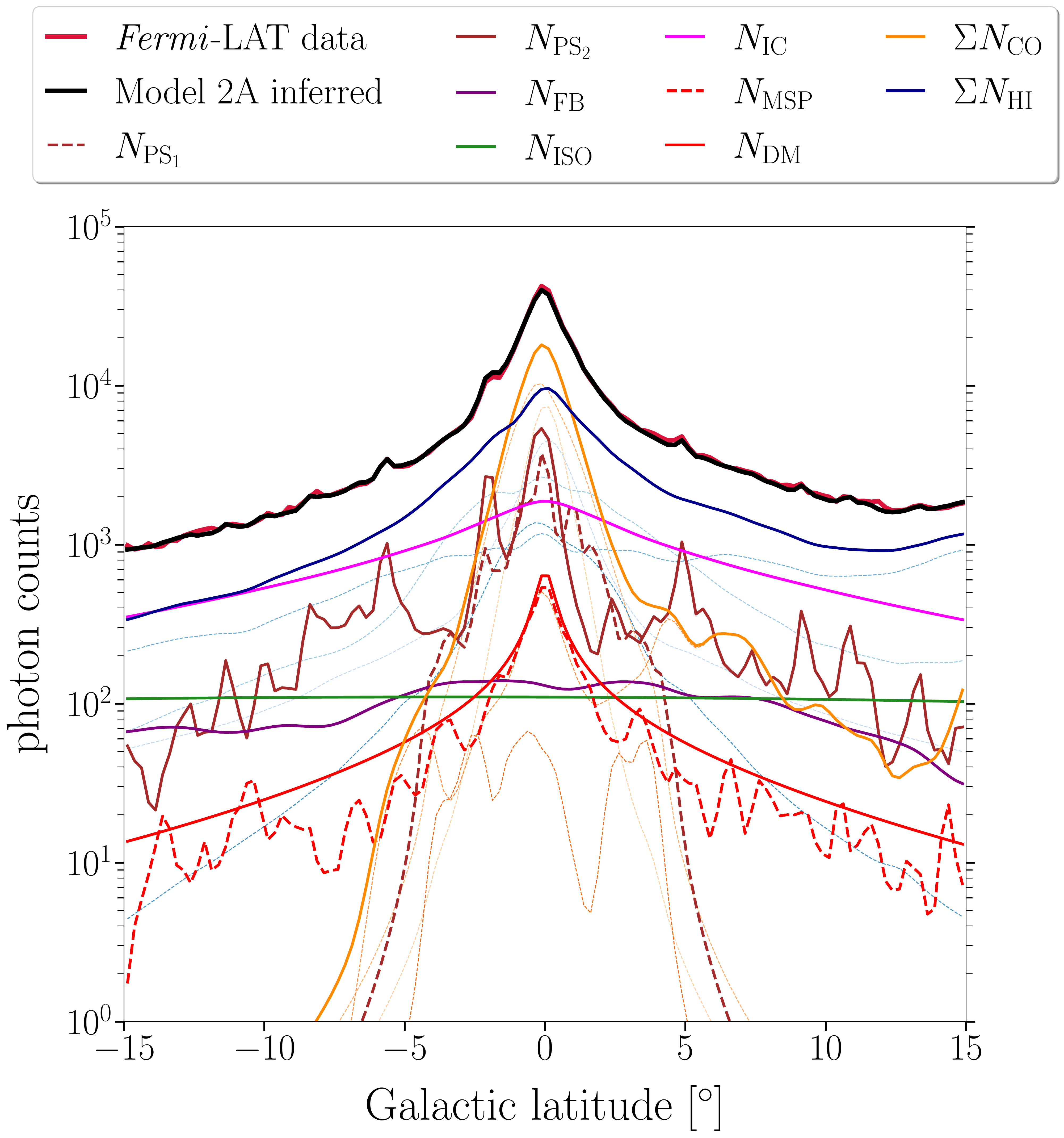}
\caption{Longitude (\emph{left panel}) and latitude profile (\emph{right panel}) of the selected \emph{Fermi}-LAT data set and the gamma-ray templates used in our baseline setup to model the gamma-ray emission in the GC region. The templates are the output of the Fermi Science Tools routine \texttt{gtmodel} and hence display the expected events from the respective flux model for the given \textit{Fermi}-LAT observation time in the infinite statistics limit. The total contribution of all four gas rings (CO, orange) and (HI, blue) is shown is a thick line, while individual gas rings are displayed as dashed lines in shades of the respective color. For definiteness, we have selected the second energy bin (1 GeV to 2 GeV) for all shown components. The relative normalisation parameters of all \gem templates as well as the spectral properties of the DM and MSP templates in particular follow the best-fit results for Model 2A as stated in the text (c.f.~Fig.~\ref{fig:results_scenarioA_model2}, which fixes $\gamma=1.25$ in Eq.~\ref{eq:gNFW_profile}. The adjacent MSP template is based on the same spatial profile whereas the spectral parameters read $\sigma_L = 0.76$ and $F_{\mathrm{MSP}}= 4.1\times10
^{-7}\;$ph/cm$^2$/s. For comparison, we show as a black line the longitude and latitude profiles of the sum of all templates according to their inferred values in Model 2A. \label{fig:lon_lat_profiles}}
\end{figure}

These templates are part of the definition of our {\it Model 1-3} setups. We show the data's and templates' two-dimensional spatial morphology and spectra in Fig.~\ref{fig:templates} while Fig.~\ref{fig:lon_lat_profiles} displays the templates' longitude and latitude profiles in comparison to the selected gamma-ray data set. We comment on the exact composition of our \gem s in Sec.~\ref{sec:model1} (Model 1), Sec.~\ref{subsec:extended} (Model 2) and Sec.~\ref{sec:newFB} (Model 3) guided by the principle of increasing the \gem's complexity after inspecting the performance of the neural network in terms of its parameter prediction. 


\subsubsection{Modelling of the GCE} 
\label{sec:GCEdata}

On top of the conventional astrophysical emission, we also add templates for the GCE. As we aim to distinguish between the DM  and MSP (point source) origin of the GCE we prepare two sets  of  GCE templates that either reflect the expectations for DM pair annihilation or a population of faint millisecond pulsars in the Galactic center. 
We assume the spatial profile of the GCE emission is identical for both the DM and the MSP templates.
We do this in an  attempt to keep the number of parameters minimal  and focus on the algorithm's performance to distinguish between truly diffuse and templates comprised of many point sources, all other ingredients being equal. We note that this assumption 
might not be realistic. Indeed, there are indications that the morphology of the GCE follows the box-shaped  bulge emission closely, which would be a strong argument  for the MSP origin of the excess \cite{Bartels:2017vsx, Storm:2017arh}, however, we leave the exploration of this effect for follow-up work. Despite being a significant signal in \textit{Fermi}-LAT data of the GC, mismodeling of the GCE has only a minor impact on the overall severity of the reality gap of our \gem s as it is mainly created by our insufficient knowledge of the bright astrophysical components. Hence, we impose spherical symmetry of the GCE and test whether the resulting signal should preferably be smooth or granular given the \textit{Fermi}-LAT data set as done in other works applying ML to the GCE puzzle \cite{Mishra-Sharma:2020kjb,List:2020mzd, List:2021aer}.

\vspace{12pt}

\noindent {\bf Spatial profile.}
Motivated by the results of previous studies of this region we adopt a spherically symmetric profile and utilize the so-called generalized NFW DM density profile \cite{Navarro:1995iw,Navarro:1996gj,Zavala:2019gpq} for the density distribution, 

\begin{equation}
\label{eq:gNFW_profile}
\rho_{\mathrm{gNFW}}(r) =  \frac{\rho_s}{(\frac{r}{r_s})^\gamma (1+\frac{r}{r_s})^{3-\gamma}},
\end{equation}
where $r_s = 20$ kpc, and $\rho_s$ is such that $\rho_{\mathrm{gNFW}}(r_\odot) = 0.4$ GeV cm$^{-1}$ at $r_\odot = 8.5$ kpc and we treat the inner slope $\gamma$, as a free parameter. Previous fits to the GCE have resulted in best-fit values $\gamma \in \left[1.1, 1.3\right]$ \cite{Calore:2014xka, DiMauro:2021raz, Cholis:2021rpp}, however, we use a significantly wider range of this parameter $\gamma \in \left[0.8, ~1.3\right]$ to avoid  potential biases. We  do that guided by the preliminary work \cite{Caron:2017udl}, which demonstrated the sensitivity of the outcomes with respect to the $\gamma$ parameter.  

\vspace{12pt}

\noindent {\bf Dark matter template.}
For the DM gamma-ray spectrum, we assume the $b {\bar b}$ annihilation channel (as implemented in the Fermi Science Tools table). We  assume a DM particle candidate of mass $m_{\mathrm{DM}} = 42$ GeV and thermally averaged, velocity-weighted self-annihilation  cross section $\langle \sigma v \rangle = 2.2 \times 10^{-26}$ cm$^3$s$^{-1}$ according to the best-fit in \cite{Daylan:2014rsa}. However, we treat $\langle \sigma v\rangle$ as a free parameter by introducing an additional normalization parameter that may vary between $10^{-3}$ and 4 (using a log-uniform prior) to sufficiently sample the expected total GCE luminosity given the assumed range of slope parameters $\gamma$. 

\vspace{12pt}

\noindent {\bf Millisecond pulsar template.}
To generate the MSP template together with the spatial distribution (adopted as above) we need to specify the spectral shape and the  luminosity function of MSPs.\\
{\it MSP gamma-ray spectrum:} We chose the parameter shape as obtained in the analysis of the stacked MSP spectrum derived in \cite{McCann:2014dea},
\begin{equation}
\frac{dF}{dE}= F_0 \left( \frac{E_\gamma}{1 {\rm GeV}} \right)^{- 1.46} \exp{\left(-\frac{E_\gamma}{3.6\,{\rm GeV}}\right)}.
\end{equation}
$F_0$ is  determined by the luminosity of the specific MSP drawn from the MSP luminosity function,  as specified in the following.\\


{\it MSP luminosity function:} We assume a log-normal luminosity distribution (following \cite{Bartels:2018xom}, consistent also with earlier works, e.g. \cite{Fermi-LAT:2017yoi,Ploeg:2017vai,Winter:2016wmy,Hooper:2015jlu,Acero:2015gva}), as 
\begin{equation}
\frac{dN}{dL} \sim \frac{1}{L} \exp \left[ \frac{-\left(\log_{10}L - \log_{10}L_0\right)^2} {2\sigma_L ^2} \right],
\end{equation}
where $L$ is the integrated luminosity in the 0.1 - 100 GeV range, and $\log_{10} L_0 = 32.61$ (as  found as the best fit in \cite{Bartels:2018xom}). We simulate MSPs in the luminosity range $L \in \left[10^{30}, 10^{37}\right]$ erg/s (which encompasses the expected fluxes of objects placed at the Galactic center distance that are well below the detection threshold of the Fermi-LAT). We leave $\sigma_L$ as a free parameter with values $\sigma_L \in [0.3, 1.2]$. The selected range for $\sigma_L$ is larger than the $5\sigma$ confidence interval around the best-fitting value reported in \cite{Bartels:2018xom}. With this broad prior we exhaustively sample the MSP luminosity function as to generate populations that are almost entirely below the \textit{Fermi}-LAT's detection threshold or that produce a large number of bright pulsars. Hence, our training data contains instances of the fringe case where a population of many dim sources is indistinguishable from smooth emission whose importance has been pointed out in \cite{List:2021aer}.
\\
{\it MSP map flux threshold:} The treatment of close-to-threshold MSPs in the training  data  set  is obviously critical for our approach which aims to detect faint (sub-threshold) sources. We, therefore, explore two different ways (scenarios) in which we treat the high flux threshold in our MSP  templates. 

 The main approach adopted in the text ({\bf `Scenario A'}), is to generate MSPs below the flux value defined as the pulsars {\it detection} threshold in the 3PC \textit{Fermi}-LAT catalog  \cite{2019HEAD...1710932L}, rescaled by the relative exposure. In particular, the 3PC catalogue is derived using the 4FGL procedure and data set (8 years of data), while we use the 10-year data in this work. We therefore use the 3PC threshold rescaled with a factor of $\sqrt{10/8}$, 
 resulting  in the flux threshold values in the range $[2.0,4.7]\times 10^{-9}$ GeV s$^{-1}$ cm$^{-2}$, in our region of interest. Since we do not match the MSP luminosity function to the number of unassociated sources in 4FGL, the issue with this approach is that it effectively breaks the luminosity function at the detection threshold, making it impossible for the network to recover the corresponding parameters (e.g. $\sigma_L$) and potentially biasing our results. In addition, the 3PC sensitivity was derived using different data selections, IRFs and diffuse models, and might not be fully applicable to our setup. Given how impactful the treatment of subthreshold sources is expected to be on our results, we also explore an alternative scenario ({\bf `Scenario B'}, further discussed in Appendix \ref{app:msp_threshold}) in  which we do not apply any flux threshold cut in the MSP template maps. The issue with  this approach is that we risk modeling (a portion) of bright sources twice, as some of the bright MSPs might already be part of the unidentified source population in 4FGL. As we will see, the results in both cases are quite similar, strengthening our general conclusions. 

\begin{figure}[t!]
\begin{center}
\includegraphics[width=0.8\linewidth]{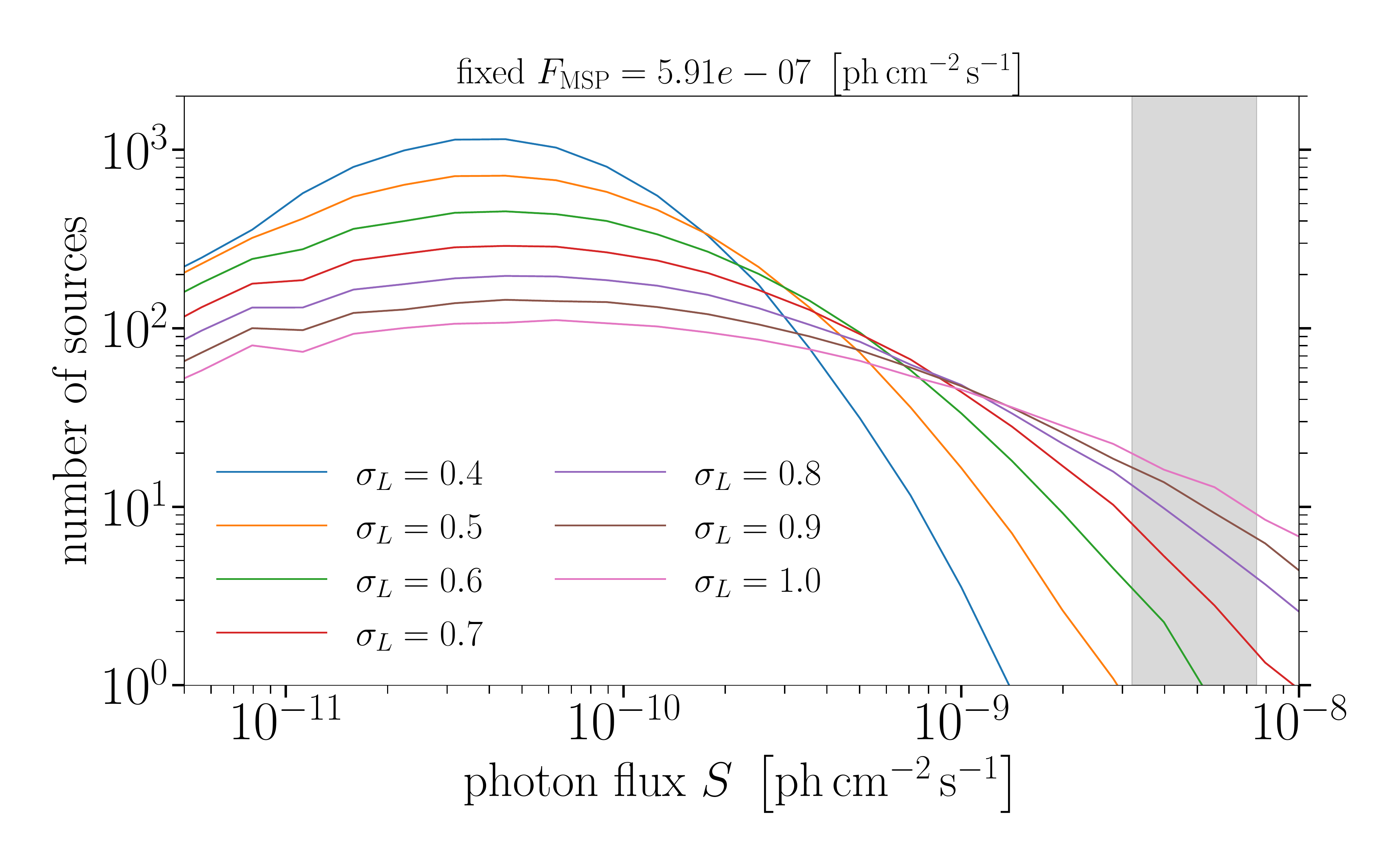}
\end{center}
\caption{Numbers of generated MSPs as a function of their photon flux $S$ (0.1 - 100 GeV) with respect to the GCE parameters $F_{\mathrm{MSP}} = 0.591\times10^{-6}\;\mathrm{ph}\,\mathrm{cm}^{-2}\,\mathrm{s}^{-1}$ and $\gamma = 1.28$. The grey band marks the MSP detection threshold (taken from 3PC catalog and rescaled to our data set).} \label{fig:MSP_populations}
\end{figure}

{\it MSP population generation:} 
We apply the following scheme to generate a realization of individual MSPs:
\begin{itemize}
    \item Randomly draw a total GCE luminosity $F_{\mathrm{MSP}}$ compatible with the GCE flux uncertainty range as visualised by the blue band in Fig.~15 of \cite{TheFermi-LAT:2017vmf} as the cumulative flux of the MSP population. Consequently, the total range of the MSP template's cumulative luminosity is $F_{\mathrm{MSP}}\in\left[0.02, 1.58\right]\times10^{-6}\;\mathrm{ph}\,\mathrm{cm}^{-2}\,\mathrm{s}^{-1}$.
    \item Draw individual pulsars: (i) draw ($\ell$, $b$) from a gNFW profile, (ii) draw an MSP luminosity from the MSP gamma-ray luminosity function, (iii) derive $F_0$ corresponding to the drawn luminosity, (iv) check if the resulting flux is less than the \textit{Fermi}-LAT detection threshold (this step is omitted in Scenario B).
    \item Stop the MSP population generation when the randomly drawn total GCE luminosity is saturated by the cumulative emission from all accepted \emph{subthreshold} MSPs.
\end{itemize}
Fig.~\ref{fig:MSP_populations} illustrates the spread of the expected characteristics of the generated MSP populations. For definiteness, we fix $F_{\mathrm{MSP}} = 0.591\times10^{-6}\;\mathrm{ph}\,\mathrm{cm}^{-2}\,\mathrm{s}^{-1}$ and $\gamma = 1.28$ to display the number of sources per photon flux $S$ (0.1 GeV - 100 GeV) for multiple values of $\sigma_L$. The shown source-count distributions highlight the fact that our choice of priors guarantees to generate enough variation in the MSP populations. The figure illustrates that for a sufficiently large part of the $\sigma_L$ parameter space, there is a significant number of MSPs just below the MSP detection threshold (denoted by the grey band taken from the 3PC catalog in our ROI). At the same time, our approach covers the extreme case of an MSP population consisting entirely of dim sources below the detection threshold, which is almost degenerate with a smooth, DM-like GCE emission. Further details about the generated MSP populations are given in App.~\ref{app:msp_threshold}.

\subsection{Definition and generation of Model 1 training data}
\label{sec:model1}

Our first \gem iteration is comprised of a minimal set of templates derived from those described in the previous Sec.~\ref{sec:astro_models}, i.e.~it exhibits a rather limited number of free parameters,  to closely follow the approach of our proof-of-principle work \cite{Caron:2017udl}. {In what follows, we refer to our \gem s and scenarios as ''Model XY`` where $\mathrm{X}\in\{\mathrm{1}, \mathrm{2}, \mathrm{3}\}$ (denoting the \gem iteration) and $\mathrm{X}\in\{\mathrm{A}, \mathrm{B}\}$ (denoting the MSP scenario).}



We derive the best-fit diffuse model via a maximum likelihood fit utilizing the astrophysical templates from Sec.~\ref{sec:astro_models} and confront them with the gamma-ray data in the chosen ROI centered on the GC (c.f.~Sec.~\ref{sec:MaxL_definition} for the technical details of the likelihood method). To prevent the astrophysical gamma-ray components degenerate with the GCE from partially accounting for it -- thus reducing its overall luminosity and/or spatial shape -- we add and fit the GCE component during the likelihood maximization.
The GCE is represented solely by a DM-like template whose normalization and slope parameter $\gamma$ are free to vary. 

{We combine the templates of the diffuse gamma-ray emission (CO, HI, IC, isotropic background and FBs) 
according to their best-fit normalization parameters in a single template, $N_{\mathrm{DIFF}}$, and repeat the same procedure for the two templates containing the 4FGL-DR2 sources, resulting in the template $N_{\mathrm{PS}}$. Consequently, Model 1 features six distinct parameters denoted as $\left\{A_{\mathrm{DIFF}}, A_{\mathrm{PS}}, A_{\mathrm{DM}}, \gamma, F_{\mathrm{MSP}}, \sigma_L \right\}$, whose values are sampled within the ranges reported in Tab.~\ref{tab:model1} from the prior distributions found in the last column of the same table and combined according to Eq.~\ref{eq:model1_parameters}
\begin{eqnarray} \label{eq:model1_parameters}
    N_{\mathrm{mock}} &=& A_{\mathrm{DIFF}} N_{\mathrm{DIFF}} + A_{\mathrm{PS}} N_{\mathrm{PS}}+ A_{\mathrm{DM}} N_{\mathrm{DM}} (\gamma)+ N_{\mathrm{MSP}, \gamma}(F_{\mathrm{MSP}} , \sigma_L, \gamma),
\end{eqnarray}
where $N_{X}$ represents the template associated to the respective \gem parameters. Since we  derived $N_{\mathrm{DIFF}}$ and $N_{\mathrm{PS}}$ from a fit to the data, their priors are chosen closely around 1 while the choice of the GCE parameters' prior distributions is explained above in Sec.~\ref{sec:GCEdata}. This way, we can generate mock sky images $N_{\mathrm{mock}}$ that serve as training data for our neural network.}

{As a further step in the training data generation
we impose a constraint that the total photon counts in the first four energy bins of our mock sky images \emph{without} the GCE components must be within $\pm10\%$ of the total counts of the \emph{Fermi}-LAT data in the respective energy bin. In total, we generated 1 million images for both scenarios A and B within this setup of Model 1.}

\begin{table}[h!]
\centering
\caption{{Summary of the parameters of Model 1 and their assumed ranges. These parameters are sampled from the prior distributions stated in the last column of this table and combined according to Eq.~\ref{eq:model1_parameters} to form a set of training data for our neural network approach. Note that an additional  constraint on the total  number of counts is applied {\it a posteriori} to the templates (see text for more details).} \label{tab:model1}}
\begin{tabular}{ l c c c } 
 \hline
 Parameter & Minimum & Maximum & Prior\\
 \hline
  \hline
  $A_{\mathrm{DIFF}}$ & 0.5 & 2.0 & uniform \\ 
  $A_{\mathrm{PS}}$ & 0.9 & 1.1 & uniform \\ 
  $A_{\mathrm{DM}}$ & $10^{-3}$ & 4 & log-uniform \\ 
  $\gamma$ & $0.8$ & $1.3$ & uniform \\ 
  $F_{\mathrm{MSP}}$ & $0.02\times10^{-6}$ & $1.58\times10^{-6}$ & log-uniform \\ 
  $\sigma_L$ & $0.3$ & $1.2$ & uniform \\ 
  \hline
\end{tabular}
\end{table}

{In this work we aim to distinguish between a DM-like (or ``smooth'') and an MSP-like (or ``granular'') GCE. We, therefore, introduce an additional quantity, $f_{\mathrm{src}}\in\left[0, 1\right]$, on which we also train the neural network. It helps characterize the composition of the emission associated with the GCE. It is defined as the fraction of the GCE that is due to MSPs, i.e. 
\begin{equation}
\label{eq:def_fsrc}
    f_{\mathrm{src}} = \frac{F_{\mathrm{MSP}}}{F_{\mathrm{MSP}} + F_{\mathrm{DM}}}\rm{,}
\end{equation}
where $F_{\mathrm{DM}}$ is the integrated gamma-ray flux (0.1\,GeV - 100\,GeV) associated with the DM template for a given value of the normalization parameter $A_{\mathrm{DM}}$. The value of $F_{\mathrm{DM}}$ hence follows from integrating the differential gamma-ray flux formula for pair-annihilating (Majorana) DM (see e.g.~\cite{Bringmann:2012ez})
\begin{equation}
\label{DMflux}
  \frac{d\Phi_{\gamma}}{d\Omega\, dE} (E,\psi, \gamma) = 
  \left(\frac{1}{4\pi} \int_\mathrm{l.o.s}
  d\ell(\psi) \rho_{\mathrm{gNFW}}^2(\mathbf{r}, \gamma)\vphantom{\frac{\langle\sigma v\rangle_\mathrm{ann}}{2 m_{\chi}^2} \sum_f
  B_f\frac{dN_\gamma^{f}}{dE}}\right) 
  \left({\frac{\langle\sigma v\rangle_\mathrm{ann}}{2 m_{\mathrm{DM}}^2} \sum_f
  B_f\frac{dN_\gamma^{f}}{dE}}\right)
\end{equation}
with respect to the chosen energy range and ROI size. The ROI size is relevant for the first term in parenthesis, the so-called $J$-factor, which is the line-of-sight integral of the squared DM density distribution from a selected portion of the sky. The second term in parenthesis is mostly fixed by our choice of DM mass and annihilation channel while $\langle\sigma v\rangle_\mathrm{ann}$ follows from the best-fit value of the normalization parameter $A_{\mathrm{DM}}$. Since $f_{\mathrm{src}}$ is entirely computed from a subset of the six \gem~parameters, it does not contain additional information that the network may use to improve its predictions. However, including $f_{\mathrm{src}}$ in the training parameter set has the advantage that the network is able to make an assessment of the uncertainty of this quantity on its own without further human intervention in contrast to, for instance, error propagation using the estimated uncertainties of the ingredients of Eq.~\ref{eq:def_fsrc}.} 

{The power of $f_{\mathrm{src}}$ to distinguish between a DM-like and MSP-like GCE depends on the number of MSPs just below the detection threshold, which is determined by $\sigma_L$.  If the bulk of the flux comes from very faint sources, the two templates become indistinguishable as, for example, explored in the context of neural networks in \cite{List:2021aer}. As visualized in Fig.~\ref{fig:MSP_populations} our choices for the MSP-related prior ranges are such that the generated populations cover the latter extreme case as well as realization with enough sources near the \textit{Fermi}-LAT detection threshold. This is also visually demonstrated in Figure~\ref{fig:templates} which shows an example of an MSP template using the network's predicted parameters.

\section{Inference techniques}

\label{sec:NN}

{While in this work we focus on the analysis  technique based on computer vision, we also compare our findings  with a traditional statistical inference method, namely the maximum likelihood template fitting technique.}


In this section, we present the general framework of both analysis setups.

\subsection{Neural network setup}
\label{sec:bayesian_networks}
For our setup, we use the deep ensembles technique \cite{lakshminarayanan2017simple}, which enables us to estimate the uncertainty in the network predictions. This is done through random initializations of neural networks that allow for exploring different modes in the network model space. This method has been shown to perform very well in terms of accuracy and uncertainty \cite{lakshminarayanan2017simple}. 

The network architecture is as follows:

\begin{itemize}
    \item The input is normalised using the batch normalisation algorithm (see \cite{batchnorm}), which performs a normalisation step (scaling to zero mean and unit variance) as part of the neural network architecture. This leads to more stable and faster network training and less dependence on preprocessing;
    \item Five convolutional ``blocks", where one block comprises the following layers: a convolutional layer with $X$ channels and a kernel size of 3, a convolutional layer with $2X$ channels, a max-pooling layer with a pooling value of 2 and 
    the number of channels $X$ of the first convolutional layer is 8;
    \item The last layer (an image that is 3x3 pixels and has 256 channels) is flattened to a vector with 2304 elements;
    \item For every output prediction, a subnetwork contains three dense layers of 128, 64 and 32 neurons and then two output neurons: one that predicts the output mean, and one that predicts the logarithm of the output variance. 
\end{itemize}

\subsubsection{Training setup}
 The neural network was trained on Tesla V100 GPUs splitting 
 the data in 80\% for training, 10\% for validation and 10\% for testing. The average training time for this setup is around a few days, while a single prediction of the trained network is in the order of hundreds of milliseconds. The network is trained using the Adam optimizer with a learning rate of 0.01 \cite{kingma2014method}, where the learning rate is halved when there is no improvement in loss on the validation set for three epochs. Training is stopped after ten of these halvings. The source code to train the network is available on \url{https://github.com/rruizbazan/GalacticCenterNet}. 
{\subsubsection{Deep Ensembles}
\label{sec:ensembles}
As mentioned above, to estimate the uncertainties of the neural network predictions, we adopt the deep ensembles method \cite{lakshminarayanan2017simple}. To do this, we train a set of five neural networks, each with a random initialisation of the network parameters, along with a random shuffling of the data following the prescription in \cite{lakshminarayanan2017simple}.

To capture uncertainty, the networks predict two outputs per output parameter: the predicted mean and its variance. Assuming that the data come from a Gaussian distribution, we define the loss function as follows
\begin{equation}
    \label{eq:aleatoric_loss}
    \mathcal{L} = \frac{1}{2} \cdot \sum_i^T e^{-\hat{z_i}} \cdot \left(\hat{y_i} - y_i\right)^2 + \frac{1}{2} \cdot \hat{z_i},
\end{equation}
where $T$ is the number of output predictions, $\hat{z}$ is the predicted logarithm of the variance, $\hat{y}$ is the predicted output value, and $y$ is the true output value. Note that this loss function is equivalent to the logarithm of a Gaussian (by substituting $z = \textrm{log}(\sigma^2)$). When $\sigma = 1$, this equation reduces to a standard mean squared error loss function.

For the final ensemble predictions, the mean and variance are derived from a mixture of Gaussian distributions. Therefore, the combined mean and variance of the predictions are
$\hat{y}_{*} = N^{-1} \sum_j \hat{y}_j$ and $\sigma^2_{*} = N^{-1} \sum_j (\sigma^2_j + \hat{y}^2_j)- \hat{y}_{*}$ respectively, where $N$ is the number of ensembles which in our case is five. 
}

{\subsection{Maximum likelihood approach}
\label{sec:MaxL_definition}}

{We employ the standard Poisson likelihood function
\begin{equation}
\mathcal{L\!}\left(\left.\bm{\mu}\right|\bm{n}\right)=\prod_{i,j}  \frac{\mu_{ij}^{n_{ij}}}{\left(n_{ij}\right)!}e^{-\mu_{ij}}
\end{equation}
for binned \gem~data (our input templates, except  the MSP one) $\bm{\mu}$ and experimental data $\bm{n}$ (where the index $i$ runs over the energy bins while the index $j$ enumerates the spatial pixels of our templates). We take the best-fit parameters of our \gem~to be the maximum likelihood estimators obtained via minimization of $-\log{\mathcal{L}}$ based on the \texttt{iminuit} python package \cite{iminuit} and the Minuit2 minimization algorithm it provides.}

{In what follows, we will present the results of the likelihood approach in combination with the corresponding predictions by the network. To stress it again, the preparation of the diffuse and point-like source templates in the definition of Model 1 relies on a prior maximum likelihood fit to obtain the best-fit parameters for the templates (except for the MSP template) shown in Fig.~\ref{fig:templates}.}

\vspace*{11pt}

\section{Results with the Model 1 set-up}
\label{sec:results_model1}

\subsection{Application to simulated data}

{We assess the trained network's performance by running it on the validation data set as a subset of the training data. The results are shown in Fig.~\ref{fig:results_scenarioA_model1}. The uncertainty bands mark the (1, 2) $\sigma$ containment intervals, while the dark grey line marks the median of the prediction $y_p$. The uncertainty bands are solely calculated from the {\it scatter} of the network's (mean) prediction for image samples that share the same true value $y_t$ of a particular \gem~parameter (see Fig.~\ref{fig:results_modelB1} for the corresponding results in scenario B). The uncertainty of predictions (as given by the DE networks) is visualized for a set of randomly chosen data points (black) with error bars. The size of the error bar reflects the epistemic and aleatoric uncertainty summed in quadrature.}


{The results demonstrate that the network is able to predict the bright astrophysical gamma-ray components with high precision. This precision is at the per cent level for the diffuse template (comprising all contributors to the diffuse emission in the GC). Such a result is expected on Monte Carlo data since the bulk of the observed photons is either due to detected sources or the Galactic diffuse emission. Moreover, the fact that we use Monte Carlo data for the validation ensures perfect agreement between the 
\gem~and mock data.

However, this accuracy is not matched by the predictive power regarding the GCE components, which are reconstructed with substantial uncertainty while retaining some predictive power in parts of the parameter space. We find that the network is capable of consistently retrieving the injected normalization of the DM and MSP components on average. 
This statement also applies to most of the tested range of the gNFW's inner slope parameter $\gamma$ except for values at the lower boundary of the sampled prior distribution. As anticipated however, the width of the MSP's luminosity function $\sigma_L$ cannot be reliably reconstructed by the network, as scenario A only features MSPs below the nominal LAT detection threshold, which leaves very little information to the network to establish a relation between the drawn MSPs in a template and their underlying luminosity function. The corresponding results on Model 1 Scenario B corroborate this interpretation. In this scenario, $\sigma_L$ can be predicted (with large uncertainties) for at least a fraction of the probed prior range\footnote{
This may be put in perspective using Fig.~\ref{fig:MSP_threshold_numbers} in Appendix \ref{app:msp_threshold} where we show the number of expected MSPs below and above the threshold for a selection of GCE model parameters. Values of $F_{\mathrm{MSP}}$ where the reconstruction of the width of the luminosity function becomes feasible seem to be already at odds with the experimentally observed number of point-like sources in the GC region so that the approach we employ here is most likely not suited to shed light on the physical parameters of a (potential) MSP population in the MW's bulge.}.}

{The prediction of $f_{\mathrm{src}}$ is also characterized by sizeable uncertainties, especially for values in the mid-range of this parameter. 
In comparison to the preliminary work in \cite{Caron:2017udl}, we obtain reasonable and even comparable results in terms of the predicted uncertainty. Overall, the containment bands derived in the present work are somewhat wider but they have been obtained with the increased complexity of our training data, i.e.~introducing an explicit dependence on $\gamma$ and $\sigma_L$. Hence, the results of this work corroborate our previous findings. Besides, we can provide an estimate of the intrinsic uncertainty for each predicted value, a piece of information that was inaccessible in \cite{Caron:2017udl}. We find large errors of $y_p$ except at both endpoints of the probed parameter range. There are further arguments why the proof-of-principle study was able to achieve a smaller uncertainty on $f_{\mathrm{src}}$: \emph{(i)} The previous work used the 3FGL catalog based on 4 years of data, but applied it on data collected over 7 years.  This inconsistency in the data could affect the flux of the MSP template because many of the sources that were unresolved in 4 years of data become resolvable with the increased exposure. \emph{(ii)} The prior range for $\sigma_L$ was also different and included more samples with many bright MSPs.  These two effects might have affected the predictive power of the $f_{\mathrm{src}}$ parameter, which was also directly predicted by the network as the only trained parameter in the previous work. }

\begin{figure}[t!]
\begin{center}
\includegraphics[width=0.70\linewidth]{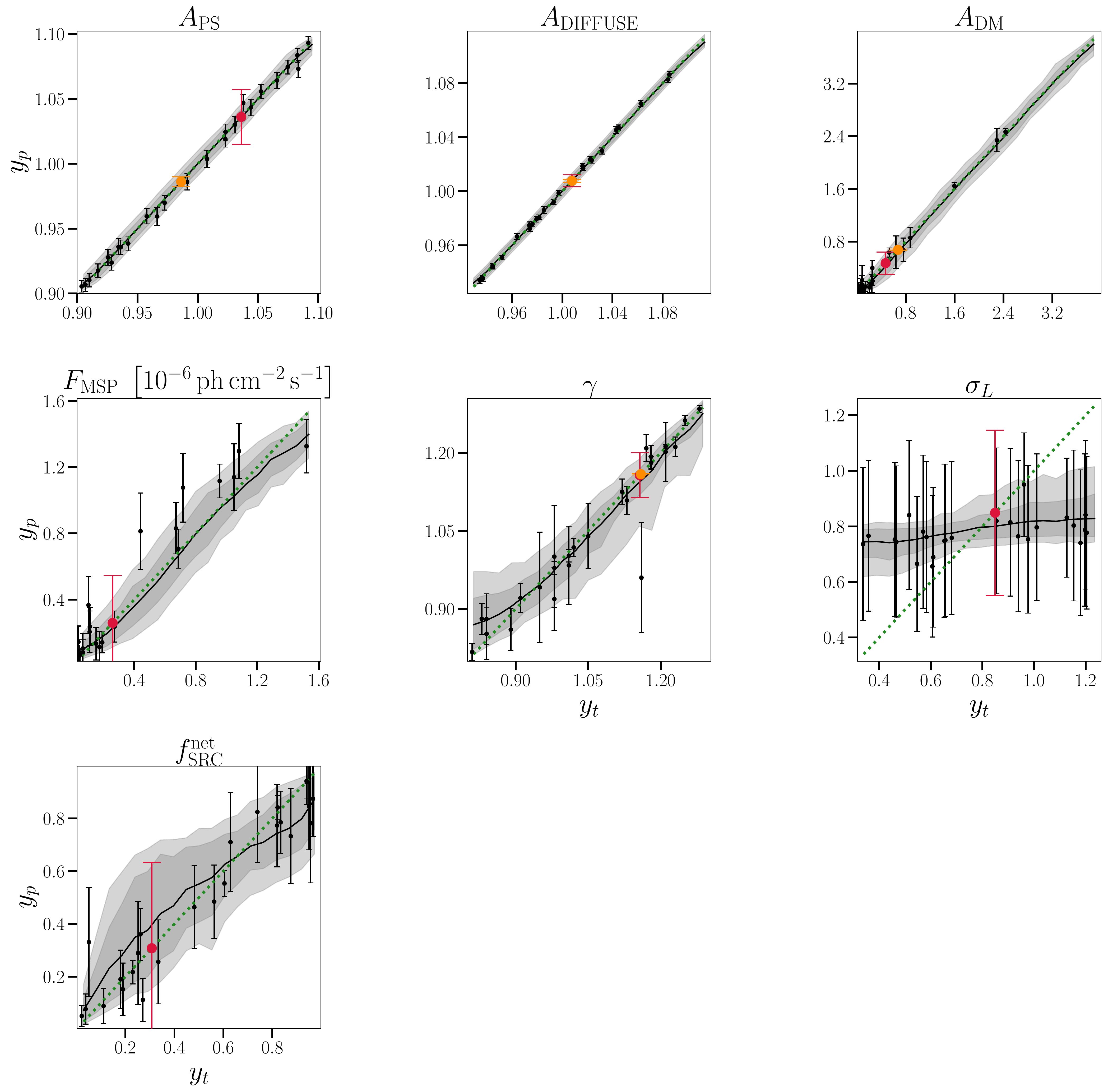}
\end{center}
\caption{{Per panel: Prediction of the neural network $y_p$ (y-axis) compared to the true parameter value $y_t$ (x-axis) based on the validation data set as part of the training data generated for Model 1A. The observed scatter of the predicted values $y_p$ is portrayed as grey-shaded bands whose opacity (dark to light) denotes the $1\sigma/2\sigma$ containment while the solid black line traces the respective median value. The black data points are randomly chosen from the validation data set with the length of the error bar reflecting the uncertainty of the prediction including the epistemic uncertainty of the network derived via the deep ensembles method and the aleatoric uncertainty summed in quadrature. In addition, we display the network's prediction on the real \textit{Fermi}-LAT data as red points while the best-fit results of the likelihood analysis on the real data are shown as orange points. The green dotted line marks the case of perfect reconstruction; a deviation from this line indicates a bias in the network's prediction.}  \label{fig:results_scenarioA_model1}}
\end{figure}


\subsection{Application to the \textit{Fermi}-LAT data}
\label{sec:realdatamodel1}

We applied the network trained on Monte Carlo data generated from Model 1A on the \textit{Fermi}-LAT data and show the findings in Fig.~\ref{fig:results_scenarioA_model1} as red points\footnote{Since the true value, in this case, is not known, the point is placed on the diagonal.} (see also Fig. \ref{fig:results_modelB1}). 
The uncertainty of the predicted values is determined by the network as described in Section~\ref{sec:bayesian_networks}.  Because the uncertainty in  general depends on the explicit position in the high-dimensional parameter space, the error bar is not identical in all cases to the size of the 68\% containment band of the validation set.  
In agreement with the validation results, the algorithm generally finds accurate results for {the astrophysical templates. This is in line with the traditional, likelihood fits (orange points, see Section \ref{subsec:likelihood} for details).}  

To further gauge the quality of the \gem~selected by the algorithm, we calculated the residuals, i.e. the difference between the selected \gem~and the LAT data, and present them in Fig. \ref{fig:spatial_res_model1_scenA}. To this end, we compute the Poisson statistical significance $\sigma$, which is given by
\begin{equation}
\label{eq:poisson_sig}
\sigma = \mathrm{sgn}\!\left(\bm{n} - \bm{\mu}\right) \sqrt{\left|2\left(\bm{n}\ln{\frac{\bm{n} + \varepsilon}{\bm{\mu}}} - \bm{n} + \bm{\mu}\right)\right|}\mathrm{,}
\end{equation}
where $\bm{n}$ denotes the LAT data set whereas $\bm{\mu}$ refers to the \gem~data following from the network's best-fit parameters for Model 1A. The value of $\varepsilon$ has been set to $10^{-100}$ to avoid numerical inconsistencies when the value of a pixel in the LAT data is zero. Besides, we resort to a different formula 
\begin{equation}
\sigma = \frac{\bm{n} - \bm{\mu}}{\sqrt{\bm{\mu}}}
\end{equation}
when the difference between \gem~and real data is small ($\left|\bm{n} - \bm{\mu}\right| \ll 1$) but $\bm{n} > 0$ to reduce the impact of possibly occurring numerical instabilities.


\begin{figure}[t!]
\begin{center}
\includegraphics[width=0.45\linewidth]{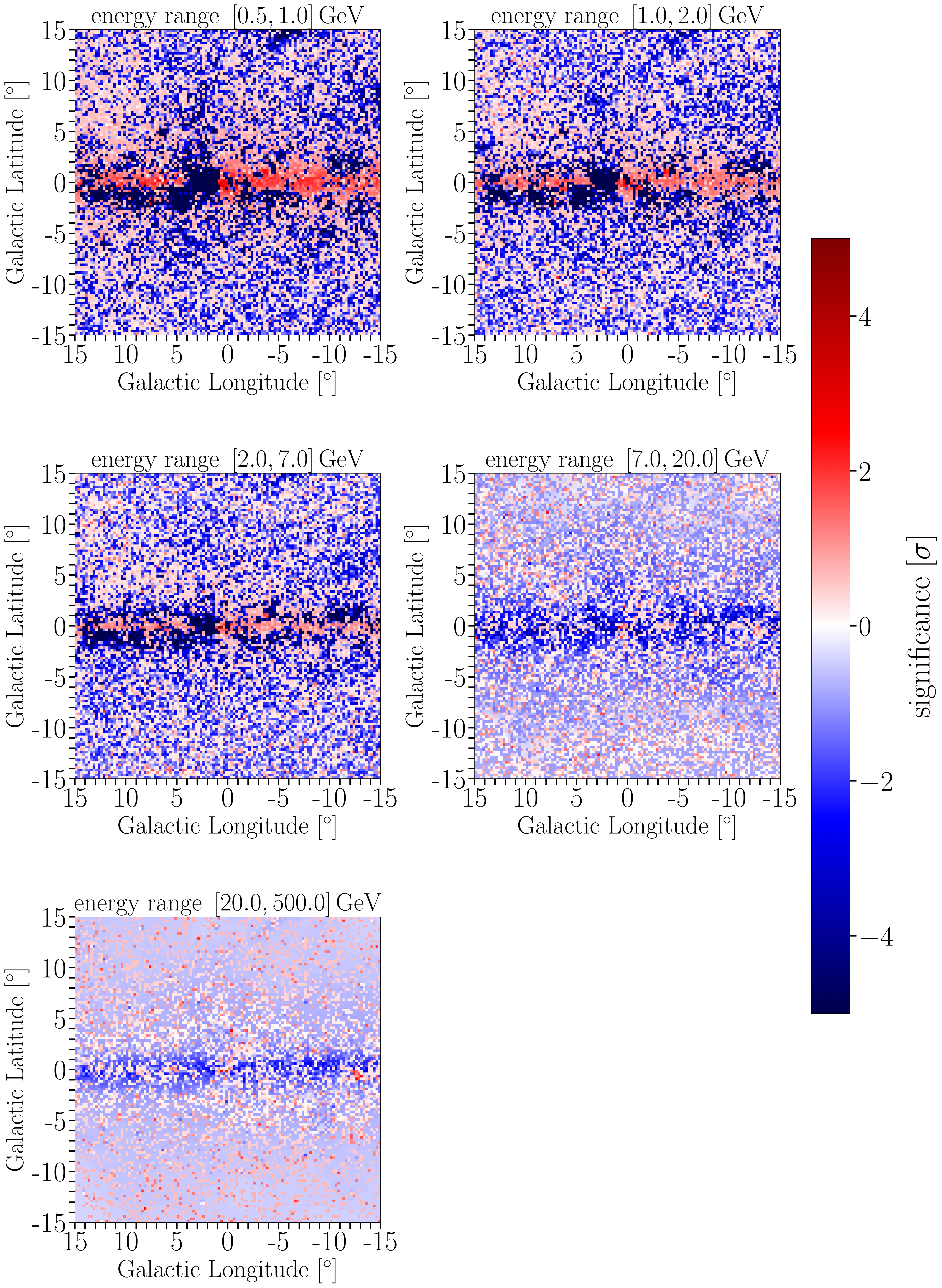}
\includegraphics[width=0.44\linewidth]{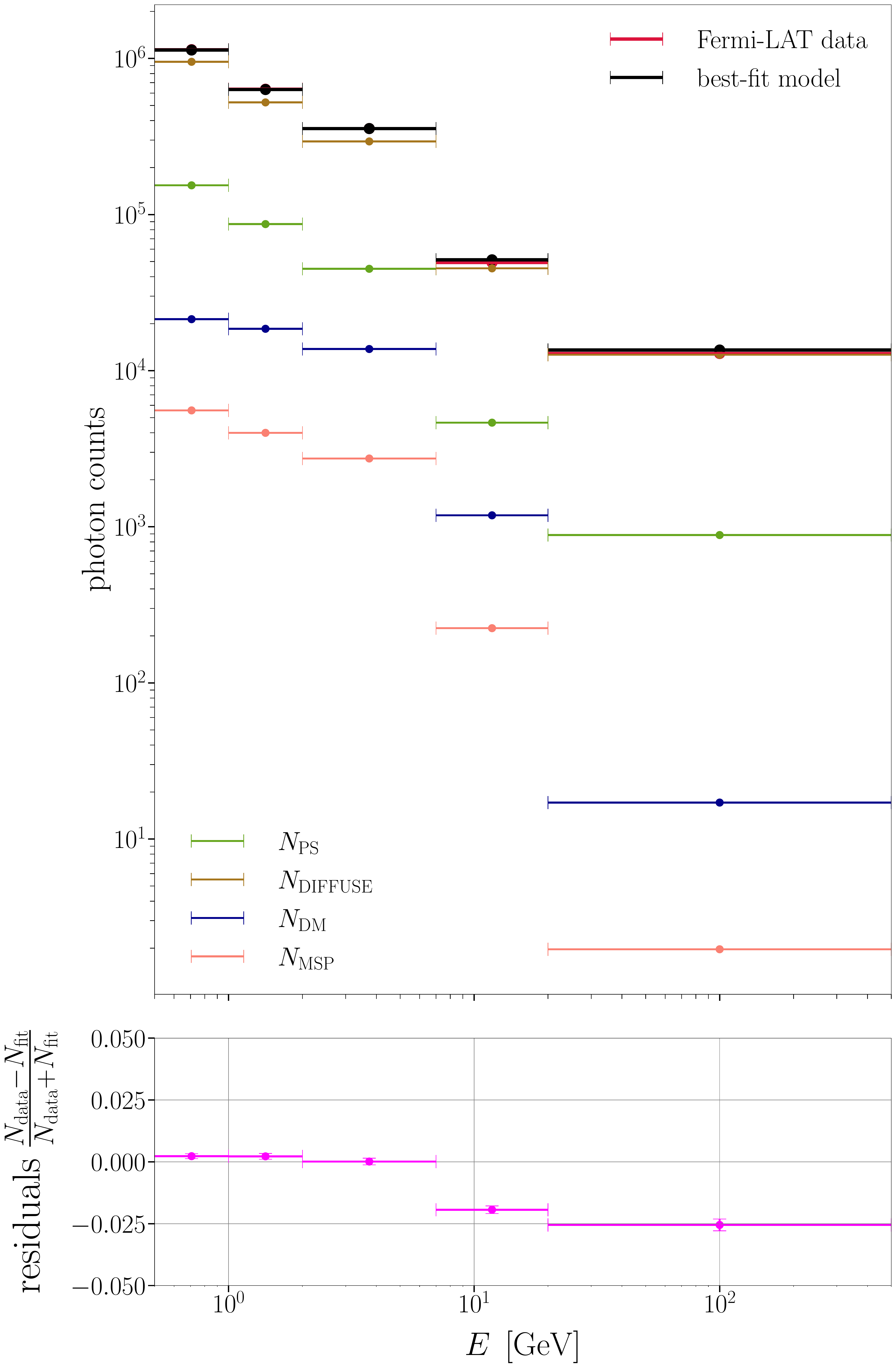}
\end{center}
\caption{Left: Poisson statistical significance (c.f.~Eq.~\ref{eq:poisson_sig}) per energy bin regarding the network's prediction (c.f.~Fig.~\ref{fig:results_scenarioA_model1}) based on training data comprised of the templates of Model 1 and scenario A. All images have been convolved with a Gaussian smoothing kernel of $0.75^{\circ}$ width. Right: Best-fit \gem~spectra and fractional spectral residuals regarding the network's prediction. Note that the contributions of all four rings per gas component have been combined in a single item. The error bars of the total spectral residuals have been computed using standard error propagation rules using the statistical errors and network prediction errors for the relevant parameters.
\label{fig:spatial_res_model1_scenA}}
\end{figure}


Significant negative residuals are present in the region of the Central Molecular Zone (CMZ), left from the GC centered approximately on $(\ell, b) = (2.0^{\circ}, 0.0^{\circ})$.  This is a known issue and is a consequence of the rapid variations in the $X_{CO}$ factor in the inner Galaxy that is caused by tidal forces acting on the molecular gas near the GC.  These show up as elongated features in velocity in the CO emission, resulting in an artificial increase in the integrated emission that is used to estimate the molecular hydrogen column density.  

There are also positive residuals away from the disk and towards the border of our ROI.  These indicate that the templates contributing at high latitudes might need more freedom. Of those, the FBs template as a part of the diffuse template $N_{\mathrm{DIFFUSE}}$ is the most uncertain and may warrant the inclusion of alternatives to the currently assumed shape of the FBs in Model 1.

{While the spectral residuals in the first three energy bins are almost negligible -- hence, pointing to a good agreement between LAT data and network prediction -- the consistency is less striking in the last two energy bins. Here, the best-fitting \gem~overestimates the gamma-ray emission, although the discrepancy is only at the per cent level. At these energies, the majority of detected photons originated from the Galactic plane so these regions of the sky are providing the bulk of the constraining power. As a consequence, the normalization of the 4FGL source template is dominated by the bright sources in the Galactic disc which may deteriorate the goodness-of-fit regarding point-like sources at higher latitudes. In combination with the relative scarcity of events at latitudes below and above the plane, such a finding is within the expectations.}

Since the {\it reality  gap}, i.e. the fact that the original set of templates might not capture all features of the real data  presents the real limitation that might bias our results, 
section \ref{sec:discusion} is devoted to extending the background model in an attempt to shrink the reality gap and explore the effect it has on our results for the GCE nature.

\section{Extending the gamma-ray emission model}
\label{sec:discusion}

The spatial and spectral residuals shown in Fig.~\ref{fig:spatial_res_model1_scenA} indicate that our model of the gamma-ray emission within our chosen ROI has well-defined limitations. 
We therefore iteratively increase the complexity of our \gems~to improve the agreement with the LAT data and to test the robustness of our results.  
Here we summarize our modification of the \gem~which are discussed and reasoned in more detail in dedicated subsections.  We end up iterating twice, defining the following two new \gems:

{
\begin{itemize}
    \item {Model 2} is created by freeing all normalization parameters related to diffuse emission in the ROI previously combined in $N_{\mathrm{DIFFUSE}}$ as well as those two of the single 4FGL-DR2 template $N_{\mathrm{PS}}$. The treatment of the GCE components remains unaltered. Consequently, the overall normalization of all templates shown in Fig.~\ref{fig:templates} are now individually varied 
    so that the number of parameters grows to 17\footnote{Note that, while we do have five spectral bins, the spectral shape of each of the templates is fixed.}. We create a training data set analogously to Model 1.
    
    \item Model 3 is created in the spirit to limit the extent by which the spectrum of the IEM templates biases our results. To this end, we add a power-law normalization for the gas components per ring and likewise for the IC template.  We use the same power-law adjustment for both gas components and still allow for independent normalization of the HI and CO templates.  This adds five new parameters with respect to Model 2. To reduce the pronounced overfitting of the interstellar emission to the left of the GC (see Fig.~\ref{fig:spatial_res_model1_scenA}) -- which is known to be related to the preparation of the gas maps -- we introduce two new templates as well as a second template describing the FBs. More details are provided in Sec.~\ref{sec:newFB}. In total, Model 3 features 24 parameters.
   \end{itemize}
 }
The \gems~are summarized in table~\ref{tab:models}.  In addition to applying the \gems~to the LAT data, we also test cross-domain applications of the trained networks, applying the network trained on Model 1 data to Model 2 data and also the network trained on Model 2 data to Model 3 data.  This allows us to determine any possible bias due to the additional freedom in the more complex \gems~in a controlled way.
Note that, as discussed in Section \ref{sec:intro}, we made sure that Model 1  is embedded in Model 2, which again is embedded in Model 3, to guarantee an appropriate realization of the domain adaptation approach (see Fig. \ref{fig:domainadaptation}). 
{We stress, again, that our goal is to use the more complex \gems~(rather than many different low-dimensional \gems) to learn those distributions and representations shared by possible sub-\gems~by means of the network itself.}

\begin{table}[h!]
\centering
{
\begin{tabular}{|p{0.12\linewidth} | p{0.8\linewidth}|}
 \hline
Scenario A & Sources above the detection threshold of 4FGL are excluded from the MSP template.  \\ 
 \hline
 Scenario B & All sources included in the MSP template, even those above the detection threshold.  \\ 
 \hline
  \hline
Model 1 & Reference \gem, see Sec.~\ref{sec:model1} for details (6 parameters, 4 linear and 2 non-linear).  \\ 
  \hline
Model 2 & 
Based on the templates used in the creation of Model 1. All templates are varied independently without prior constraints obtained via a maximum likelihood fit (17 parameters). \\ 
  \hline
Model 3 & 
The templates of Model 2 are kept, but now the IEM templates are normalized with a power law and a scale factor. In addition, the $CO_1$ template is split into two and there are now two templates for the FBs (24 parameters, see Sec.~\ref{sec:newFB} for all details).  \\ 
 \hline
\end{tabular}
\caption{Summary of the \gems~that we explore in this work, see Fig. \ref{fig:domainadaptation}.}
\label{tab:models}
}
\end{table}

\subsection{Model 2}
\label{subsec:extended}

As discussed above, Model 2 introduces a level of complexity by freeing all normalization parameters of the components that constitute the astrophysical background model. This training data generated within the framework of Model 2 no longer include prior information obtained via established statistical inference methods like a maximum likelihood fit and enables an independent comparison of the performance of a neural network to traditional methods. Note that, to obtain the best-fit \gem~parameters, the network does not minimize a Poisson log-likelihood function but rather an internal loss function (see Eq.~\ref{eq:aleatoric_loss}) whose relation to the likelihood function is not a priori known. It is therefore not expected to recover the same best-fit values.

{To train the neural network with Model 2 data we generated 1.8 million images for both scenarios A and B. The GCE parameter space is sampled from the same prior distributions specified in Tab.~\ref{tab:model1} whereas all parameters related to conventional astrophysics are drawn from uniform priors between 0 and 3. We show the results of a neural network trained on Model 2 data and applied to the real data in Fig.~\ref{fig:results_scenarioA_model2} for scenario A. The corresponding results for scenario B are displayed in Fig.~\ref{fig:results_modelB2}. The color coding of the panels follows the respective plots for Model 1. For completeness, the observed spatial and spectral residuals with respect to both scenarios of Model 2 are shown in Appendix \ref{app:residuals} since they do not change significantly and thus provide only a marginal additional value to the discussion.\\
Despite the potentially different reasoning of a maximum likelihood fit and supervised deep learning algorithms to derive best-fit parameters, the correspondence between both methods is quite striking for the majority of the astrophysical parameters not related to the GCE. The exceptions from this general trend are those components whose main contribution is not necessarily falling into the chosen ROI of this study, namely gamma-ray components that are present at high Galactic latitudes, i.e.~the third and fourth CO ring, the fourth HI ring and the isotropic gamma-ray background. These components are likewise predicted with a larger uncertainty by the network.\\
Regarding the GCE parameters, the predictions for Model 2 exhibit larger uncertainties both with respect to the validation data set and the real LAT data. Since the complexity of the \gem~has increased, a growth of the network's uncertainty on its own predictions can be understood in this regard. On top of this, this deterioration of the precision of the network is reinforced by the aforementioned astrophysical components that are difficult to learn.\\
The predictions for the GCE parameters of the likelihood and ML approach are not the same. As already witnessed in the case of Model 1, the ML approach is capable of detecting a non-Poissonian component like the MSP population in the MW's bulge whereas it may seem impossible for all practical matters for the likelihood fit\footnote{To illustrate this statement: The images have a size of $120\times120 = 14400$ pixels. A typical number of MSPs needed to saturate the GCE given the reconstructed parameters of Model 2 is around 3000. Hence, there are $\left(\begin{array}{c} 17399\\ 3000 \end{array}\right)$ ways to place these sources in the image (with repetition). In numbers, there are $\sim10^{3470}$ ways to distribute these sources in the GC images while further variations like the intrinsic brightness of each of the MSPs are not yet taken into account.}. In comparison to Model 1, the network finds a larger MSP-like contribution to the GCE while at the same time, the spatial morphology of the signal is predicted with a steeper slope $\gamma > 1.2$ than obtained via the maximum likelihood method. This in return increases the parameter $f_{\mathrm{src}}$, which is about $50\%$ in this \gem~and scenario.\\
The differences between the likelihood and ML approach are the same in scenario B but there, we find a smaller MSP contribution to the GCE of around $20\%$ (although being higher than in Model 1B). Hence, dropping constraints on the template normalizations imposed in Model 1 leads to an increase in the reconstructed fractional contribution of unresolved point-like sources to the GCE.}

\begin{figure}[t!]
\begin{center}
\includegraphics[width=0.8\linewidth]{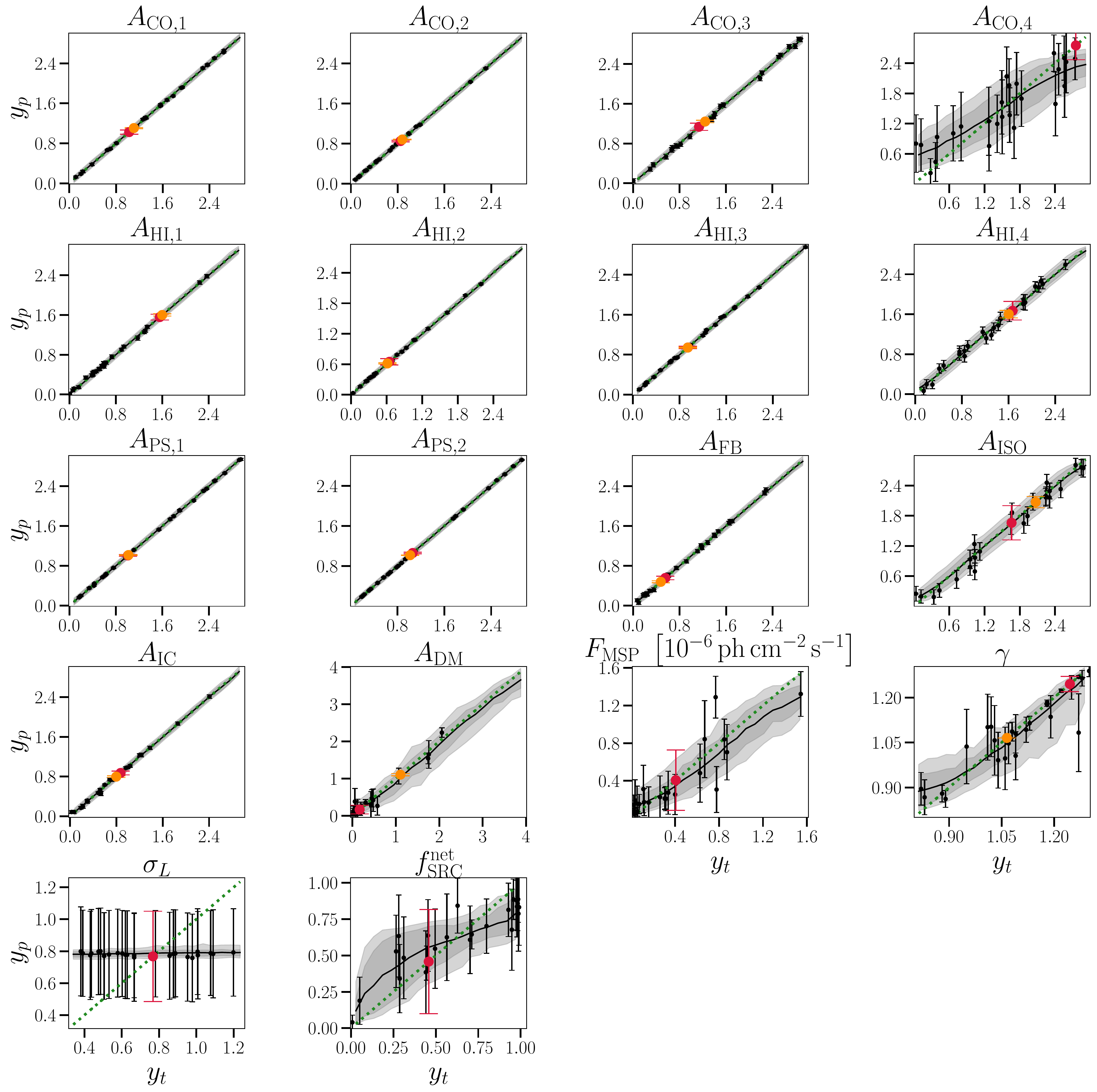}
\end{center}
\caption{The same as Fig. \ref{fig:results_scenarioA_model1}, but in the context of Model 2 (Scenario A).
}\label{fig:results_scenarioA_model2}
\end{figure}

\subsection{Model 3}
\label{sec:newFB}

In our final \gem~iteration, we enlarge our template selection (see Fig.~\ref{fig:templates_extended_setup_1}) to address the aforementioned issue with the central CO ring. To further alleviate the positive residuals at high latitudes, we add a template quantifying the emission of the FBs because this component is the least constrained and potentially one of the most degenerate with the DM template. In particular,
\begin{itemize}
    \item We split the innermost CO ring -- whose emission shows significant left/right asymmetry -- in two independent templates. For this rather ad-hoc solution, we define a gamma-ray flux threshold at 25$\%$ of the maximal value found in the input gamma-ray flux model of the first CO ring. Each pixel above this threshold is part of a $N_{\mathrm{CO}, 1, \mathrm{high}}$ template whereas the remainder of pixels is cast into an independent template to which we assign the already used label $N_{\mathrm{CO}, 1}$.
    \item We add a second FBs template $N_{\mathrm{FBcatenary}}$, taken from \cite{Acero:2016qlg}. This template is built by extracting residual intensity maps from the LAT data, after the application of the Fermi diffuse model.  The edges of the bubbles' residuals are found to be well reproduced by two catenary curves (see \cite{Acero:2016qlg} for parameter values), which also reproduce correctly the structures observed close to the GC in the ROSAT X-ray observations. While this template is quite similar in morphology to the original FB template from Model 1, it is centered closer to the GC. An additional difference is that we assume that it exhibits a uniform intensity while the spectrum is the one that we use for the FB template in Model 1.
    \item We vary the spectral indices of four gas components, in such a way that we  keep the  same spectral index per ring for CO and HI, but allow for their normalization to change independently. Similarly, we also allow for variations of the spectral index of the IC component. In all of the cases we allow for the change of the spectral index in the $\Delta \gamma \in \left[-0.3, 0.3\right]$ range, where we define the new spectral index as $\gamma - \Delta \gamma$, i.e.  negative values of $\Delta \gamma$ imply spectral {\it softening}.  
\end{itemize}
The normalizations of these two templates yield two additional parameters in our fits, while the spectral indices add another five parameters, see Table~\ref{tab:models}.

\begin{figure}[h!]
\begin{center}
\includegraphics[width=0.8\linewidth]{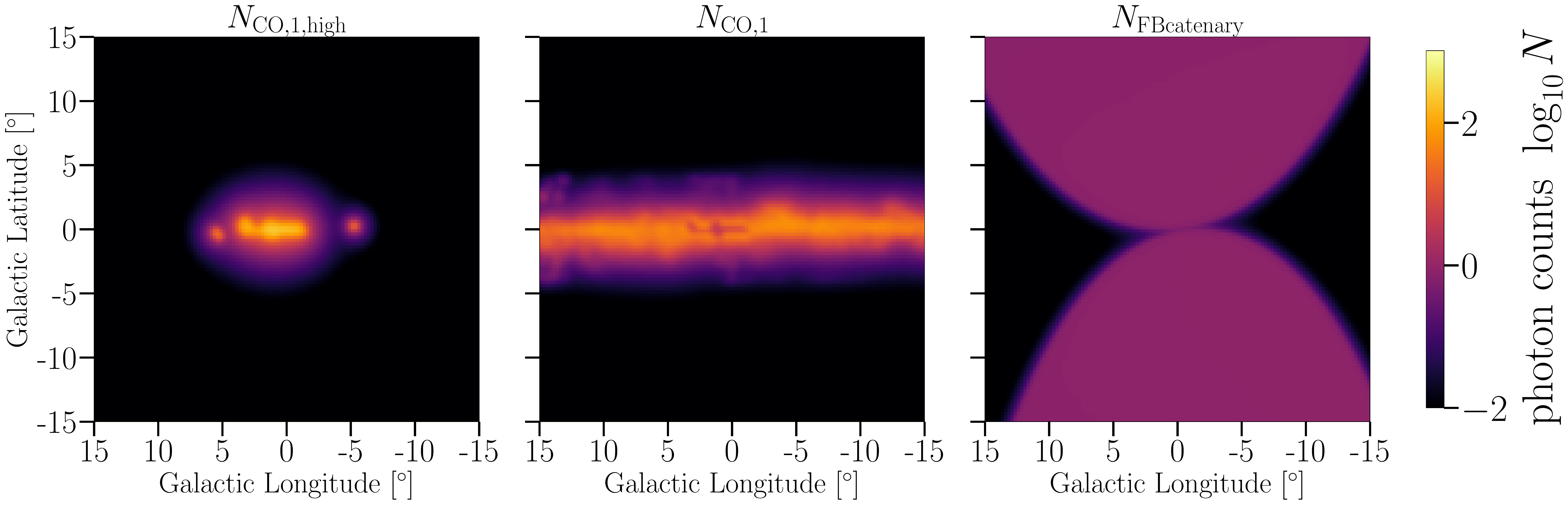}
\end{center}
\caption{
Compilation of the additional gamma-ray templates used in Models 2 and 3. The templates are the output of the Fermi Science Tools routine \texttt{gtmodel} and hence display the number of expected events from the respective flux model for the given \textit{Fermi}-LAT observation time in the infinite statistics limit. The color indicates the decadic logarithm of the number of expected gamma-ray events per spatial pixel. For definiteness, we have selected the second energy bin (1 GeV to 2 GeV) for all shown templates. \label{fig:templates_extended_setup_1}}
\end{figure}

\begin{table}[t!]
\centering
\begin{tabular}{ c c c c } 
 \hline
 Parameter & Minimum & Maximum & Prior \\
 \hline
  \hline
  $A_{\mathrm{CO,1,high}}$ & 0 & 3 & uniform \\ 
  $A_{\mathrm{FBcatenary}}$ & 0 & 3 & uniform \\ 
  $PL_1$ & -0.3 & 0.3 & uniform \\ 
  $PL_2$ & -0.3 & 0.3 & uniform \\
  $PL_3$ & -0.3 & 0.3 & uniform \\
  $PL_4$ & -0.3 & 0.3 & uniform \\
  $PL_\mathrm{IC}$ & -0.3 & 0.3 & uniform \\
\end{tabular}
\caption{Random sampling prior distributions for the extra parameters of Model 3. \label{tab:model3}} 
\end{table}

\vspace*{10pt}

To train the neural network with Model 3 data we generated 2.4 million images for both scenarios A and B. The priors for the additional parameters are stated in Tab.~\ref{tab:model3} while the remaining parameters are sampled as in Model 2. The predictions of the neural network trained on data created within the Model 3A framework are shown in Fig.~\ref{fig:results_scenarioA_model3} together with the corresponding maximum likelihood results, while the residuals can be found in Fig.~\ref{fig:spatial_residuals_extended_2_sc2} in Appendix \ref{app:residuals} (see also Figs. \ref{fig:results_modelB3} and \ref{fig:residuals_modelB3}, for Scenario B). To conduct the maximum likelihood fit, we have used astrophysical templates corrected by the network's best-fit spectral indices. 

{The left-right asymmetry of the central residuals as well as those in the Galactic plane, are reduced in this iteration (compare Figs.~\ref{fig:spatial_res_model1_scenA} and \ref{fig:spatial_residuals_extended_2_sc2}), with the new CO template normalization ($A_{\mathrm{CO}, 1, \mathrm{high}}$) found to be smaller than one, thus reducing the \gem~intensity in that region. The algorithm finds the two FBs templates {to be} present in the data, with comparable intensities. 
Regarding the reconstructed spectral indices, we note that despite the added freedom, most $\Delta \gamma$ values are predicted close to zero, meaning little spectral correction is required. A noteworthy exception from this observation is the spectral correction for the first gas ring, which is close to $\Delta \gamma = 0.1$ for both scenarios indicating a hardening of the default gamma-ray spectrum. Such a spectral hardening might explain the increase of the residuals in Model 3 with respect to the last two energy bins (see Figs.~\ref{fig:spatial_residuals_extended_2_sc2} and \ref{fig:residuals_modelB3}). }

The parameters for the GCE are still compatible with those from the network trained on Model 1 data, although there is a small trend towards higher values of $F_{\mathrm{MSP}}$ and therefore a higher $f_{\mathrm{src}}$ value (Model 2A: $f_{\mathrm{src}} = (46\pm36)\%$; Model 3A: $f_{\mathrm{src}} = (93\pm20)\%$).



\begin{figure}[t!]
\begin{center}
\includegraphics[width=0.8\linewidth]{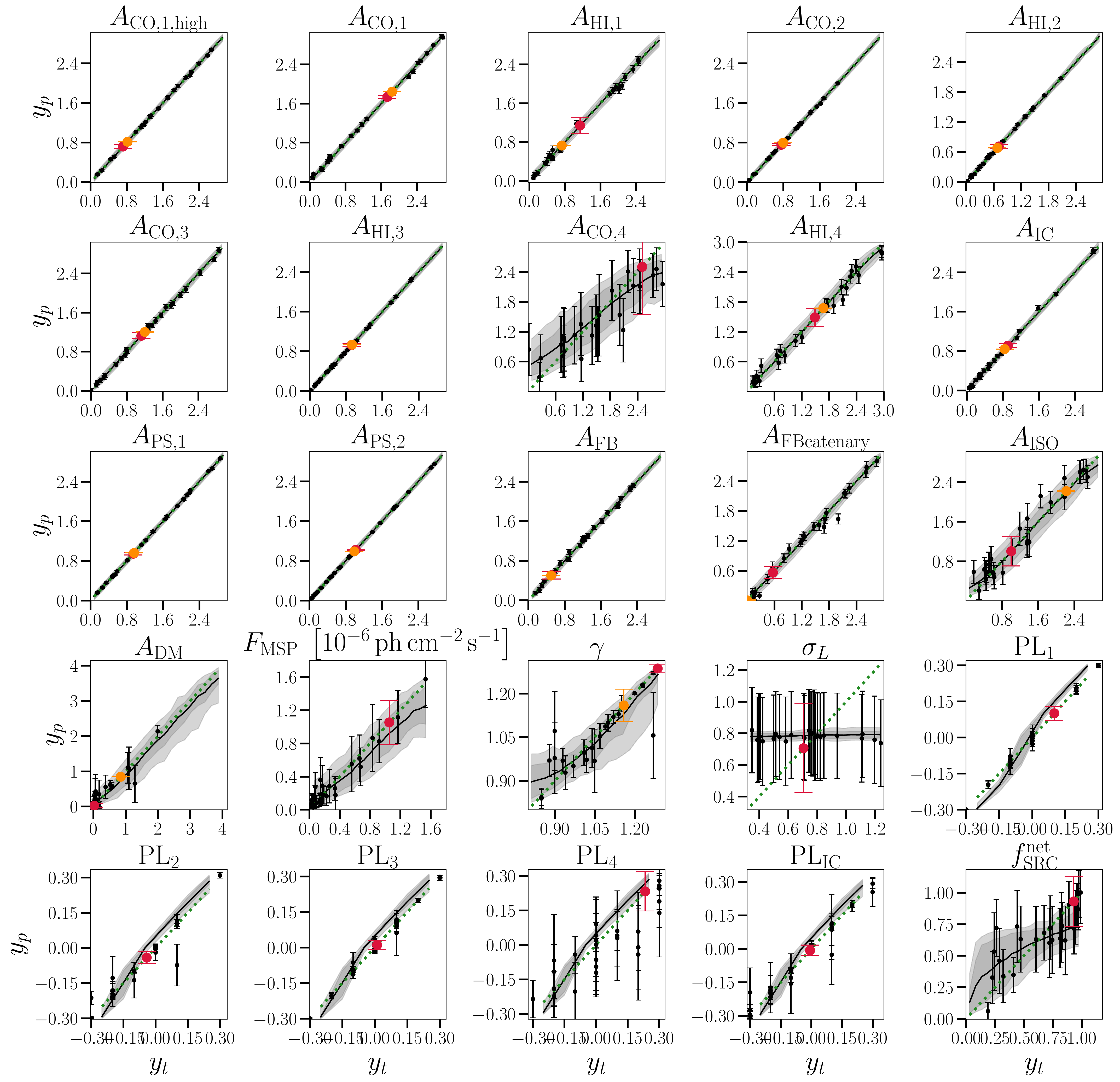}
\end{center}
\caption{The same as Fig. \ref{fig:results_scenarioA_model1}, but in the context of Model 3 (Scenario A).
}\label{fig:results_scenarioA_model3} 
\end{figure}

\section{Discussion}
\label{sec:robustness}

\subsection{Comparison with traditional likelihood template fitting}
\label{subsec:likelihood}

The main caveat is that when using the likelihood fitting, it is challenging to include the MSP template(s) in the fitting stage due to their probabilistic character. In addition, we limit our fitting to {\it linear} parameters only  (except for the $\gamma$ parameter), while the rest of the nonlinear parameters are marginalized over. The nature of uncertainties in the two analysis methods is different as well: while the likelihood analysis is based on the statistical (Poisson) uncertainty, the mapping of the network's aleatoric uncertainty of the input to the predicted uncertainty is complex and depends on several parameters (for example on the loss function of the network, see  discussion in Section \ref{sec:bayesian_networks}). Despite a lack of a one-to-one comparison, we find that the results obtained from these two methods are consistent, except in a few cases discussed below.

To  gauge the impact of an unmodeled MSP contribution in the likelihood analysis, we use simulated data based on {Model 3 with $\Delta \gamma = 0.0$} and draw in total 500 Poisson realizations where all other parameters are set to 1, except that $F_{\mathrm{MSP}} = 5\times10^{-7}$ ph/cm$^{2}$/s.  In Table~\ref{tab:likelihood_fit_baseline} we compare the likelihood analysis results in which the MSPs are present in the mock data but {\it not} in the {simulated gamma-ray data} 
(the first column, labeled 'DM+MSP'), with the case where that template is not present in the {mock} data (the second column, 'DM only'). In addition, we show the results of the likelihood fit when the {mock} data contain only an MSP component which is then fitted via the smooth DM template. We observe that the majority of parameters corresponding to bright templates are unaffected by the omission of the MSP template in the modeling step. However,  the flux of the MSP template,  calculated from the mean best-fit results under the assumption that the original DM setup is known (i.e. $A_{DM}=1$ and $\gamma=1$), and shown in  the last row, is recovered  only partially.  

\begin{table}
\begin{centering}
\begin{tabular}{|c|c|c|c|}
\hline 
 & DM + MSP & DM only & MSP only\tabularnewline
\hline 
\hline 
$A_{\mathrm{CO}, 1, \mathrm{high}}$ & $0.99 \pm 0.01$ & $1.0 \pm 0.01$ & $1.0 \pm 0.01$\tabularnewline
\hline 
$A_{\mathrm{CO}, 1}$ & $1.0 \pm 0.02$ & $1.0 \pm 0.02$ & $1.0 \pm 0.02$ \tabularnewline
\hline 
$A_{\mathrm{CO}, 2}$ & $1.0 \pm 0.01$ & $1.0 \pm 0.01$ & $1.0 \pm 0.01$ \tabularnewline
\hline 
$A_{\mathrm{CO}, 3}$ & $0.99 \pm 0.03$ & $1.0 \pm 0.03$ & $0.98 \pm 0.03$ \tabularnewline
\hline 
$A_{\mathrm{CO}, 4}$ & $0.44 \pm 0.38$ & $0.92 \pm 0.44$ & $0.5 \pm 0.39$ \tabularnewline
\hline 
$A_{\mathrm{HI}, 1}$ & $1.0 \pm 0.02$ & $1.0 \pm 0.02$ & $1.0 \pm 0.02$ \tabularnewline
\hline 
$A_{\mathrm{HI}, 2}$ & $1.0 \pm 0.01$ & $1.0 \pm 0.01$ & $1.01 \pm 0.01$ \tabularnewline
\hline 
$A_{\mathrm{HI}, 3}$ & $1.01 \pm 0.01$ & $1.0 \pm 0.01$ & $1.01 \pm 0.01$ \tabularnewline
\hline 
$A_{\mathrm{HI}, 4}$ & $1.02 \pm 0.08$ & $0.99 \pm 0.08$ & $1.04 \pm 0.07$ \tabularnewline
\hline 
$A_{\mathrm{PS}, 1}$ & $1.0 \pm 0.0$ & $1.0 \pm 0.0$ & $1.0 \pm 0.0$ \tabularnewline
\hline 
$A_{\mathrm{PS}, 2}$ & $1.0 \pm 0.01$ & $1.0 \pm 0.01$ & $1.0 \pm 0.01$ \tabularnewline
\hline 
$A_{\mathrm{ISO}}$ & $0.74 \pm 0.17$ & $0.97 \pm 0.14$ & $0.92 \pm 0.15$ \tabularnewline
\hline 
$A_{\mathrm{IC}}$ & $1.03 \pm 0.03$ & $1.01 \pm 0.02$ & $0.99 \pm 0.02$ \tabularnewline
\hline 
$A_{\mathrm{FB}}$ & $1.0 \pm 0.03$ & $1.0 \pm 0.02$ & $1.0 \pm 0.03$ \tabularnewline
\hline 
$A_{\mathrm{FBcatenary}}$ & $1.06 \pm 0.06$ & $1.01 \pm 0.06$ & $1.0 \pm 0.06$ \tabularnewline
\hline 
$A_{\mathrm{DM}}$ & $1.26 \pm 0.13$ & $0.98 \pm 0.09$ & $0.51 \pm 0.1$ \tabularnewline
\hline 
$\gamma$ & $1.03 \pm 0.02$ & $1.0 \pm 0.02$ & $1.0 \pm 0.03$ \tabularnewline
\hline 
$F_{\mathrm{MSP}}$ & $2.29 \times10^{-7}$ ph/cm$^{2}$/s & / & $2.71\times10^{-7}$ ph/cm$^{2}$/s\tabularnewline
\hline 
\end{tabular}
\par\end{centering}
\caption{Likelihood fit results for a mock data set based on Model 3 where all normalization
constants in the mock data are equal to unity, except for the MSP template which features a total flux of $5\times10^{-7}$ ph/cm$^{2}$/s as well as all spectral corrections $\Delta \gamma$ being set to zero. In the header row of the table, we list the GCE templates that are used to create the mock data utilized to perform the fit with the results listed in the respective column. In all cases, the spatial GCE profile follows a gNFW density distribution with $\gamma=1.0$. 
Note that the reported statistical error is the standard deviation of the sample of the respective best-fit parameter resulting from 500 Poisson realisations of the prepared gamma-ray mock data. The 'recovered' MSP total flux value shown in the last row is calculated from the mean best-fit results for $\gamma$ and the DM normalization under the assumption that the original DM setup is known (i.e.~$A_{\mathrm{DM}} = 1$ and $\gamma = 1.0$). \label{tab:likelihood_fit_baseline}}
\end{table}

The optimal parameters derived from our likelihood fit to the real LAT data are incorporated into our findings in Figs. \ref{fig:results_scenarioA_model1}, \ref{fig:results_scenarioA_model2}, and \ref{fig:results_scenarioA_model3} as orange points. We note that the likelihood outcomes align with those generated by the network, particularly for the luminous gas emission components. However, for certain components, the network predictions fall outside the one-sigma statistical uncertainty, which is typically narrower in the likelihood outcomes.

The concurrence is less satisfactory for the fainter and less organized IC, FBs, and ISO components, where the likelihood outcomes reveal a systematically elevated normalization for the ISO component and a consistently reduced normalization for the IC and FBs components. While not flawless, the degree of agreement illustrates that the networks can effectively disentangle the data into the astrophysical components they were trained on.

In terms of 
the GCE decomposition (i.e. DM normalization), there is a significant discrepancy, which is expected, since the MSP component is not included in the likelihood approach.  In particular, the value of $A_{\mathrm{DM}}$ is systematically larger to accommodate for the missing MSP component, but also because the value of $\gamma$ is slightly smaller, affecting the total flux from the smooth DM component.

\subsection{Robustness of the results}
\label{sec:crossdomain-maintext}

{In Fig.~\ref{fig:summaryplot} we {summarize our networks' results for the parameters controlling the GCE excess, plotting them for all \gems~and scenarios.}}
{The parameter values are mostly consistent considering the large uncertainty, but there seems to be a trend towards lower values for $A_{DM}$ (or $F_{DM}$, respectively), and higher values for $F_{MSP}$, $\gamma$, and $f_{src}$ when going from the lower complexity of Model 1 to the higher complexity present in Model 3.}
{The value of $f_{src}$ is lower for Scenario B by construction. {This observation is in line with our expectations since in Scenario B a higher MSP-like luminosity is linked to a larger number of MSPs above the detection threshold in the simulated gamma-ray images (c.f.~Fig.~\ref{fig:MSP_threshold_numbers} in App.~\ref{app:msp_threshold}). The room in the real data for such artificially injected, bright sources in the GC is rather limited, thus lowering the value of $f_{src}$.}}
{The network thus consistently finds an excess over our astrophysical background models while the characterization of the results into a smooth and point-source-like component depends on the selected background model.  It is tempting to conclude that the trend of an increasing $f_{src}$ will continue as the \gem's complexity increases, but we caution against that because there is no guarantee that the increased complexity of the background model did not adversely affect the characterization of the GCE.  }

To make a direct comparison with our previous  work \cite{Caron:2017udl}, we retrain the Model 3 networks to predict merely $f_{src}$ while treating the remaining set of parameters as nuisance parameters, and label such networks '1D'. We observe that the predicted network uncertainty significantly shrinks in this case (as shown in Fig.~\ref{fig:summaryplot} middle panel), consistently with what is observed in \cite{Caron:2017udl}. This finding explains why the old results were more precise but not necessarily more accurate.

The trend of increasing $f_{src}$ from Model 1 to Model 3 rather suggests that the astrophysical background modeling is the dominating source of systematic uncertainty, an issue we focus on in the rest of this section.

\begin{figure}[h!]
\begin{center}
\includegraphics[width=\textwidth]{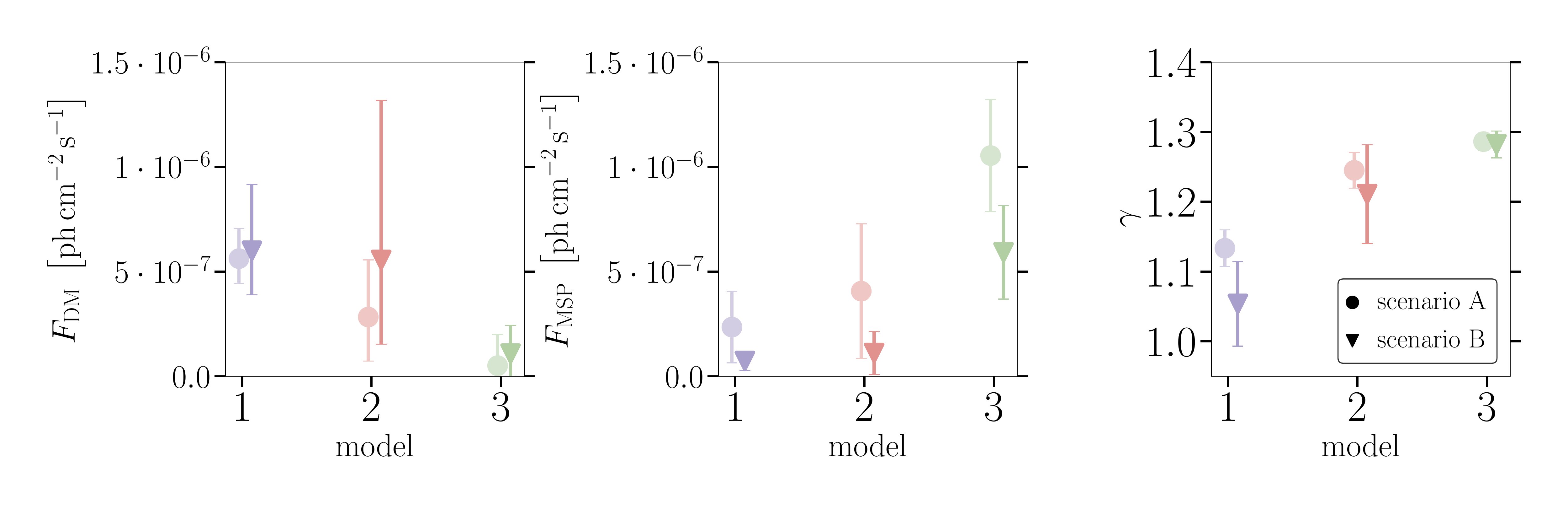}\\
\includegraphics[width=0.50\textwidth]{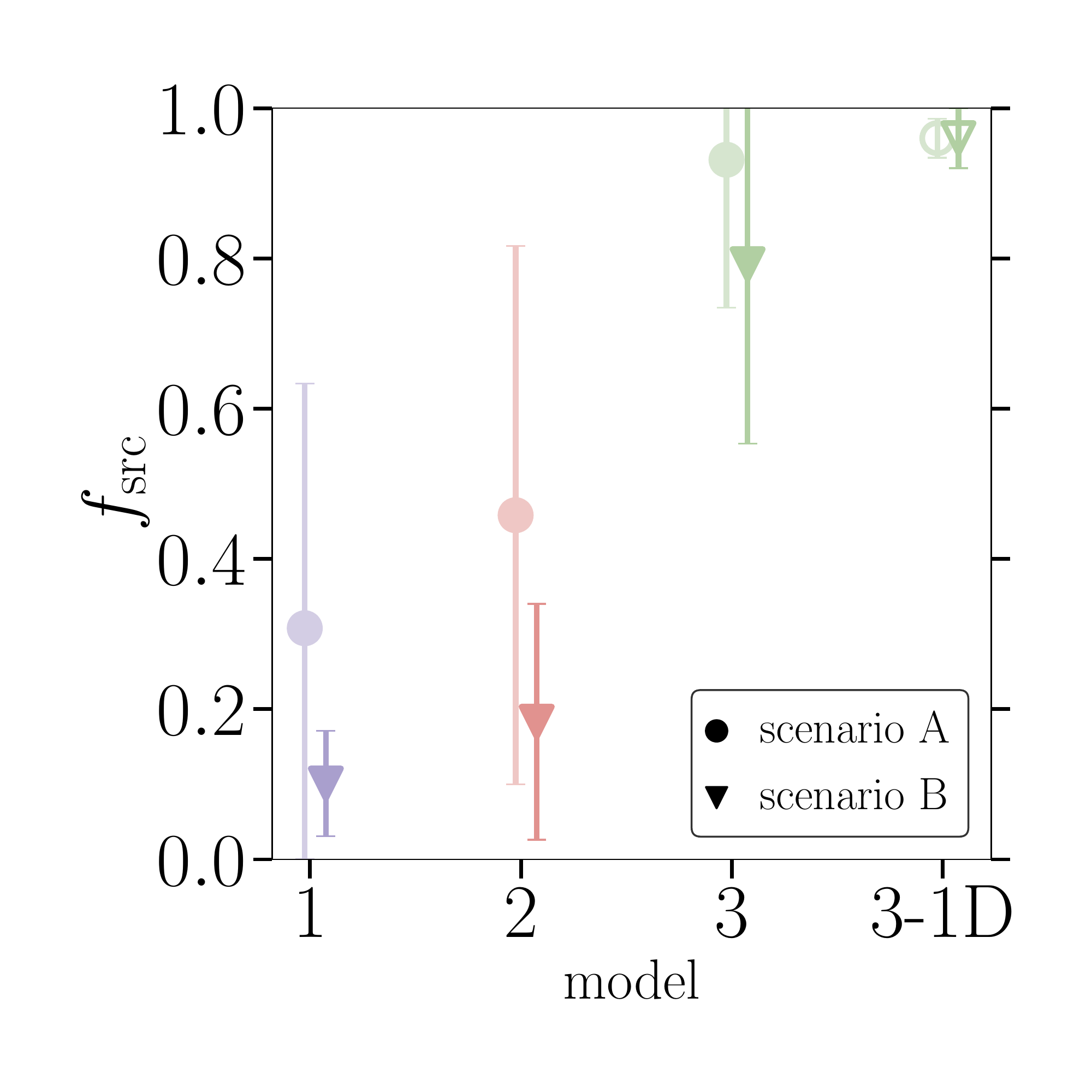}
\end{center}
\caption{{\it Top:} Results of the network prediction on the real data in our three Models and two Scenarios (circle: A; triangle: B) with respect to the parameters of the GCE. 
{\it Bottom:} Comparison of the network’s prediction for $f_{\rm src}$ per \gem~and scenario. 3-1D shows the result of networks trained only on one ($f_{\rm src}$) parameter (treating the  rest as nuisance parameters) when applied to Models 3.
\label{fig:summaryplot}}
\end{figure}


{\textbf{Outline of the robustness checks.}} To quantify its impact, we apply the network trained on simulated data created by one \gem~to data simulated by another \gem. {We do that in two steps. Firstly, we focus on cross-domain application, for various \gems~and scenario combinations of our {\gems~and scenarios}} (see Fig.~\ref{fig:domainadaptation}). {After that, we  extend our test by applying the networks trained on our \gems, to an 'independent' verification data set that is created under a range of substantially different assumptions. For this independent \gem, we chose the Galactic Interstellar Emission Model for the 4FGL Catalog Analysis (so-called Pass 8 Fermi diffuse model, \texttt{gll\_iem\_v07.fits})\footnote{See \url{https://fermi.gsfc.nasa.gov/ssc/data/analysis/software/aux/4fgl/Galactic_Diffuse_Emission_Model_for_the_4FGL_Catalog_Analysis.pdf}}, which is not based on GALPROP calculation of CR propagation but uses a different, data-driven approach to calculate gas emissivity (see \cite{Fermi-LAT:2016zaq} for details). It also uses 'patches' of selected sky regions when deriving the model, which is derived from the residual $\gamma$-ray intensity from fitting the overall \gem~in an iterative procedure.
As a consequence, the large-scale residual emission (including the (smoothed) GCE and FBs) is part of the diffuse model itself, in stark difference from the IEM our network was trained on. Despite these differences, the overall emission in the verification data set (that is built by summing up our usual PS, ISO, FB, and GCE templates together with this IEM), contains the emission components the network is expected to be broadly familiar with and should be able to quantify}. We note that the diffuse model \texttt{gll\_iem\_v07.fits} is not meant to be used for the analysis of large-scale gamma-ray sources (like the GCE) in real \textit{Fermi}-LAT data\footnote{See \url{https://fermi.gsfc.nasa.gov/ssc/data/analysis/software/aux/4fgl/Galactic_Diffuse_Emission_Model_for_the_4FGL_Catalog_Analysis.pdf} for a list of caveats compiled by the \textit{Fermi}-LAT collaboration.}. However, this caveat is not relevant for the purpose of our robustness check since it is solely based on Monte Carlo data.

We detail the outcomes of both of these approaches below. 

\begin{figure}[t!]
\begin{center}
\includegraphics[width=0.98\linewidth]{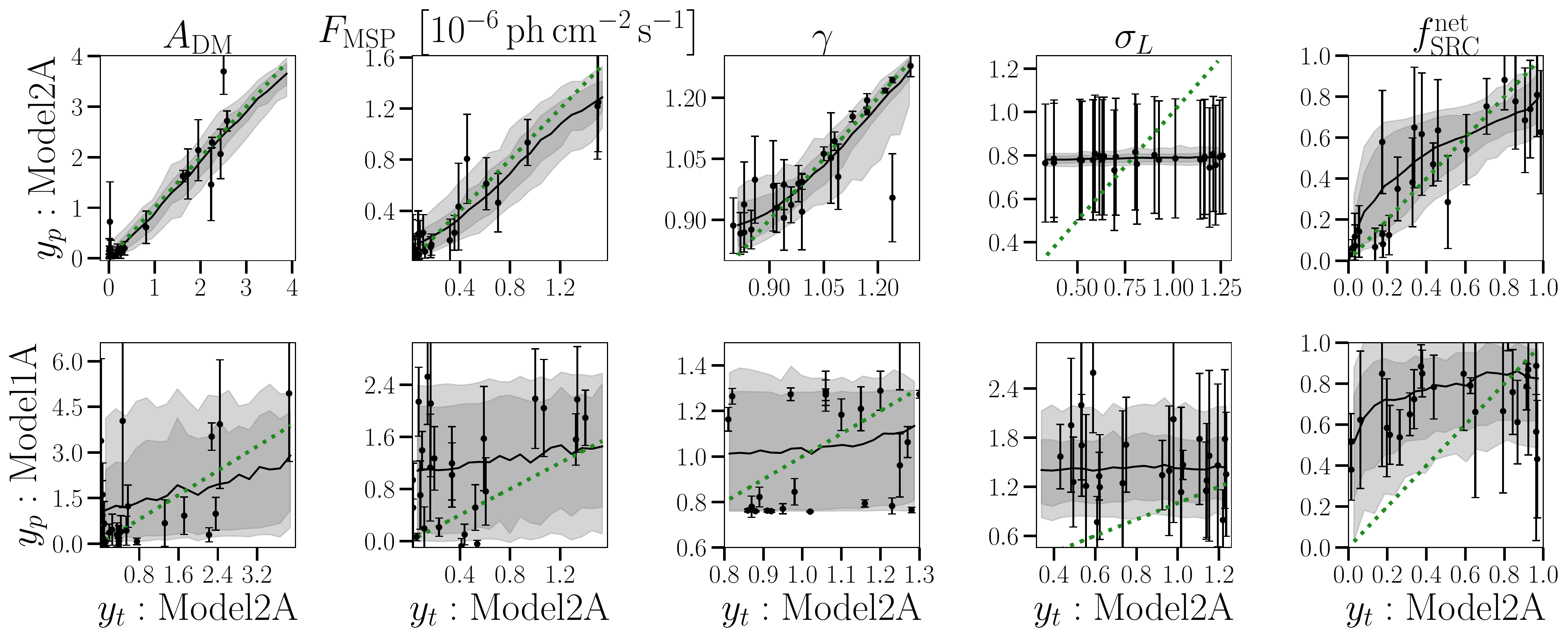}
\end{center}
\caption{Results of the cross-domain application with respect to validation data prepared within the framework of Model 2A applied to a fully trained network whose parameter reconstruction is tuned to Model 1A. (\emph{Top}:) Same as the results for Model 2A shown in Fig.~\ref{fig:results_scenarioA_model2} reduced to the five parameters characterizing the GCE. (\emph{Bottom}:) Performance of the network trained on Model 1A data ($y_p$) when given the validation data set of Model 2A ($y_t$). The color coding is adopted from Fig.~\ref{fig:results_scenarioA_model2}.
\label{fig:crossdomain_A1A2}}
\end{figure}

{\bf Cross domain applications.} Since our \gems~are nested, it is only necessary to consider a small subset of all the possible combinations of \gems~and scenarios. For us, it suffices to explore the cross-domain results obtained from using a network trained on a low number of parameters to data that exhibits a higher complexity.

The outcome of this cross-check -- focusing here on the GCE parameters -- is visualized in Fig.~\ref{fig:crossdomain_A1A2}. The top row of the plot re-states the results of network A2 on validation data of the same \gem~and scenario shown in Fig.~\ref{fig:results_scenarioA_model2}. The cross-domain application of the same validation data set to a network trained on Model 1A data is displayed in the lower row. 
We observe that in the lower panel (cross-domain results) the network results are clearly biased as all prediction means depart from the dotted, green diagonal line representing perfect reconstruction. Moreover, the predicted uncertainties are  underestimated in a significant number of cases. This shows very clearly that the networks are very sensitive to the background model in the training data, demonstrating nicely the relevance of the reality gap. {We stress that this observation is not an inherent limitation of our inference method based on NNs but rather the impact of the reality gap. A template-based approach using a maximum likelihood fit would be likewise affected.}

Besides the explicit cross-model robustness check in Fig.~\ref{fig:crossdomain_A1A2}, we have also quantitatively examined the impact of cross-domain analyses on data from different MSP scenarios (Scenarios A and B). Here, the induced bias of the network predictions is much less pronounced than before. The cross-domain results imply that a \gem~trained with data from scenario B is fairly robust against changing the MSP scenario to A. Almost all parameters are reconstructed without any additional biases. The reverse exercise, i.e.~letting a network trained on scenario A data reconstruct parameters of scenario B data sets, leads to stronger biases within the $1\sigma$ containment band of the scatter of the predicted values. Hence, the systematic model uncertainty introduced via the two MSP scenarios has substantially less impact on the accuracy and precision of the network's predictions.

{\bf New domain applications.}  We confirm this trend also by applying a network to a significantly different \gem, namely the Pass 8 Fermi diffuse model, where we observe that the performance of the networks {degrades when it comes to estimating the GCE components}. Fig. \ref{fig:Diffmod_3B} shows the results for Model 3B, while the rest of the cases are shown in Appendix \ref{sec:diffuse_bkg_check}. The bright emission components remain well predicted also in this case and predictions for some of the faint templates, namely that of the FBs behave also as expected. In particular, our networks tend to overpredict the normalization of the FB template, which is reasonable given the fact that the FBs are now both, part of the diffuse model itself and also added as a separate template. At the same time, the networks have a significantly harder time predicting the DM and MSP components, indicating possible internal biases. In Fig.~\ref{fig:Diffmod_3B} we observe, for example, that the MSP and ISO components are significantly over-predicted while the DM component is under-predicted. {This was \textit{a priori} not expected because the residual component of the Pass 8 Fermi diffuse model was deliberately smoothed so it would not contain emission from individual point sources.  We expected that the MSP component would be accurately predicted while $A_{DM}$ would be over-predicted.  A possible reason is that the smoothed residual in the Pass 8 Fermi model is not perfectly symmetric around the GC and therefore looks more similar to the MSP component.}
This corroborates our conclusion of a lack of robustness of the features {that our DENs can reconstruct due to the presence of the reality gap.} This lack of robustness was also present in all other attempts to determine the nature of GCE {as evident by the large and sometimes contradicting body of scientific articles on the nature of the GCE. In particular, the history of NPTF results described in Sec.~\ref{sec:gce_short_review} illustrates the impact of background mis-modeling and the presence of a reality gap in general. 

The work we present in this paper goes beyond the exploration of previous analyses that are subject to the reality gap/background mis-modeling. We lend more freedom to the astrophysical contribution -- a nested multi-parametric modeling with progressively increasing complexity -- to the GCE  compared to \cite{Leane:2019xiy, Chang:2019ars} where the astrophysical background composition and flexibility are essentially the same for the entire analysis (except for a few sanity checks). As seen in the results of the `new domain application', an artificially injected signal (here: smooth, DM-like) may be mis-attributed to the MSP contribution in presence of a substantial reality gap depending on the modeling of the diffuse astrophysical backgrounds. However, we did not observe a prominent influence of the 4FGL-DR2 templates on this matter. In the following section, we propose a method to quantify and explore the severity of the reality gap of a gamma-ray model of the GC region to minimize its impact in future works on the GCE.
}

\begin{figure}[t!]
\begin{center}
\includegraphics[width=0.95\linewidth]{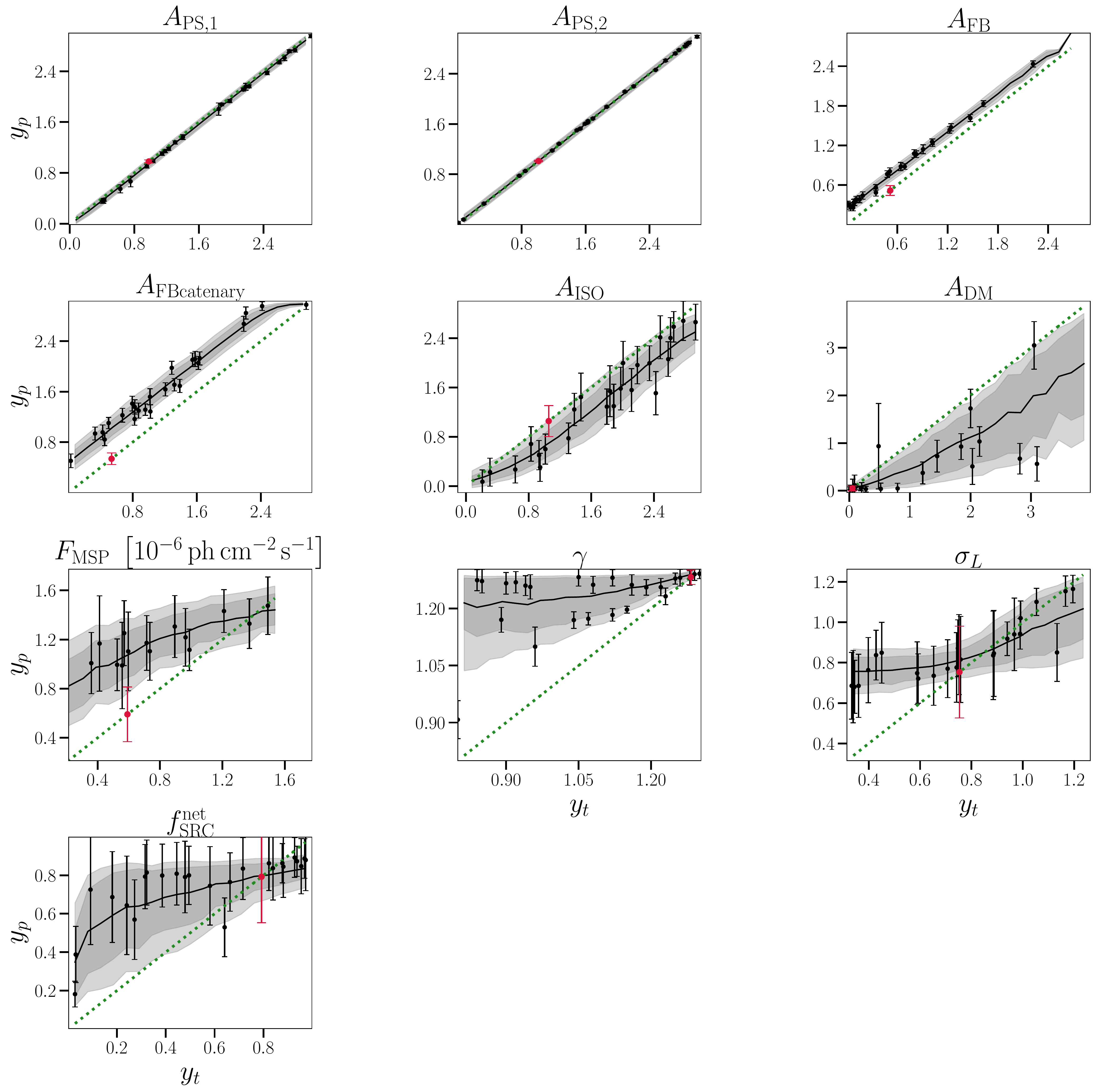}
\end{center}
\caption{Robustness check of the network trained on Model 3B data by applying a validation data set that contains the Fermi diffuse background model instead of the gas and IC templates of Model 3's set of diffuse components. The line and color style is adopted from Fig.~\ref{fig:results_scenarioA_model1}. We only show those parameters of Model 3B that are in common with the validation data set containing the Fermi diffuse background model.} \label{fig:Diffmod_3B}
\end{figure}

\subsection{Quantifying the reality gap}

To quantify the gap between our \gems~and a more accurate \gem~of the GCE, we used an ensemble of network models \cite{pmlr-v80-ruff18a, Caron_2022}. The Deep SVDD model is an encoding network that is trained to map the input, in our case a [120 x 120 x 5] tensor, to a constant vector with a certain dimensionality. This vector essentially defines a manifold onto which we compress the data.

The loss is defined as 
\begin{equation}
s(x) = \left [O_n^d - {\rm Model}~(x) \right ]^2,
\end{equation}
where the model maps the input x to the same tensor shape as the manifold $O$. In our case, it is a vector of identical scalar values, with the subscript $n$ defining the scalar value and superscript $d$ the number of elements in the vector. 

When optimizing the SVDD model, the loss function between the predictions for the training data set and the target is minimised. 
Therefore, data coming from the same joint distribution can be expected to have a lower reconstruction loss than data coming from a different distribution. In other words, a loss out of distribution with respect to the test data would quantify how different the real data is with respect to our \gem.

To verify this, we follow \cite{Caron_2022} and train an ensemble of SVDD models with different combinations of target output values $n = \left[10, 25, 50, 100\right]$ and dimensionalities $d = \left[5, 89, 144, 233\right]$. The SVDD architecture used is the network described in Sec. \ref{sec:NN}, where the last layer was replaced by a layer with as many neurons as the dimensionality of the output vector. The ML model was trained with the Adam optimiser with a learning rate of 0.01, halving the learning rate if the loss in the validation set did not improve over five epochs.

Finally, we used the trained SVDD models to predict the distance in a d-dimensional latent space between the encoded vector for the test data and the Fermi diffuse model data with respect to the target vector in the case of Model 3 \footnote{This is equivalent to compute the $L_2$ norm or Euclidean distance between two vectors in an n-dimensional space.}. We combined the results of the trained SVDD models resulting from the combination of n and d values, by adding them together, as shown in Fig.~\ref{fig:svdd} in the form of histograms. There one can see how the real data (red) are clearly out of distribution not only with respect to Model 3, but also to the Fermi diffuse model. We also checked the consistency of our results by using noise as input, which leads to the much larger prediction of distances (green line). Note that the plots are normalized in such a way that the distance to the maximally different \gem~is set to the value of 1.

\begin{figure}[t!]
\begin{center}
\includegraphics[width=0.49\columnwidth]{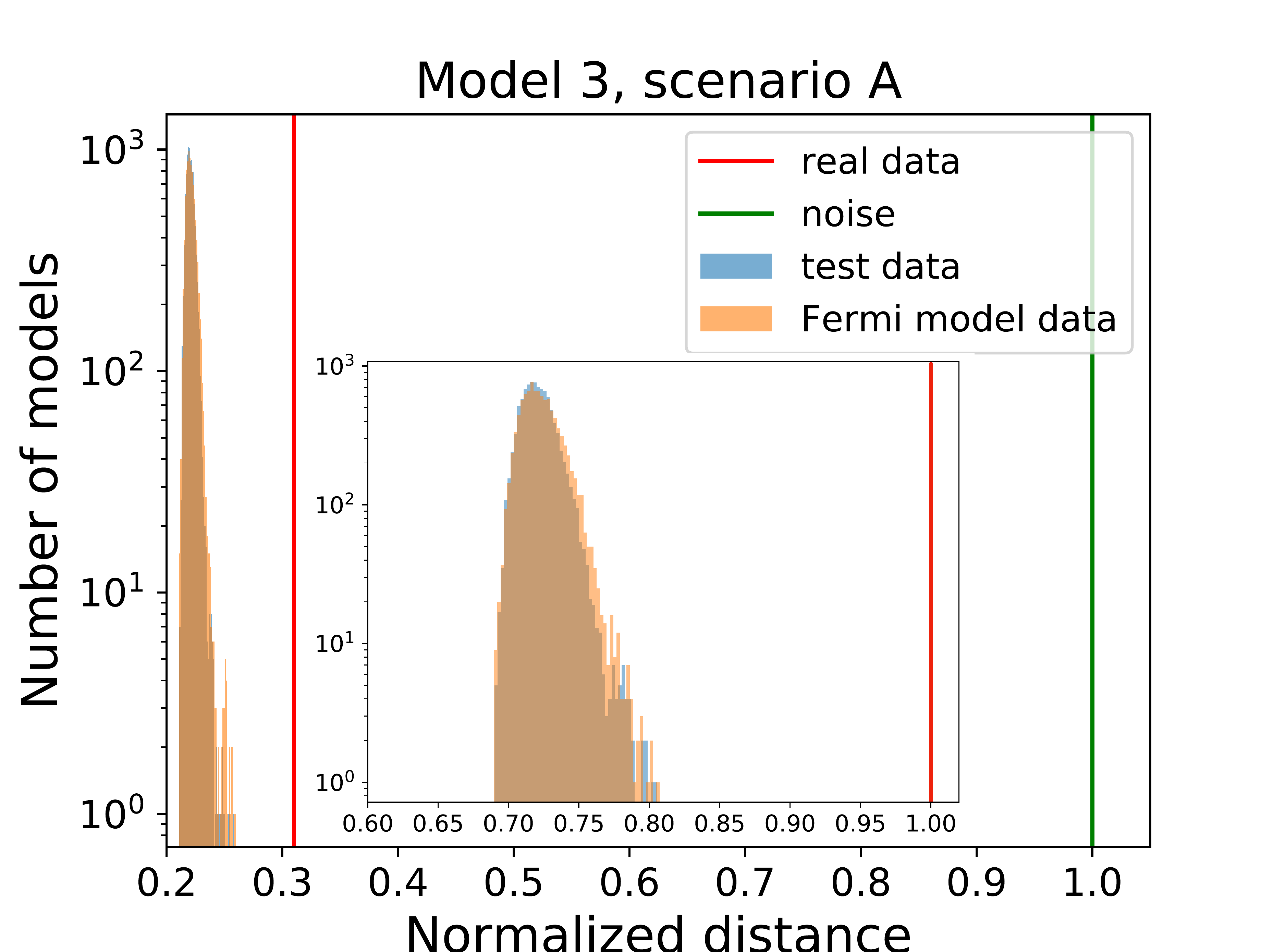} \hspace{-0.5cm}
\includegraphics[width=0.49\columnwidth]{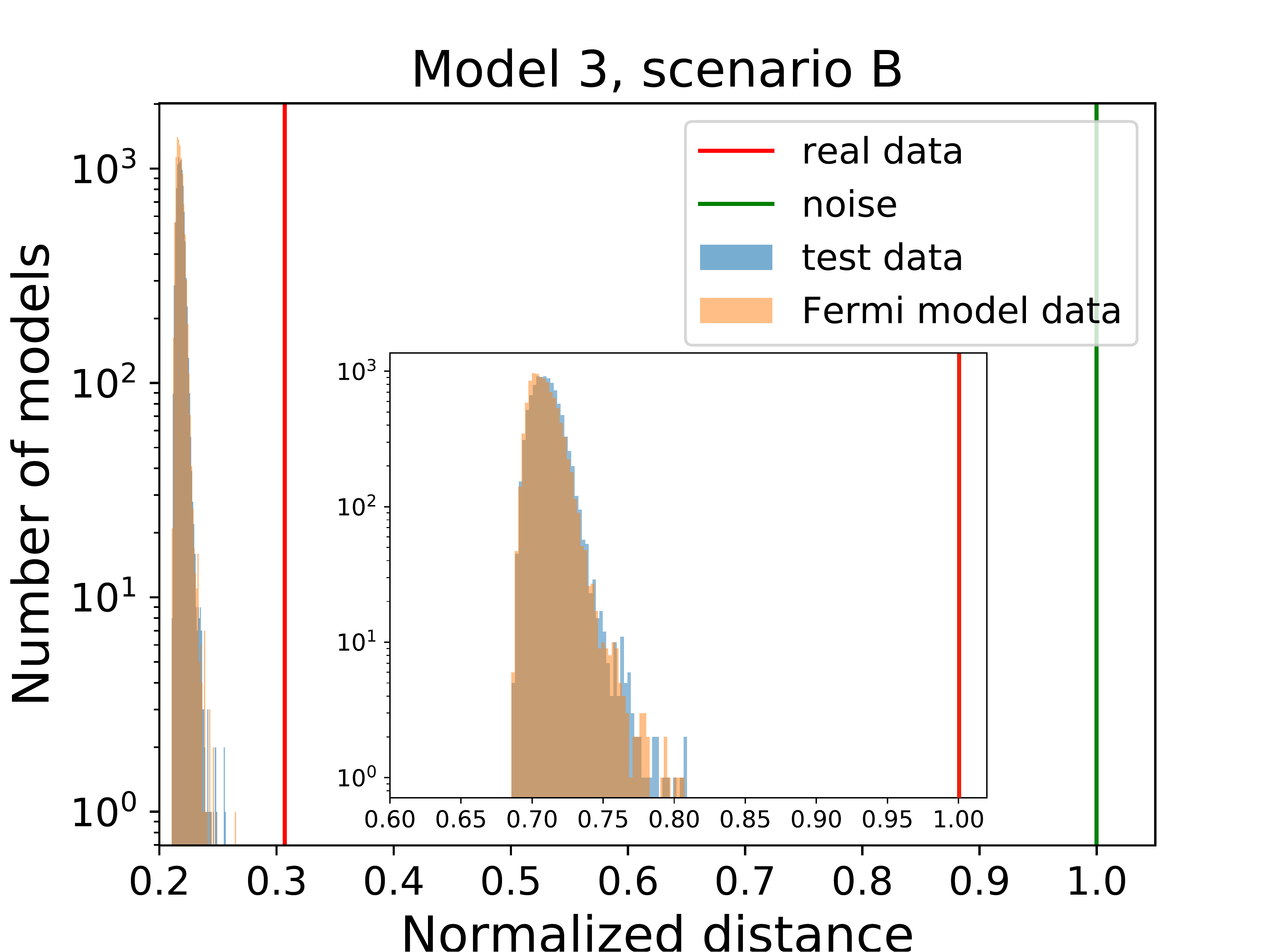}
\end{center}
\caption{SVDD prediction of the distance in latent space between the encoded vector for the test data (blue, Model 3A (\emph{left}) and Model 3B (\emph{right})), the Fermi diffuse model (orange), the real data (red line) and an image of the same dimensions filled only with Gaussian noise (green line). Note that the normalized distance in both inset figures is adjusted to better reflect the proximity of the real data to the training data compared to the pure noise.} \label{fig:svdd}
\end{figure}

\section{Summary and Future Prospects}
\label{sec:conclusion}


We use DeepEnsemble Networks to perform a detailed gamma-ray analysis of the complex Galactic center region. By applying the trained networks on the verification data set we demonstrate that they are capable of recovering components of {background} emission and the presence of the GCE. In particular:

\begin{itemize}
\item Bright components associated with conventional astrophysical processes are detected robustly and consistently between our \gem s, see Fig. \ref{fig:results_network_pred_rest}. They are also largely consistent with the prediction from the  traditional likelihood method and exhibit comparable uncertainties for most templates. 

\item The networks robustly detect the presence of the GCE in all our \gem s, with the properties (flux and spatial distribution) consistent with other works (see Fig. \ref{fig:GCE_spectrs_summary}).  

\item The nature of the GCE, however, while well predicted within each \gem, does not appear to be robust when the networks are applied outside of their training domain. We conduct a dedicated study of the limitation of our networks to successfully generalize the GCE nature and conclude that the reality gap remains the final obstacle in addressing the nature of GCE in our framework.

\item We further quantify the reality  gap by adopting Deep  SVDD architecture, which allows  us to compare the distance between the encoded vector for our \gem s and the real data rendering it possible to conclude that the real data is out-of-distribution even for the most complex \gem.

\item While this lack of robustness proved to be present in many other attempts to determine the nature of GCE -- as evidenced by the large and sometimes contradicting body of scientific articles on the nature of the GCE -- our work provides the first detailed study of such limitations for nested \gem s aimed at extracting the small-scale clustering of photons in gamma-ray data. More generally, whenever the NN application is strongly dependent on training data based on theoretical or data-driven models, it is imperative to study and quantify the reality gap before the results can be claimed robust.


\end{itemize}

In summary, this work represents a dedicated study of the often encountered issue of \gem s being an, at best, incomplete description of the full physics of the GC region. We analyzed the consequences of this reality gap in the setting of nested \gem s with increasing complexity to derive a quantitative statement about the contribution of point sources to the GCE; an approach never attempted in the literature. We obtain $f_{\mathrm{src}} = 0.10 \pm 0.07$ for Model 1B while Model 3B yields $f_{\mathrm{src}} = 0.79 \pm 0.24$. We consider Model 3 as the most reliable description of reality in our study. However, in light of the persistent reality gap, we are unable to faithfully quantify the error on $f_{\mathrm{src}}$ because our network's predictions do not include the additional uncertainty arising from the fact that the real data is outside of the investigated \gem~space. 
While our analysis employs NN as the means to infer the properties of the GCE  -- as it allows us to efficiently examine a range of \gem s and to account for various levels of uncertainties associated with background and signal (GCE) components -- the implications of our work hold in more generality irrespective of the chosen inference technique. We, therefore, argue that bridging the reality gap or understanding the limitations and biases it introduces, is an imperative step before being able to derive a verdict on the nature of the GCE. We invite the authors of all previous studies aiming at assessing the nature of the GCE to verify that their \gem~can bridge the reality gap to the LAT gamma-ray sky.


\begin{figure}[h!]
\begin{center}
\includegraphics[width=0.48\linewidth]{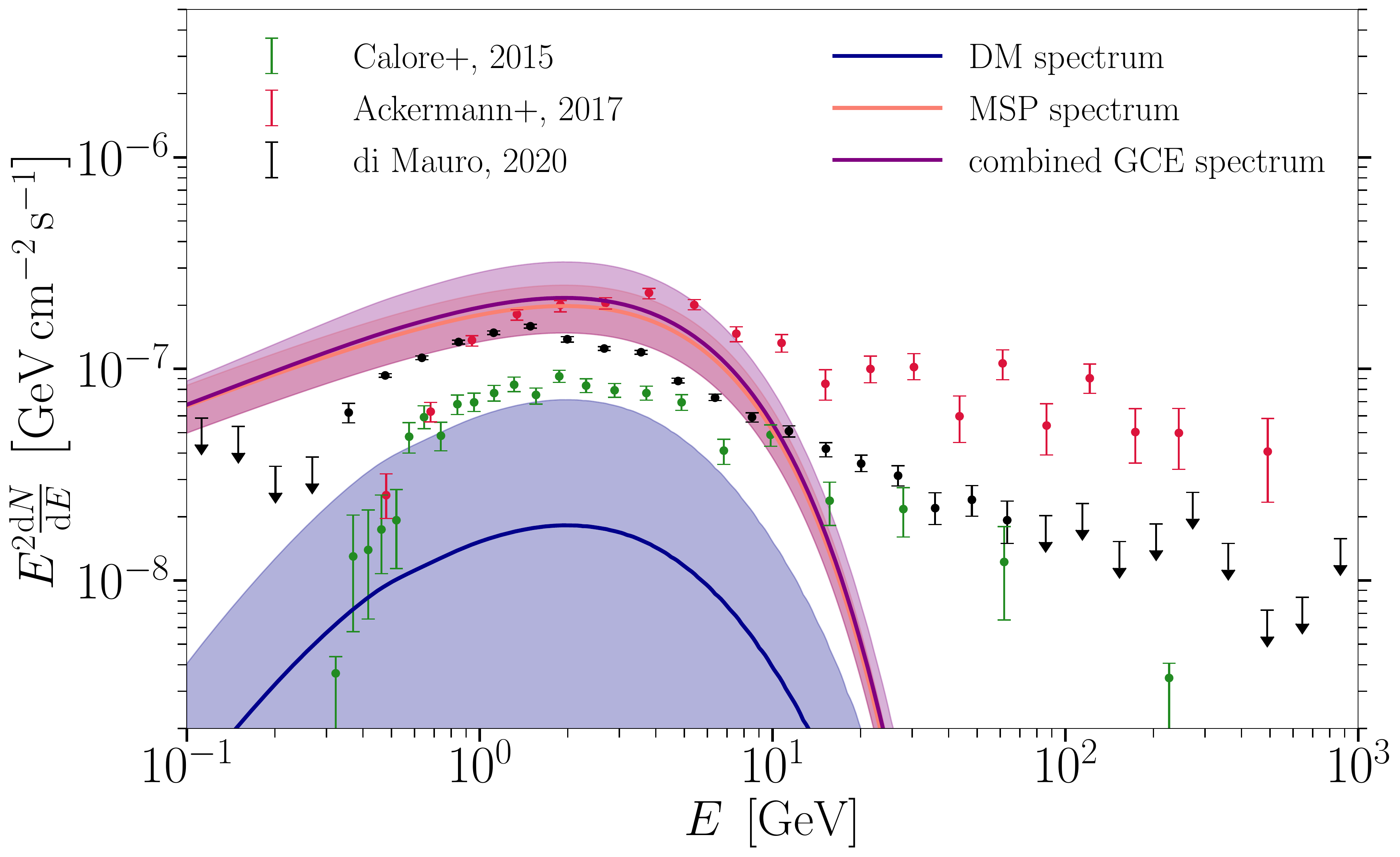}\hfill
\includegraphics[width=0.48\linewidth]{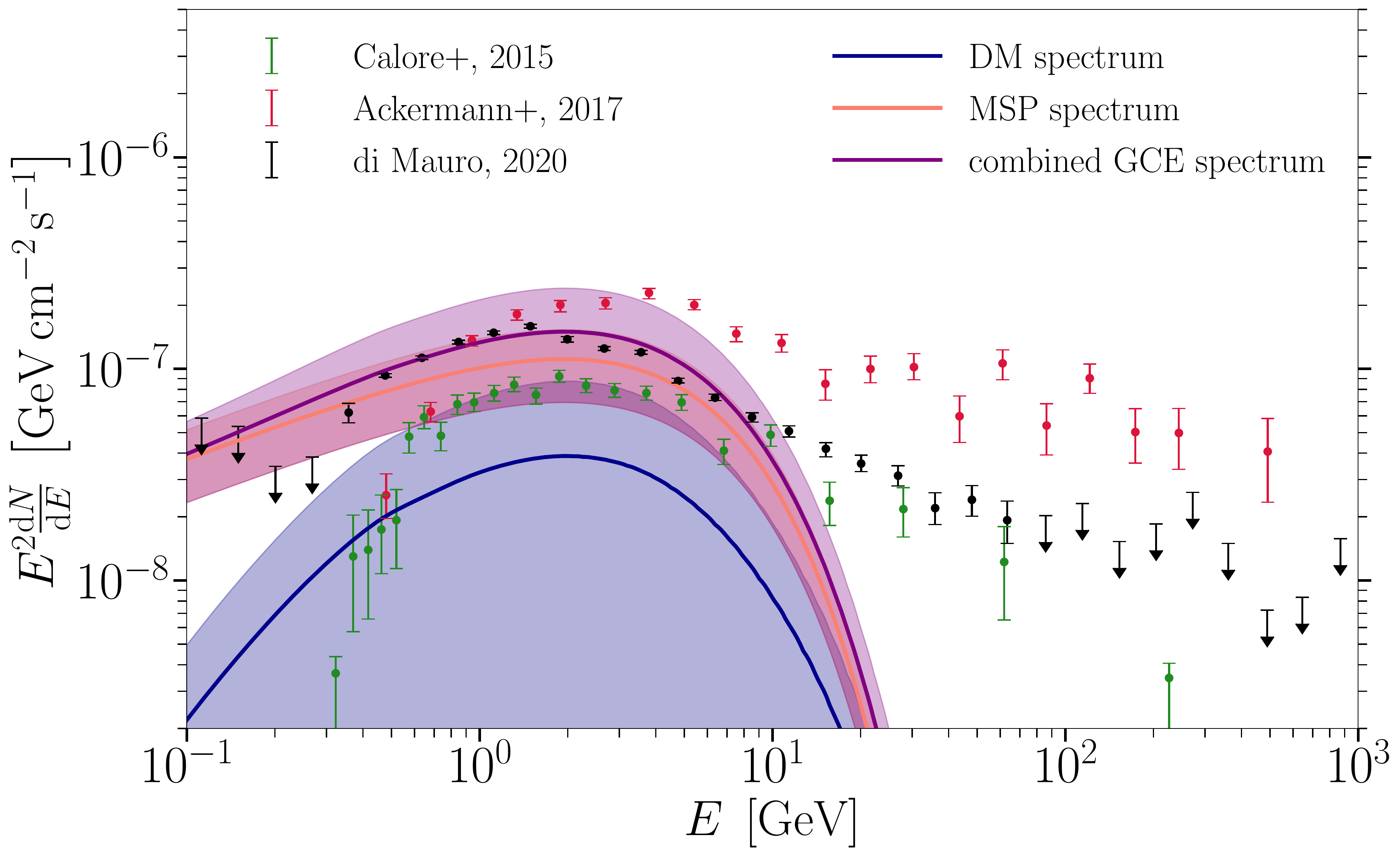}
\end{center}
\caption{Reconstructed spectrum of the GCE (purple) according to the network's prediction for the DM (blue) and MSP (light red) components based on training data of Model 3 + Scenario A ({\it Left}) and Model 3 + Scenario B ({\it Right}) compared with the spectrum as derived in previous analyses \cite{TheFermi-LAT:2017vmf, Calore:2014xka, DiMauro:2021raz}. The colour-shaded bands around each prediction denote the network's error estimate. For definiteness, we have fixed $\gamma = 1.28$, which is the best-fit value of our network regarding Model 3 in both scenarios. 
\label{fig:GCE_spectrs_summary}}
\end{figure}

\clearpage

\acknowledgments
The author(s) gratefully acknowledges the computer resources at Artemisa, funded by the European Union ERDF and Comunitat Valenciana as well as the technical support provided by the Instituto de Fisica Corpuscular, IFIC (CSIC-UV). R. RdA is supported by PID2020-113644GB-I00 from the Spanish Ministerio de Ciencia e Innovación. G.~Z.~and C.~E. acknowledge the financial support from the Slovenian Research Agency (grants P1-0031, I0-0033, J1-1700 and the Young Researcher program). C.~E.~further acknowledges support by the ``Agence Nationale de la Recherche'', grant n. ANR-19-CE31-0005-01 (PI: F. Calore). The work of C.~E.~has been supported by the EOSC Future project which is co-funded by the European Union Horizon Programme call INFRAEOSC-03-2020, Grant Agreement 101017536.

The \textit{Fermi} LAT Collaboration acknowledges generous ongoing support from a number of agencies and institutes that have supported both the development and the operation of the LAT as well as scientific data analysis. These include the National Aeronautics and Space Administration and the Department of Energy in the United States, the Commissariat \`a l'Energie Atomique and the Centre National de la Recherche Scientifique / Institut National de Physique Nucl\'eaire et de Physique des Particules in France, the Agenzia Spaziale Italiana and the Istituto Nazionale di Fisica Nucleare in Italy, the Ministry of Education, Culture, Sports, Science and Technology (MEXT), High Energy Accelerator Research Organization (KEK) and Japan Aerospace Exploration Agency (JAXA) in Japan, and the K.~A.~Wallenberg Foundation, the Swedish Research Council and the Swedish National Space Board in Sweden.
 
Additional support for science analysis during the operations phase is gratefully acknowledged from the Istituto Nazionale di Astrofisica in Italy and the Centre National d'\'Etudes Spatiales in France. This work performed in part under DOE Contract DE-AC02-76SF00515.

\bibliographystyle{JHEP}
\bibliography{GCEwML}

\providecommand{\href}[2]{#2}\begingroup\raggedright\begin{thebibliography}{100}

\bibitem{Ackermann:2012pya}
{\scshape Fermi-LAT} collaboration, M.~Ackermann et~al., \emph{{Fermi-LAT
  Observations of the Diffuse Gamma-Ray Emission: Implications for Cosmic Rays
  and the Interstellar Medium}},
  \href{http://dx.doi.org/10.1088/0004-637X/750/1/3}{\emph{Astrophys. J.} {\bf
  750} (2012) 3}, [\href{http://arxiv.org/abs/1202.4039}{{\tt 1202.4039}}].

\bibitem{Zhou:2017lgv}
{\scshape HAWC} collaboration, H.~Zhou, C.~D. Rho and G.~Vianello,
  \emph{{Probing Galactic Diffuse TeV Gamma-Ray Emission with the HAWC
  Observatory}}, \href{http://dx.doi.org/10.22323/1.301.0689}{\emph{PoS} {\bf
  ICRC2017} (2018) 689}, [\href{http://arxiv.org/abs/1709.03619}{{\tt
  1709.03619}}].

\bibitem{Aharonian:2006au}
{\scshape H.E.S.S.} collaboration, F.~Aharonian et~al., \emph{{Discovery of
  very-high-energy gamma-rays from the galactic centre ridge}},
  \href{http://dx.doi.org/10.1038/nature04467}{\emph{Nature} {\bf 439} (2006)
  695--698}, [\href{http://arxiv.org/abs/astro-ph/0603021}{{\tt
  astro-ph/0603021}}].

\bibitem{Abramowski:2014vox}
{\scshape H.E.S.S.} collaboration, A.~Abramowski et~al., \emph{{Diffuse
  Galactic gamma-ray emission with H.E.S.S}},
  \href{http://dx.doi.org/10.1103/PhysRevD.90.122007}{\emph{Phys. Rev.} {\bf
  D90} (2014) 122007}, [\href{http://arxiv.org/abs/1411.7568}{{\tt
  1411.7568}}].

\bibitem{Abdalla:2017xja}
{\scshape HESS} collaboration, H.~Abdalla et~al., \emph{{Characterising the VHE
  diffuse emission in the central 200 parsecs of our Galaxy with H.E.S.S}},
  \href{http://dx.doi.org/10.1051/0004-6361/201730824}{\emph{Astron.
  Astrophys.} {\bf 612} (2018) A9},
  [\href{http://arxiv.org/abs/1706.04535}{{\tt 1706.04535}}].

\bibitem{2006ApJ...640L.155M}
I.~V. {Moskalenko}, T.~A. {Porter} and A.~W. {Strong}, \emph{{Attenuation of
  Very High Energy Gamma Rays by the Milky Way Interstellar Radiation Field}},
  \href{http://dx.doi.org/10.1086/503524}{\emph{\apjl} {\bf 640} (Apr., 2006)
  L155--L158}, [\href{http://arxiv.org/abs/astro-ph/0511149}{{\tt
  astro-ph/0511149}}].

\bibitem{2018PhRvD..98d1302P}
T.~A. {Porter}, G.~P. {Rowell}, G.~{J{\'o}hannesson} and I.~V. {Moskalenko},
  \emph{{Galactic PeVatrons and helping to find them: Effects of galactic
  absorption on the observed spectra of very high energy {\ensuremath{\gamma}}
  -ray sources}},
  \href{http://dx.doi.org/10.1103/PhysRevD.98.041302}{\emph{\prd} {\bf 98}
  (Aug., 2018) 041302}, [\href{http://arxiv.org/abs/1808.07596}{{\tt
  1808.07596}}].

\bibitem{Genzel:2010zy}
R.~Genzel, F.~Eisenhauer and S.~Gillessen, \emph{{The Galactic Center Massive
  Black Hole and Nuclear Star Cluster}},
  \href{http://dx.doi.org/10.1103/RevModPhys.82.3121}{\emph{Rev. Mod. Phys.}
  {\bf 82} (2010) 3121--3195}, [\href{http://arxiv.org/abs/1006.0064}{{\tt
  1006.0064}}].

\bibitem{Barnes_2017}
A.~T. Barnes, S.~N. Longmore, C.~Battersby, J.~Bally, J.~M.~D. Kruijssen, J.~D.
  Henshaw et~al., \emph{Star formation rates and efficiencies in the galactic
  centre}, \href{http://dx.doi.org/10.1093/mnras/stx941}{\emph{Monthly Notices
  of the Royal Astronomical Society} {\bf 469} (Apr, 2017) 2263–2285}.

\bibitem{Petrovic:2014uda}
J.~Petrovi\'c, P.~D. Serpico and G.~Zaharija\v{s}, \emph{{Galactic Center
  gamma-ray ''excess'' from an active past of the Galactic Centre?}},
  \href{http://dx.doi.org/10.1088/1475-7516/2014/10/052}{\emph{JCAP} {\bf 10}
  (2014) 052}, [\href{http://arxiv.org/abs/1405.7928}{{\tt 1405.7928}}].

\bibitem{Gaggero:2017jts}
D.~Gaggero, D.~Grasso, A.~Marinelli, M.~Taoso and A.~Urbano, \emph{{Diffuse
  cosmic rays shining in the Galactic center: A novel interpretation of
  H.E.S.S. and Fermi-LAT $\gamma$-ray data}},
  \href{http://dx.doi.org/10.1103/PhysRevLett.119.031101}{\emph{Phys. Rev.
  Lett.} {\bf 119} (2017) 031101}, [\href{http://arxiv.org/abs/1702.01124}{{\tt
  1702.01124}}].

\bibitem{Carlson:2016iis}
E.~Carlson, T.~Linden and S.~Profumo, \emph{{Improved Cosmic-Ray Injection
  Models and the Galactic Center Gamma-Ray Excess}},
  \href{http://dx.doi.org/10.1103/PhysRevD.94.063504}{\emph{Phys. Rev. D} {\bf
  94} (2016) 063504}, [\href{http://arxiv.org/abs/1603.06584}{{\tt
  1603.06584}}].

\bibitem{2010ApJ...717..825D}
G.~{Dobler}, D.~P. {Finkbeiner}, I.~{Cholis}, T.~{Slatyer} and N.~{Weiner},
  \emph{{The Fermi Haze: A Gamma-ray Counterpart to the Microwave Haze}},
  \href{http://dx.doi.org/10.1088/0004-637X/717/2/825}{\emph{\apj} {\bf 717}
  (July, 2010) 825--842}, [\href{http://arxiv.org/abs/0910.4583}{{\tt
  0910.4583}}].

\bibitem{TheFermi-LAT:2017vmf}
{\scshape Fermi-LAT} collaboration, M.~Ackermann et~al., \emph{{The Fermi
  Galactic Center GeV Excess and Implications for Dark Matter}},
  \href{http://dx.doi.org/10.3847/1538-4357/aa6cab}{\emph{Astrophys. J.} {\bf
  840} (2017) 43}, [\href{http://arxiv.org/abs/1704.03910}{{\tt 1704.03910}}].

\bibitem{2009arXiv0910.2998G}
L.~{Goodenough} and D.~{Hooper}, \emph{{Possible Evidence For Dark Matter
  Annihilation In The Inner Milky Way From The Fermi Gamma Ray Space
  Telescope}}, {\emph{arXiv e-prints} (Oct., 2009) arXiv:0910.2998},
  [\href{http://arxiv.org/abs/0910.2998}{{\tt 0910.2998}}].

\bibitem{Vitale:2009hr}
{\scshape Fermi-LAT} collaboration, V.~Vitale and A.~Morselli, \emph{{Indirect
  Search for Dark Matter from the center of the Milky Way with the Fermi-Large
  Area Telescope}},  in \emph{{2009 Fermi Symposium}}, 12, 2009.
\newblock \href{http://arxiv.org/abs/0912.3828}{{\tt 0912.3828}}.

\bibitem{Hooper:2010mq}
D.~Hooper and L.~Goodenough, \emph{{Dark Matter Annihilation in The Galactic
  Center As Seen by the Fermi Gamma Ray Space Telescope}},
  \href{http://dx.doi.org/10.1016/j.physletb.2011.02.029}{\emph{Phys.Lett.}
  {\bf B697} (2011) 412--428}, [\href{http://arxiv.org/abs/1010.2752}{{\tt
  1010.2752}}].

\bibitem{Abazajian:2010zy}
K.~N. Abazajian, \emph{{The Consistency of Fermi-LAT Observations of the
  Galactic Center with a Millisecond Pulsar Population in the Central Stellar
  Cluster}}, \href{http://dx.doi.org/10.1088/1475-7516/2011/03/010}{\emph{JCAP}
  {\bf 03} (2011) 010}, [\href{http://arxiv.org/abs/1011.4275}{{\tt
  1011.4275}}].

\bibitem{Hooper:2011ti}
D.~Hooper and T.~Linden, \emph{{On The Origin Of The Gamma Rays From The
  Galactic Center}},
  \href{http://dx.doi.org/10.1103/PhysRevD.84.123005}{\emph{Phys. Rev. D} {\bf
  84} (2011) 123005}, [\href{http://arxiv.org/abs/1110.0006}{{\tt 1110.0006}}].

\bibitem{Abazajian:2012pn}
K.~N. Abazajian and M.~Kaplinghat, \emph{{Detection of a Gamma-Ray Source in
  the Galactic Center Consistent with Extended Emission from Dark Matter
  Annihilation and Concentrated Astrophysical Emission}},
  \href{http://dx.doi.org/10.1103/PhysRevD.86.083511}{\emph{Phys.Rev.} {\bf
  D86} (2012) 083511}, [\href{http://arxiv.org/abs/1207.6047}{{\tt
  1207.6047}}].

\bibitem{Hooper:2013nhl}
D.~Hooper, I.~Cholis, T.~Linden, J.~Siegal-Gaskins and T.~Slatyer,
  \emph{{Pulsars Cannot Account for the Inner Galaxy's GeV Excess}},
  \href{http://dx.doi.org/10.1103/PhysRevD.88.083009}{\emph{Phys. Rev. D} {\bf
  88} (2013) 083009}, [\href{http://arxiv.org/abs/1305.0830}{{\tt 1305.0830}}].

\bibitem{Gordon:2013vta}
C.~Gordon and O.~Macias, \emph{{Dark Matter and Pulsar Model Constraints from
  Galactic Center Fermi-LAT Gamma Ray Observations}},
  \href{http://dx.doi.org/10.1103/PhysRevD.88.083521}{\emph{Phys. Rev. D} {\bf
  88} (2013) 083521}, [\href{http://arxiv.org/abs/1306.5725}{{\tt 1306.5725}}].

\bibitem{Macias:2013vya}
O.~Macias and C.~Gordon, \emph{{The Contribution of Cosmic Rays Interacting
  With Molecular Clouds to the Galactic Center Gamma-Ray Excess}},
  \href{http://dx.doi.org/10.1103/PhysRevD.89.063515}{\emph{Phys.Rev.} {\bf
  D89} (2014) 063515}, [\href{http://arxiv.org/abs/1312.6671}{{\tt
  1312.6671}}].

\bibitem{Daylan:2014rsa}
T.~Daylan, D.~P. Finkbeiner, D.~Hooper, T.~Linden, S.~K.~N. Portillo et~al.,
  \emph{{The Characterization of the Gamma-Ray Signal from the Central Milky
  Way: A Compelling Case for Annihilating Dark Matter}},
  \href{http://arxiv.org/abs/1402.6703}{{\tt 1402.6703}}.

\bibitem{Abazajian:2014fta}
K.~N. Abazajian, N.~Canac, S.~Horiuchi and M.~Kaplinghat, \emph{{Astrophysical
  and Dark Matter Interpretations of Extended Gamma Ray Emission from the
  Galactic Center}},  \href{http://arxiv.org/abs/1402.4090}{{\tt 1402.4090}}.

\bibitem{Zhou:2014lva}
B.~Zhou, Y.-F. Liang, X.~Huang, X.~Li, Y.-Z. Fan, L.~Feng et~al., \emph{{GeV
  excess in the Milky Way: The role of diffuse galactic gamma-ray emission
  templates}}, \href{http://dx.doi.org/10.1103/PhysRevD.91.123010}{\emph{Phys.
  Rev. D} {\bf 91} (2015) 123010}, [\href{http://arxiv.org/abs/1406.6948}{{\tt
  1406.6948}}].

\bibitem{Huang:2015rlu}
X.~Huang, T.~En\ss{}lin and M.~Selig, \emph{{Galactic dark matter search via
  phenomenological astrophysics modeling}},
  \href{http://dx.doi.org/10.1088/1475-7516/2016/04/030}{\emph{JCAP} {\bf 04}
  (2016) 030}, [\href{http://arxiv.org/abs/1511.02621}{{\tt 1511.02621}}].

\bibitem{Fermi-LAT:2015sau}
{\scshape Fermi-LAT} collaboration, M.~Ajello et~al., \emph{{Fermi-LAT
  Observations of High-Energy $\gamma$-Ray Emission Toward the Galactic
  Center}},
  \href{http://dx.doi.org/10.3847/0004-637X/819/1/44}{\emph{Astrophys. J.} {\bf
  819} (2016) 44}, [\href{http://arxiv.org/abs/1511.02938}{{\tt 1511.02938}}].

\bibitem{Calore:2014nla}
F.~Calore, I.~Cholis, C.~McCabe and C.~Weniger, \emph{{A Tale of Tails: Dark
  Matter Interpretations of the Fermi GeV Excess in Light of Background Model
  Systematics}},
  \href{http://dx.doi.org/10.1103/PhysRevD.91.063003}{\emph{Phys. Rev. D} {\bf
  91} (2015) 063003}, [\href{http://arxiv.org/abs/1411.4647}{{\tt 1411.4647}}].

\bibitem{Calore:2014xka}
F.~Calore, I.~Cholis and C.~Weniger, \emph{{Background Model Systematics for
  the Fermi GeV Excess}},
  \href{http://dx.doi.org/10.1088/1475-7516/2015/03/038}{\emph{JCAP} {\bf 03}
  (2015) 038}, [\href{http://arxiv.org/abs/1409.0042}{{\tt 1409.0042}}].

\bibitem{DiMauro:2021raz}
M.~Di~Mauro, \emph{{The characteristics of the Galactic center excess measured
  with 11 years of Fermi-LAT data}},
  \href{http://arxiv.org/abs/2101.04694}{{\tt 2101.04694}}.

\bibitem{Cholis:2021rpp}
I.~Cholis, Y.-M. Zhong, S.~D. McDermott and J.~P. Surdutovich, \emph{{Return of
  the templates: Revisiting the Galactic Center excess with multimessenger
  observations}},
  \href{http://dx.doi.org/10.1103/PhysRevD.105.103023}{\emph{Phys. Rev. D} {\bf
  105} (2022) 103023}, [\href{http://arxiv.org/abs/2112.09706}{{\tt
  2112.09706}}].

\bibitem{McDermott:2022zmq}
S.~D. McDermott, Y.-M. Zhong and I.~Cholis, \emph{{A Phantom Menace: On the
  Morphology of the Galactic Center Excess}},
  \href{http://arxiv.org/abs/2209.00006}{{\tt 2209.00006}}.

\bibitem{Bartels:2015aea}
R.~Bartels, S.~Krishnamurthy and C.~Weniger, \emph{{Strong support for the
  millisecond pulsar origin of the Galactic center GeV excess}},
  \href{http://dx.doi.org/10.1103/PhysRevLett.116.051102}{\emph{Phys. Rev.
  Lett.} {\bf 116} (2016) 051102}, [\href{http://arxiv.org/abs/1506.05104}{{\tt
  1506.05104}}].

\bibitem{Lee:2015fea}
S.~K. Lee, M.~Lisanti, B.~R. Safdi, T.~R. Slatyer and W.~Xue, \emph{{Evidence
  for Unresolved $\gamma$-Ray Point Sources in the Inner Galaxy}},
  \href{http://dx.doi.org/10.1103/PhysRevLett.116.051103}{\emph{Phys. Rev.
  Lett.} {\bf 116} (2016) 051103}, [\href{http://arxiv.org/abs/1506.05124}{{\tt
  1506.05124}}].

\bibitem{Ploeg:2017vai}
H.~Ploeg, C.~Gordon, R.~Crocker and O.~Macias, \emph{{Consistency Between the
  Luminosity Function of Resolved Millisecond Pulsars and the Galactic Center
  Excess}}, \href{http://dx.doi.org/10.1088/1475-7516/2017/08/015}{\emph{JCAP}
  {\bf 08} (2017) 015}, [\href{http://arxiv.org/abs/1705.00806}{{\tt
  1705.00806}}].

\bibitem{Eckner:2017oul}
C.~Eckner et~al., \emph{{Millisecond pulsar origin of the Galactic center
  excess and extended gamma-ray emission from Andromeda - a closer look}},
  \href{http://dx.doi.org/10.3847/1538-4357/aac029}{\emph{Astrophys. J.} {\bf
  862} (2018) 79}, [\href{http://arxiv.org/abs/1711.05127}{{\tt 1711.05127}}].

\bibitem{Fragione:2017rsp}
G.~Fragione, F.~Antonini and O.~Y. Gnedin, \emph{{Disrupted Globular Clusters
  and the Gamma-Ray Excess in the Galactic Centre}},
  \href{http://dx.doi.org/10.1093/mnras/sty183}{\emph{Mon. Not. Roy. Astron.
  Soc.} {\bf 475} (2018) 5313--5321},
  [\href{http://arxiv.org/abs/1709.03534}{{\tt 1709.03534}}].

\bibitem{Fragione:2018jxd}
G.~Fragione, V.~Pavl\'\i{}k and S.~Banerjee, \emph{{Neutron stars and
  millisecond pulsars in star clusters: implications for the diffuse
  $\gamma$-radiation from the Galactic Centre}},
  \href{http://dx.doi.org/10.1093/mnras/sty2234}{\emph{Mon. Not. Roy. Astron.
  Soc.} {\bf 480} (2018) 4955--4962},
  [\href{http://arxiv.org/abs/1804.04856}{{\tt 1804.04856}}].

\bibitem{Gonthier:2018ymi}
P.~L. Gonthier, A.~K. Harding, E.~C. Ferrara, S.~E. Frederick, V.~E. Mohr and
  Y.-M. Koh, \emph{{Population syntheses of millisecond pulsars from the
  Galactic Disk and Bulge}},
  \href{http://dx.doi.org/10.3847/1538-4357/aad08d}{\emph{Astrophys. J.} {\bf
  863} (2018) 199}, [\href{http://arxiv.org/abs/1806.11215}{{\tt 1806.11215}}].

\bibitem{Ploeg:2020jeh}
H.~Ploeg, C.~Gordon, R.~Crocker and O.~Macias, \emph{{Comparing the Galactic
  Bulge and Galactic Disk Millisecond Pulsars}},
  \href{http://dx.doi.org/10.1088/1475-7516/2020/12/035}{\emph{JCAP} {\bf 12}
  (2020) 035}, [\href{http://arxiv.org/abs/2008.10821}{{\tt 2008.10821}}].

\bibitem{deBoer:2017sxb}
W.~de~Boer, L.~Bosse, I.~Gebauer, A.~Neumann and P.~L. Biermann,
  \emph{{Molecular clouds as origin of the Fermi gamma-ray GeV excess}},
  \href{http://dx.doi.org/10.1103/PhysRevD.96.043012}{\emph{Phys. Rev. D} {\bf
  96} (2017) 043012}, [\href{http://arxiv.org/abs/1707.08653}{{\tt
  1707.08653}}].

\bibitem{Carlson:2014cwa}
E.~Carlson and S.~Profumo, \emph{{Cosmic Ray Protons in the Inner Galaxy and
  the Galactic Center Gamma-Ray Excess}},
  \href{http://dx.doi.org/10.1103/PhysRevD.90.023015}{\emph{Phys. Rev. D} {\bf
  90} (2014) 023015}, [\href{http://arxiv.org/abs/1405.7685}{{\tt 1405.7685}}].

\bibitem{Gaggero:2015nsa}
D.~Gaggero, M.~Taoso, A.~Urbano, M.~Valli and P.~Ullio, \emph{{Towards a
  realistic astrophysical interpretation of the gamma-ray Galactic center
  excess}}, \href{http://dx.doi.org/10.1088/1475-7516-2015-12-056}{\emph{JCAP}
  {\bf 12} (2015) 056}, [\href{http://arxiv.org/abs/1507.06129}{{\tt
  1507.06129}}].

\bibitem{Goodenough:2009gk}
L.~Goodenough and D.~Hooper, \emph{{Possible Evidence For Dark Matter
  Annihilation In The Inner Milky Way From The Fermi Gamma Ray Space
  Telescope}},  \href{http://arxiv.org/abs/0910.2998}{{\tt 0910.2998}}.

\bibitem{Dinsmore:2021nip}
J.~T. Dinsmore and T.~R. Slatyer, \emph{{Luminosity functions consistent with a
  pulsar-dominated Galactic Center excess}},
  \href{http://dx.doi.org/10.1088/1475-7516/2022/06/025}{\emph{JCAP} {\bf 06}
  (2022) 025}, [\href{http://arxiv.org/abs/2112.09699}{{\tt 2112.09699}}].

\bibitem{Pohl:2022nnd}
M.~Pohl, O.~Macias, P.~Coleman and C.~Gordon, \emph{{Assessing the Impact of
  Hydrogen Absorption on the Characteristics of the Galactic Center Excess}},
  \href{http://dx.doi.org/10.3847/1538-4357/ac6032}{\emph{Astrophys. J.} {\bf
  929} (2022) 136}, [\href{http://arxiv.org/abs/2203.11626}{{\tt 2203.11626}}].

\bibitem{Macias:2016nev}
O.~Macias, C.~Gordon, R.~M. Crocker, B.~Coleman, D.~Paterson, S.~Horiuchi
  et~al., \emph{{Galactic bulge preferred over dark matter for the Galactic
  centre gamma-ray excess}},
  \href{http://dx.doi.org/10.1038/s41550-018-0414-3}{\emph{Nature Astron.} {\bf
  2} (2018) 387--392}, [\href{http://arxiv.org/abs/1611.06644}{{\tt
  1611.06644}}].

\bibitem{Bartels:2017vsx}
R.~Bartels, E.~Storm, C.~Weniger and F.~Calore, \emph{{The Fermi-LAT GeV excess
  as a tracer of stellar mass in the Galactic bulge}},
  \href{http://dx.doi.org/10.1038/s41550-018-0531-z}{\emph{Nature Astron.} {\bf
  2} (2018) 819--828}, [\href{http://arxiv.org/abs/1711.04778}{{\tt
  1711.04778}}].

\bibitem{Macias:2019omb}
O.~Macias, S.~Horiuchi, M.~Kaplinghat, C.~Gordon, R.~M. Crocker and D.~M.
  Nataf, \emph{{Strong Evidence that the Galactic Bulge is Shining in Gamma
  Rays}}, \href{http://dx.doi.org/10.1088/1475-7516/2019/09/042}{\emph{JCAP}
  {\bf 09} (2019) 042}, [\href{http://arxiv.org/abs/1901.03822}{{\tt
  1901.03822}}].

\bibitem{Abazajian:2020tww}
K.~N. Abazajian, S.~Horiuchi, M.~Kaplinghat, R.~E. Keeley and O.~Macias,
  \emph{{Strong constraints on thermal relic dark matter from Fermi-LAT
  observations of the Galactic Center}},
  \href{http://dx.doi.org/10.1103/PhysRevD.102.043012}{\emph{Phys. Rev. D} {\bf
  102} (2020) 043012}, [\href{http://arxiv.org/abs/2003.10416}{{\tt
  2003.10416}}].

\bibitem{Calore:2021jvg}
F.~Calore, F.~Donato and S.~Manconi, \emph{{Dissecting the Inner Galaxy with
  \ensuremath{\gamma}-Ray Pixel Count Statistics}},
  \href{http://dx.doi.org/10.1103/PhysRevLett.127.161102}{\emph{Phys. Rev.
  Lett.} {\bf 127} (2021) 161102}, [\href{http://arxiv.org/abs/2102.12497}{{\tt
  2102.12497}}].

\bibitem{Cao:2013dwa}
L.~Cao, S.~Mao, D.~Nataf, N.~J. Rattenbury and A.~Gould, \emph{{A New
  Photometric Model of the Galactic Bar using Red Clump Giants}},
  \href{http://dx.doi.org/10.1093/mnras/stt1045}{\emph{Mon. Not. Roy. Astron.
  Soc.} {\bf 434} (2013) 595--605}, [\href{http://arxiv.org/abs/1303.6430}{{\tt
  1303.6430}}].

\bibitem{Portail:2016vei}
M.~Portail, O.~Gerhard, C.~Wegg and M.~Ness, \emph{{Dynamical modelling of the
  galactic bulge and bar: the Milky Way's pattern speed, stellar and dark
  matter mass distribution}},
  \href{http://dx.doi.org/10.1093/mnras/stw2819}{\emph{Mon. Not. Roy. Astron.
  Soc.} {\bf 465} (2017) 1621--1644},
  [\href{http://arxiv.org/abs/1608.07954}{{\tt 1608.07954}}].

\bibitem{Launhardt:2002tx}
R.~Launhardt, R.~Zylka and P.~G. Mezger, \emph{{The nuclear bulge of the
  galaxy. 3. Large scale physical characteristics of stars and interstellar
  matter}}, \href{http://dx.doi.org/10.1051/0004-6361:20020017}{\emph{Astron.
  Astrophys.} {\bf 384} (2002) 112--139},
  [\href{http://arxiv.org/abs/astro-ph/0201294}{{\tt astro-ph/0201294}}].

\bibitem{1998MNRAS.301...15D}
M.~B. {Davies} and B.~M.~S. {Hansen}, \emph{{Neutron star retention and
  millisecond pulsar production in globular clusters}},
  \href{http://dx.doi.org/10.1046/j.1365-8711.1998.01923.x}{\emph{\mnras} {\bf
  301} (Nov., 1998) 15--24}.

\bibitem{Hui:2010vt}
C.~Y. Hui, K.~S. Cheng and R.~E. Taam, \emph{{Dynamical Formation of
  Millisecond Pulsars in Globular Clusters}},
  \href{http://dx.doi.org/10.1088/0004-637X/714/2/1149}{\emph{Astrophys. J.}
  {\bf 714} (2010) 1149--1154}, [\href{http://arxiv.org/abs/1003.4332}{{\tt
  1003.4332}}].

\bibitem{Mirabal:2013rba}
N.~Mirabal, \emph{{Dark matter vs. Pulsars: Catching the impostor}},
  \href{http://dx.doi.org/10.1093/mnras/stt1740}{\emph{Mon. Not. Roy. Astron.
  Soc.} {\bf 436} (2013) 2461}, [\href{http://arxiv.org/abs/1309.3428}{{\tt
  1309.3428}}].

\bibitem{Bartels:2016uxz}
R.~Bartels and C.~Weniger, \emph{{Millisecond Pulsars in the Galactic Bulge? An
  Extended Discussion on the Wavelet Analysis of the Fermi-LAT data}},
  \href{http://dx.doi.org/10.1017/S174392131601200X}{\emph{IAU Symp.} {\bf 322}
  (2016) 193--196}.

\bibitem{Zhong:2019ycb}
Y.-M. Zhong, S.~D. McDermott, I.~Cholis and P.~J. Fox, \emph{{Testing the
  Sensitivity of the Galactic Center Excess to the Point Source Mask}},
  \href{http://dx.doi.org/10.1103/PhysRevLett.124.231103}{\emph{Phys. Rev.
  Lett.} {\bf 124} (2020) 231103}, [\href{http://arxiv.org/abs/1911.12369}{{\tt
  1911.12369}}].

\bibitem{Leane:2019xiy}
R.~K. Leane and T.~R. Slatyer, \emph{{Revival of the Dark Matter Hypothesis for
  the Galactic Center Gamma-Ray Excess}},
  \href{http://dx.doi.org/10.1103/PhysRevLett.123.241101}{\emph{Phys. Rev.
  Lett.} {\bf 123} (2019) 241101}, [\href{http://arxiv.org/abs/1904.08430}{{\tt
  1904.08430}}].

\bibitem{Chang:2019ars}
L.~J. Chang, S.~Mishra-Sharma, M.~Lisanti, M.~Buschmann, N.~L. Rodd and B.~R.
  Safdi, \emph{{Characterizing the nature of the unresolved point sources in
  the Galactic Center: An assessment of systematic uncertainties}},
  \href{http://dx.doi.org/10.1103/PhysRevD.101.023014}{\emph{Phys. Rev. D} {\bf
  101} (2020) 023014}, [\href{http://arxiv.org/abs/1908.10874}{{\tt
  1908.10874}}].

\bibitem{Buschmann:2020adf}
M.~Buschmann, N.~L. Rodd, B.~R. Safdi, L.~J. Chang, S.~Mishra-Sharma,
  M.~Lisanti et~al., \emph{{Foreground Mismodeling and the Point Source
  Explanation of the Fermi Galactic Center Excess}},
  \href{http://dx.doi.org/10.1103/PhysRevD.102.023023}{\emph{Phys. Rev. D} {\bf
  102} (2020) 023023}, [\href{http://arxiv.org/abs/2002.12373}{{\tt
  2002.12373}}].

\bibitem{Leane:2020nmi}
R.~K. Leane and T.~R. Slatyer, \emph{{Spurious Point Source Signals in the
  Galactic Center Excess}},
  \href{http://dx.doi.org/10.1103/PhysRevLett.125.121105}{\emph{Phys. Rev.
  Lett.} {\bf 125} (2020) 121105}, [\href{http://arxiv.org/abs/2002.12370}{{\tt
  2002.12370}}].

\bibitem{Leane:2020pfc}
R.~K. Leane and T.~R. Slatyer, \emph{{The enigmatic Galactic Center excess:
  Spurious point sources and signal mismodeling}},
  \href{http://dx.doi.org/10.1103/PhysRevD.102.063019}{\emph{Phys. Rev. D} {\bf
  102} (2020) 063019}, [\href{http://arxiv.org/abs/2002.12371}{{\tt
  2002.12371}}].

\bibitem{Mishra-Sharma:2021oxe}
S.~Mishra-Sharma and K.~Cranmer, \emph{{Neural simulation-based inference
  approach for characterizing the Galactic Center \ensuremath{\gamma}-ray
  excess}}, \href{http://dx.doi.org/10.1103/PhysRevD.105.063017}{\emph{Phys.
  Rev. D} {\bf 105} (2022) 063017},
  [\href{http://arxiv.org/abs/2110.06931}{{\tt 2110.06931}}].

\bibitem{Storm:2017arh}
E.~Storm, C.~Weniger and F.~Calore, \emph{{SkyFACT: High-dimensional modeling
  of gamma-ray emission with adaptive templates and penalized likelihoods}},
  \href{http://dx.doi.org/10.1088/1475-7516/2017/08/022}{\emph{JCAP} {\bf 1708}
  (2017) 022}, [\href{http://arxiv.org/abs/1705.04065}{{\tt 1705.04065}}].

\bibitem{Zechlin:2015wdz}
H.-S. Zechlin, A.~Cuoco, F.~Donato, N.~Fornengo and A.~Vittino,
  \emph{{Unveiling the Gamma-ray Source Count Distribution Below the Fermi
  Detection Limit with Photon Statistics}},
  \href{http://dx.doi.org/10.3847/0067-0049/225/2/18}{\emph{Astrophys. J.
  Suppl.} {\bf 225} (2016) 18}, [\href{http://arxiv.org/abs/1512.07190}{{\tt
  1512.07190}}].

\bibitem{Caron:2017udl}
S.~Caron, G.~A. Gómez-Vargas, L.~Hendriks and R.~Ruiz~de Austri,
  \emph{{Analyzing $\gamma$-rays of the Galactic Center with Deep Learning}},
  \href{http://dx.doi.org/10.1088/1475-7516/2018/05/058}{\emph{JCAP} {\bf 05}
  (2018) 058}, [\href{http://arxiv.org/abs/1708.06706}{{\tt 1708.06706}}].

\bibitem{List:2020mzd}
F.~List, N.~L. Rodd, G.~F. Lewis and I.~Bhat, \emph{{The GCE in a New Light:
  Disentangling the $\gamma$-ray Sky with Bayesian Graph Convolutional Neural
  Networks}},  \href{http://arxiv.org/abs/2006.12504}{{\tt 2006.12504}}.

\bibitem{Mishra-Sharma:2020kjb}
S.~Mishra-Sharma and K.~Cranmer, \emph{{Semi-parametric $\gamma$-ray modeling
  with Gaussian processes and variational inference}},  in \emph{{34th
  Conference on Neural Information Processing Systems}}, 10, 2020.
\newblock \href{http://arxiv.org/abs/2010.10450}{{\tt 2010.10450}}.

\bibitem{List:2021aer}
F.~List, N.~L. Rodd and G.~F. Lewis, \emph{{Extracting the Galactic Center
  excess\textquoteright{} source-count distribution with neural nets}},
  \href{http://dx.doi.org/10.1103/PhysRevD.104.123022}{\emph{Phys. Rev. D} {\bf
  104} (2021) 123022}, [\href{http://arxiv.org/abs/2107.09070}{{\tt
  2107.09070}}].

\bibitem{2021A&A...655A..64M}
P.~{Mertsch} and A.~{Vittino}, \emph{{Bayesian inference of three-dimensional
  gas maps. I. Galactic CO}},
  \href{http://dx.doi.org/10.1051/0004-6361/202141000}{\emph{A\&A} {\bf 655}
  (Nov., 2021) A64}, [\href{http://arxiv.org/abs/2012.15770}{{\tt
  2012.15770}}].

\bibitem{Mertsch:2022oee}
P.~Mertsch and V.~H.~M. Phan, \emph{{Bayesian inference of three-dimensional
  gas maps: II. Galactic HI}},  \href{http://arxiv.org/abs/2202.02341}{{\tt
  2202.02341}}.

\bibitem{Karwin:2022xgn}
C.~M. Karwin, A.~Broughton, S.~Murgia, A.~Shmakov, M.~Tavakoli and P.~Baldi,
  \emph{{Improved Modeling of the Discrete Component of the Galactic
  Interstellar $\gamma$-ray Emission and Implications for the $Fermi$
  \textendash{}LAT Galactic Center Excess}},
  \href{http://arxiv.org/abs/2206.02809}{{\tt 2206.02809}}.

\bibitem{Shmakov:2022vuc}
A.~Shmakov, M.~Tavakoli, P.~Baldi, C.~M. Karwin, A.~Broughton and S.~Murgia,
  \emph{{Deep learning models of the discrete component of the Galactic
  interstellar \ensuremath{\gamma}-ray emission}},
  \href{http://dx.doi.org/10.1103/PhysRevD.107.063018}{\emph{Phys. Rev. D} {\bf
  107} (2023) 063018}, [\href{http://arxiv.org/abs/2206.02819}{{\tt
  2206.02819}}].

\bibitem{Bartels:2018xom}
R.~Bartels, T.~Edwards and C.~Weniger, \emph{{Bayesian model comparison and
  analysis of the Galactic disc population of gamma-ray millisecond pulsars}},
  \href{http://dx.doi.org/10.1093/mnras/sty2529}{\emph{Mon. Not. Roy. Astron.
  Soc.} {\bf 481} (2018) 3966--3987},
  [\href{http://arxiv.org/abs/1805.11097}{{\tt 1805.11097}}].

\bibitem{2013ApJ...774...76A}
W.~B. {Atwood}, L.~{Baldini}, J.~{Bregeon}, P.~{Bruel}, A.~{Chekhtman},
  J.~{Cohen-Tanugi} et~al., \emph{{New Fermi-LAT Event Reconstruction Reveals
  More High-energy Gamma Rays from Gamma-Ray Bursts}},
  \href{http://dx.doi.org/10.1088/0004-637X/774/1/76}{\emph{\apj} {\bf 774}
  (Sept., 2013) 76}, [\href{http://arxiv.org/abs/1307.3037}{{\tt 1307.3037}}].

\bibitem{2018arXiv181011394B}
P.~{Bruel}, T.~H. {Burnett}, S.~W. {Digel}, G.~{Johannesson}, N.~{Omodei} and
  M.~{Wood}, \emph{{Fermi-LAT improved Pass\raisebox{-0.5ex}\textasciitilde8
  event selection}}, {\emph{arXiv e-prints} (Oct., 2018) arXiv:1810.11394},
  [\href{http://arxiv.org/abs/1810.11394}{{\tt 1810.11394}}].

\bibitem{Acero:2015prw}
{\scshape Fermi-LAT} collaboration, F.~Acero et~al., \emph{{The First Fermi LAT
  Supernova Remnant Catalog}},
  \href{http://dx.doi.org/10.3847/0067-0049/224/1/8}{\emph{Astrophys. J.
  Suppl.} {\bf 224} (2016) 8}, [\href{http://arxiv.org/abs/1511.06778}{{\tt
  1511.06778}}].

\bibitem{2006MNRAS.372..777L}
D.~R. {Lorimer}, A.~J. {Faulkner}, A.~G. {Lyne}, R.~N. {Manchester},
  M.~{Kramer}, M.~A. {McLaughlin} et~al., \emph{{The Parkes Multibeam Pulsar
  Survey - VI. Discovery and timing of 142 pulsars and a Galactic population
  analysis}},
  \href{http://dx.doi.org/10.1111/j.1365-2966.2006.10887.x}{\emph{\mnras} {\bf
  372} (Oct., 2006) 777--800},
  [\href{http://arxiv.org/abs/astro-ph/0607640}{{\tt astro-ph/0607640}}].

\bibitem{Herold:2019pei}
L.~Herold and D.~Malyshev, \emph{{Hard and bright gamma-ray emission at the
  base of the Fermi bubbles}},
  \href{http://dx.doi.org/10.1051/0004-6361/201834670}{\emph{Astron.
  Astrophys.} {\bf 625} (2019) A110},
  [\href{http://arxiv.org/abs/1904.01454}{{\tt 1904.01454}}].

\bibitem{Ballet:2020hze}
{\scshape Fermi-LAT} collaboration, J.~Ballet, T.~H. Burnett, S.~W. Digel and
  B.~Lott, \emph{{Fermi Large Area Telescope Fourth Source Catalog Data Release
  2}},  \href{http://arxiv.org/abs/2005.11208}{{\tt 2005.11208}}.

\bibitem{Fermi-LAT:2015bhf}
{\scshape Fermi-LAT} collaboration, F.~Acero et~al., \emph{{Fermi Large Area
  Telescope Third Source Catalog}},
  \href{http://dx.doi.org/10.1088/0067-0049/218/2/23}{\emph{Astrophys. J.
  Suppl.} {\bf 218} (2015) 23}, [\href{http://arxiv.org/abs/1501.02003}{{\tt
  1501.02003}}].

\bibitem{Navarro:1995iw}
J.~F. Navarro, C.~S. Frenk and S.~D.~M. White, \emph{{The Structure of cold
  dark matter halos}}, \href{http://dx.doi.org/10.1086/177173}{\emph{Astrophys.
  J.} {\bf 462} (1996) 563--575},
  [\href{http://arxiv.org/abs/astro-ph/9508025}{{\tt astro-ph/9508025}}].

\bibitem{Navarro:1996gj}
J.~F. Navarro, C.~S. Frenk and S.~D.~M. White, \emph{{A Universal density
  profile from hierarchical clustering}},
  \href{http://dx.doi.org/10.1086/304888}{\emph{Astrophys. J.} {\bf 490} (1997)
  493--508}, [\href{http://arxiv.org/abs/astro-ph/9611107}{{\tt
  astro-ph/9611107}}].

\bibitem{Zavala:2019gpq}
J.~Zavala and C.~S. Frenk, \emph{{Dark matter haloes and subhaloes}},
  \href{http://dx.doi.org/10.3390/galaxies7040081}{\emph{Galaxies} {\bf 7}
  (2019) 81}, [\href{http://arxiv.org/abs/1907.11775}{{\tt 1907.11775}}].

\bibitem{McCann:2014dea}
A.~McCann, \emph{{A stacked analysis of 115 pulsars observed by the Fermi
  LAT}}, \href{http://dx.doi.org/10.1088/0004-637X/804/2/86}{\emph{Astrophys.
  J.} {\bf 804} (2015) 86}, [\href{http://arxiv.org/abs/1412.2422}{{\tt
  1412.2422}}].

\bibitem{Fermi-LAT:2017yoi}
{\scshape Fermi-LAT} collaboration, M.~Ajello et~al., \emph{{Characterizing the
  population of pulsars in the inner Galaxy with the Fermi Large Area
  Telescope}},  \href{http://arxiv.org/abs/1705.00009}{{\tt 1705.00009}}.

\bibitem{Winter:2016wmy}
M.~Winter, G.~Zaharijas, K.~Bechtol and J.~Vandenbroucke, \emph{{Estimating the
  GeV Emission of Millisecond Pulsars in Dwarf Spheroidal Galaxies}},
  \href{http://dx.doi.org/10.3847/2041-8205/832/1/L6}{\emph{Astrophys. J.
  Lett.} {\bf 832} (2016) L6}, [\href{http://arxiv.org/abs/1607.06390}{{\tt
  1607.06390}}].

\bibitem{Hooper:2015jlu}
D.~Hooper and G.~Mohlabeng, \emph{{The Gamma-Ray Luminosity Function of
  Millisecond Pulsars and Implications for the GeV Excess}},
  \href{http://dx.doi.org/10.1088/1475-7516/2016/03/049}{\emph{JCAP} {\bf 03}
  (2016) 049}, [\href{http://arxiv.org/abs/1512.04966}{{\tt 1512.04966}}].

\bibitem{Acero:2015gva}
{\scshape Fermi-LAT} collaboration, F.~Acero et~al., \emph{{Fermi Large Area
  Telescope Third Source Catalog}},
  \href{http://dx.doi.org/10.1088/0067-0049/218/2/23}{\emph{Astrophys. J.
  Suppl.} {\bf 218} (2015) 23}, [\href{http://arxiv.org/abs/1501.02003}{{\tt
  1501.02003}}].

\bibitem{2019HEAD...1710932L}
B.~{Limyansky}, \emph{{The Third Fermi Pulsar Catalog}},  in \emph{AAS/High
  Energy Astrophysics Division}, AAS/High Energy Astrophysics Division,
  p.~109.32, Mar., 2019.

\bibitem{Bringmann:2012ez}
T.~Bringmann and C.~Weniger, \emph{{Gamma Ray Signals from Dark Matter:
  Concepts, Status and Prospects}},
  \href{http://dx.doi.org/10.1016/j.dark.2012.10.005}{\emph{Phys. Dark Univ.}
  {\bf 1} (2012) 194--217}, [\href{http://arxiv.org/abs/1208.5481}{{\tt
  1208.5481}}].

\bibitem{lakshminarayanan2017simple}
B.~Lakshminarayanan, A.~Pritzel and C.~Blundell, \emph{Simple and scalable
  predictive uncertainty estimation using deep ensembles},  2017.

\bibitem{batchnorm}
S.~Ioffe and C.~Szegedy, \emph{Batch normalization: Accelerating deep network
  training by reducing internal covariate shift},  ICML'15, p.~448–456,
  JMLR.org, 2015.

\bibitem{kingma2014method}
D.~P. Kingma and J.~Ba, \emph{Adam: A method for stochastic optimization},
  2014.

\bibitem{iminuit}
H.~Dembinski, P.~Ongmongkolkul, C.~Deil, D.~M. Hurtado, M.~Feickert,
  H.~Schreiner et~al., \emph{scikit-hep/iminuit: v1.5.2}, .

\bibitem{Acero:2016qlg}
{\scshape Fermi-LAT} collaboration, F.~Acero et~al., \emph{{Development of the
  Model of Galactic Interstellar Emission for Standard Point-Source Analysis of
  Fermi Large Area Telescope Data}},
  \href{http://dx.doi.org/10.3847/0067-0049/223/2/26}{\emph{Astrophys. J.
  Suppl.} {\bf 223} (2016) 26}, [\href{http://arxiv.org/abs/1602.07246}{{\tt
  1602.07246}}].

\bibitem{Fermi-LAT:2016zaq}
{\scshape Fermi-LAT} collaboration, F.~Acero et~al., \emph{{Development of the
  Model of Galactic Interstellar Emission for Standard Point-Source Analysis of
  Fermi Large Area Telescope Data}},
  \href{http://dx.doi.org/10.3847/0067-0049/223/2/26}{\emph{Astrophys. J.
  Suppl.} {\bf 223} (2016) 26}, [\href{http://arxiv.org/abs/1602.07246}{{\tt
  1602.07246}}].

\bibitem{pmlr-v80-ruff18a}
L.~Ruff, R.~Vandermeulen, N.~Goernitz, L.~Deecke, S.~A. Siddiqui, A.~Binder
  et~al., \emph{Deep one-class classification},  in \emph{Proceedings of the
  35th International Conference on Machine Learning} (J.~Dy and A.~Krause,
  eds.), vol.~80 of \emph{Proceedings of Machine Learning Research},
  pp.~4393--4402, PMLR, 10--15 Jul, 2018.

\bibitem{Caron_2022}
S.~Caron, L.~Hendriks and R.~Verheyen, \emph{Rare and different: Anomaly scores
  from a combination of likelihood and out-of-distribution models to detect new
  physics at the {LHC}},
  \href{http://dx.doi.org/10.21468/scipostphys.12.2.077}{\emph{{SciPost}
  Physics} {\bf 12} (feb, 2022) }.

\end{thebibliography}\endgroup

\begin{appendices}

\section{Astrophysical parameters}
\label{app:param_comparison_rest}

In this section, we present a complementary view of the results of the network findings in terms of astrophysical components. In Fig.~\ref{fig:results_network_pred_rest} we show the network's prediction for parameters not directly related to the GCE excess, that were not part of Fig.~\ref{fig:summaryplot} in the main text while retaining the style of this plot. Note that the two astrophysical parameters of Model 1 are not shown since they do not fit reasonably into this scheme.

We notice a slight anti-correlation between the evolution per model of the normalizations for the isotropic and inverse-Compton templates. Moreover, both scenarios indicate a trend of the 4FGL-DR2 template normalizations to decrease with increasing model complexity. Lastly, the agreement between scenarios A and B regarding the astrophysical parameters of our models are best for Model 3 lending further credence to the fact that it represents the best configuration among the tested cases.

\begin{figure}[t!]
\begin{center}
\includegraphics[width=0.8\linewidth]{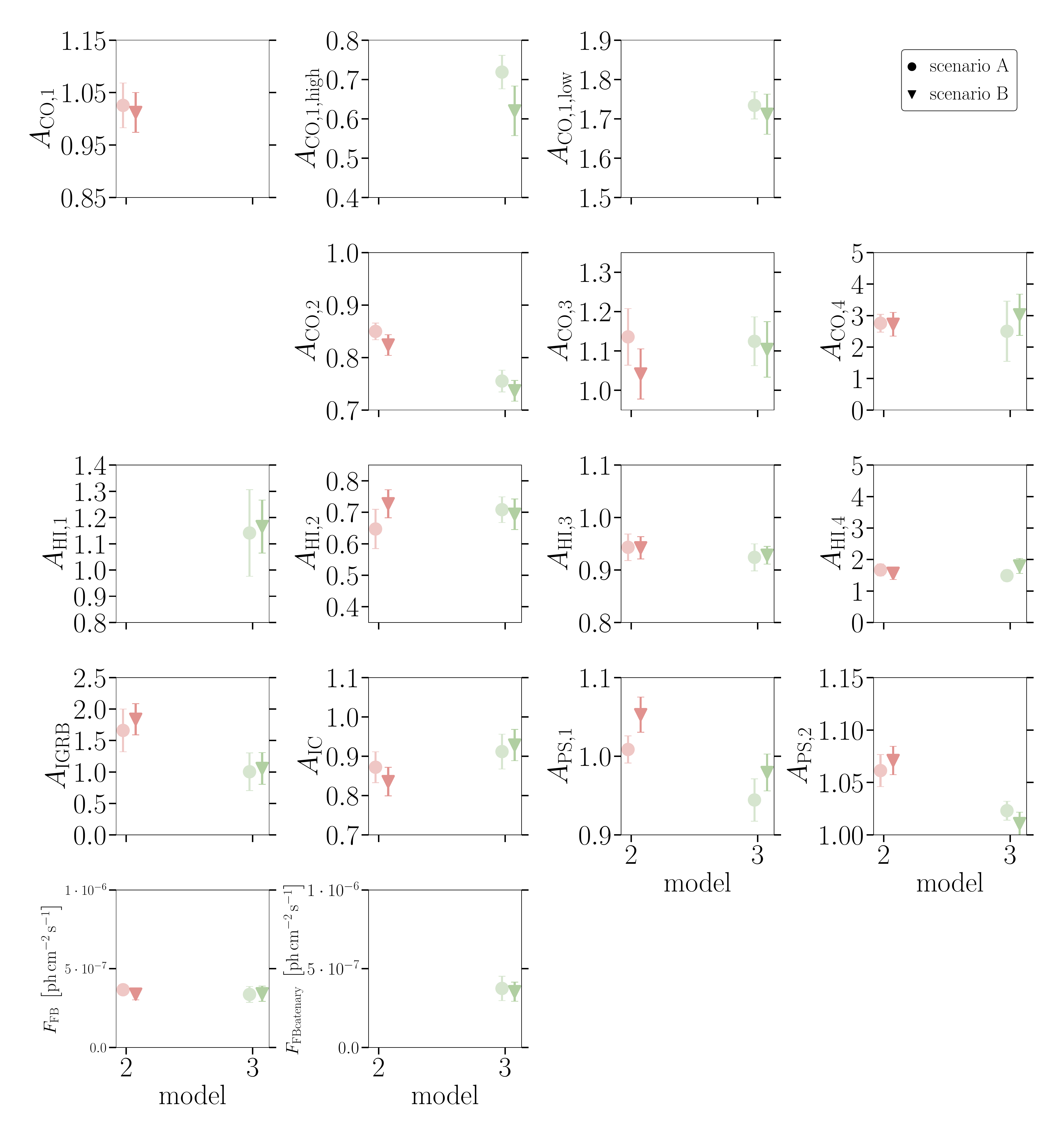}
\end{center}
\caption{Same as Fig.~\ref{fig:summaryplot} (upper panel) displaying a comparison of the network's prediction for relevant astrophysical parameters as a function of the model number. Note that the template normalisation parameters associated with the FBs templates have been converted into a photon flux over the full energy range of our analysis using the assumed FBs spectral shape (see Sec.~\ref{sec:astro_models}). We do not provide corresponding numbers for Model 1 since it does not exhibit the same structure as the other two model iterations.
\label{fig:results_network_pred_rest}}
\end{figure}

\section{MSP template without flux detection threshold - Scenario B}
\label{app:msp_threshold}

In this section, we show the results of the network in the case where the MSP template generation proceeds without a high-flux cut-off, i.e.~we generate MSPs both below and above the LAT detection threshold. This is an approach dubbed Scenario B in Section \ref{trainingdata} and illustrated schematically in Fig.~\ref{fig:domainadaptation}. 

\begin{figure}[t!]
\begin{center}
\includegraphics[width=0.8\linewidth]{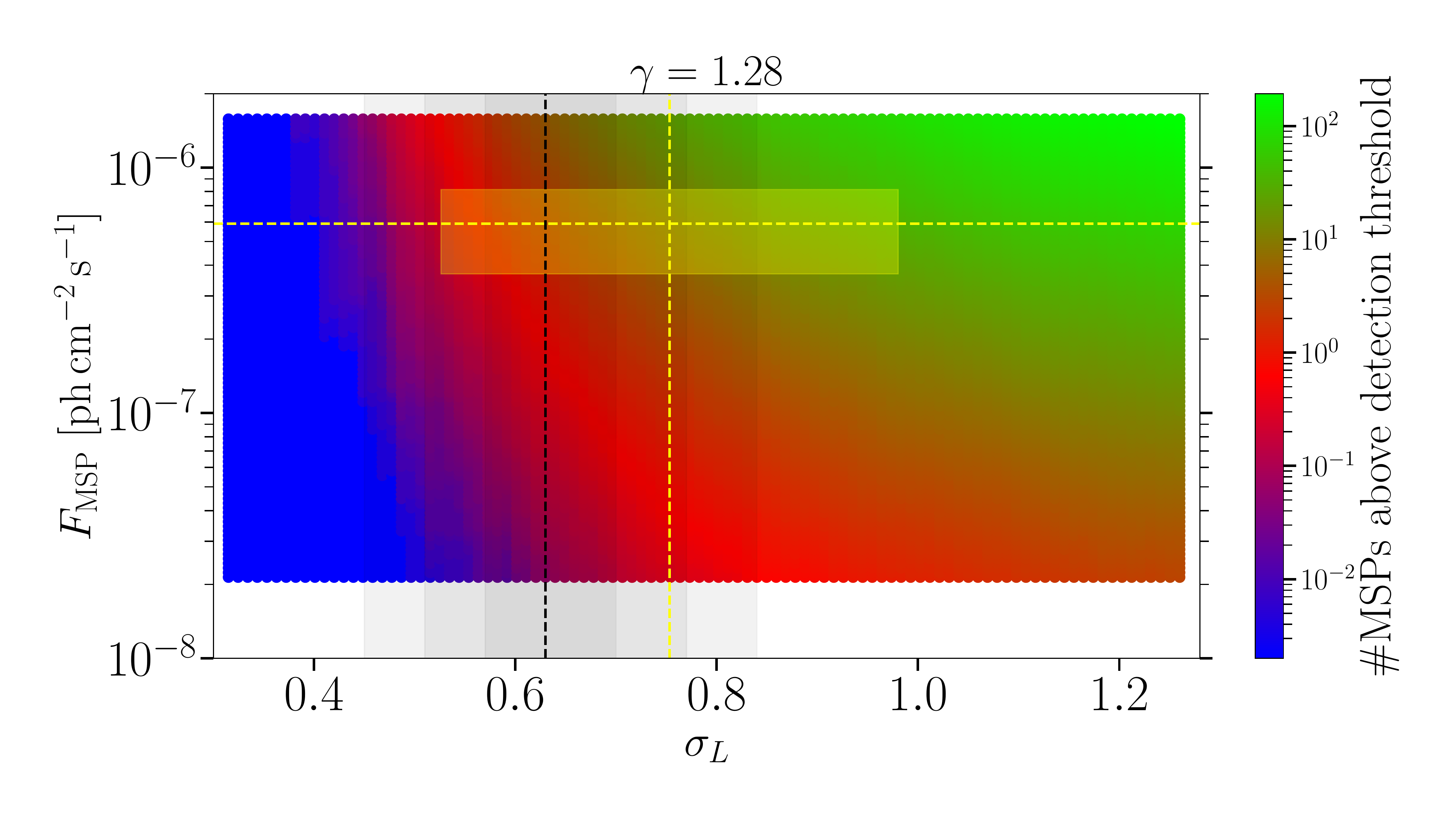}
\end{center}
\caption{The color bar marks the number of MSPs in our mock catalogue that have a flux above the 4FGL-DR2 detection threshold, for  a  given choice of $\sigma_L$ and total luminosity of the MSP template $F_{MSP}$ (see  Section \ref{trainingdata} for details). The spatial distribution is fixed to $\gamma = 1.28$. With the dashed vertical line, we plot the best-fit value of $\sigma_L$ obtained in \protect\cite{Bartels:2018xom}, while the gray zone marks $1-3$ $\sigma$ uncertainty range. The horizontal, dashed yellow line denotes the best-fit MSP cumulative flux with respect to Model 3B while the height of the yellow box displays its 1$\sigma$ error margin. Analogously, the vertical shaded yellow line refers to the best-fit value of $\sigma_L$ in this model and scenario; the width of the box shows the 1$\sigma$ uncertainty of this parameter.} \label{fig:MSP_threshold_numbers}
\end{figure}

To gauge how many sources we add above the 4FGL detection threshold in this  procedure (risking the 'double counting' issue) we show in Figure \ref{fig:MSP_threshold_numbers} the number of sources above the 4FGL threshold that  stay in our MSP  template in  this {scenario}. We see that the number of bright  sources is subdominant with respect  to  the total number of 4FGL sources in our ROI which is 243 4FGL point sources, within the angular distance to the GC of $|l,b| \leq 15^\circ$ (this includes 22 pulsars and 153 unidentified objects).

All other procedure steps are identical as  in the main text. Results are shown in Figures \ref{fig:results_modelB1}, \ref{fig:results_modelB2}, \ref{fig:results_modelB3} and the corresponding residuals of the models selected by the network in Figs. \ref{fig:residuals_modelB1}, \ref{fig:residuals_modelB2}, \ref{fig:residuals_modelB3}.

\begin{figure}[h!]
\begin{center}
\includegraphics[width=0.8\linewidth]{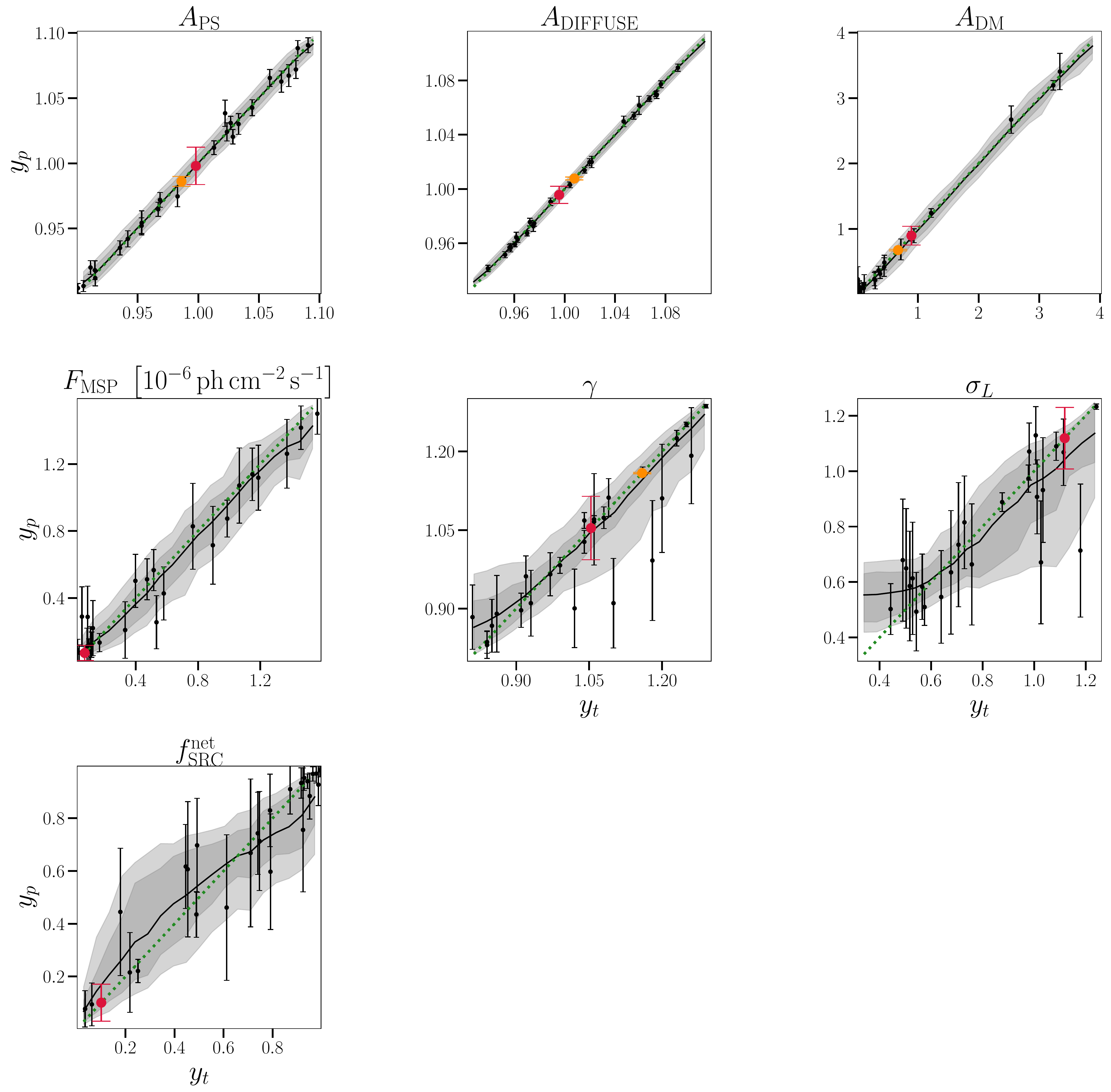}
\end{center}
\caption{The same as Fig. \ref{fig:results_scenarioA_model1} (Model 1), but for Scenario B. 
\label{fig:results_modelB1}}
\end{figure}

We observe that
\begin{itemize}
    \item All main results are unchanged in 
    Scenario B. Indeed the network seems to  predict quite reliably the total flux of the MSP template, independently of the scenario (see Fig.  \ref{fig:summaryplot}).
    \item As expected, the network has higher sensitivity on the $\sigma_L$ parameter, which is found to be in the $\sigma_L ~\epsilon ~[0.4-0.8]$ range, consistent with the findings in \cite{Bartels:2018xom}. This parameter was instead fully unconstrained in Scenario A, as expected since in that case, the luminosity function was discontinuous at the 4FGL detection threshold.
    \end{itemize}




\begin{figure}[t!]
\begin{center}
\includegraphics[width=0.8\linewidth]{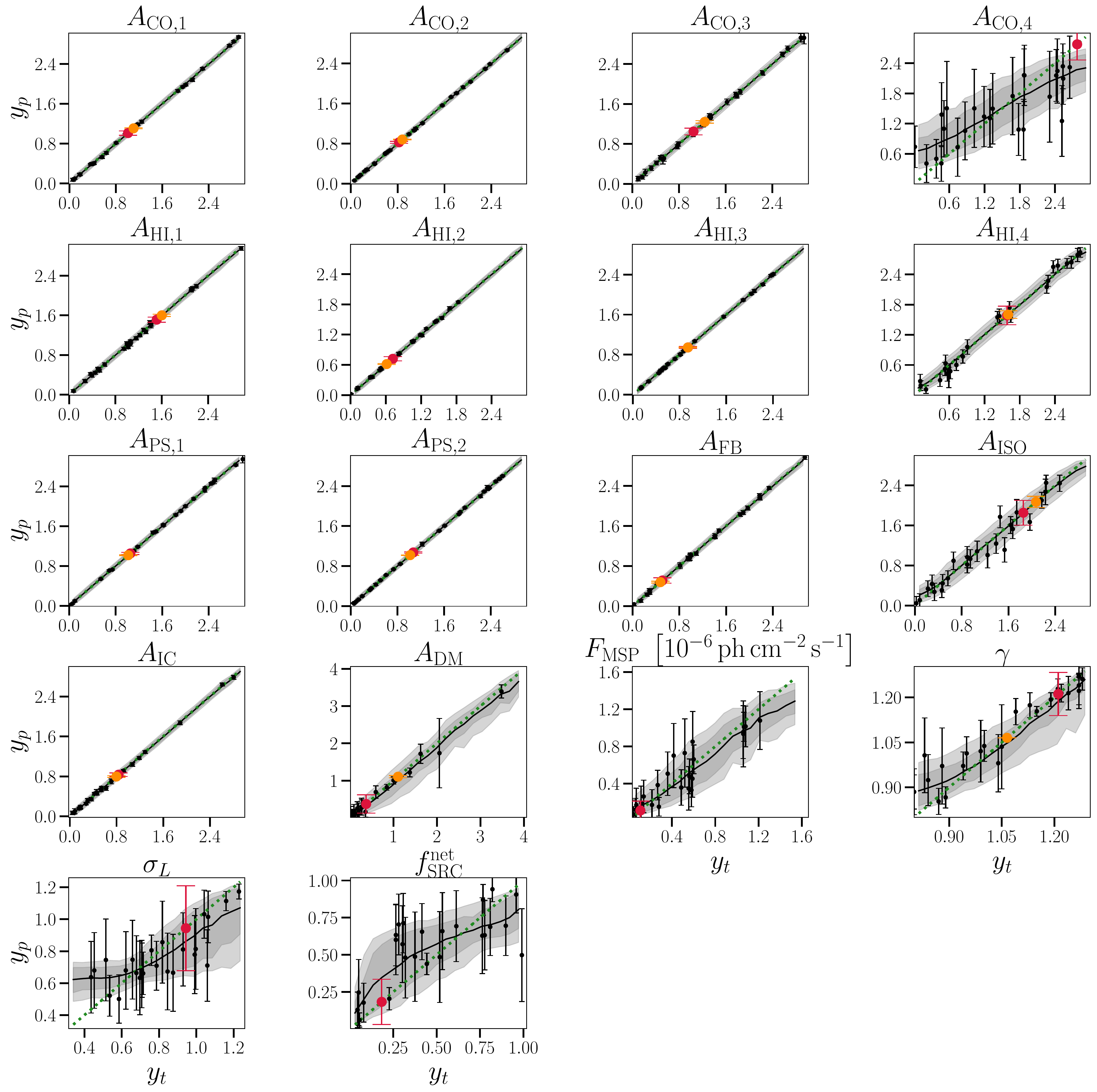}
\end{center}
\caption{Model 2,  Scenario B. 
\label{fig:results_modelB2}}
\end{figure}

\begin{figure}[t!]
\begin{center}
\includegraphics[width=0.8\linewidth]{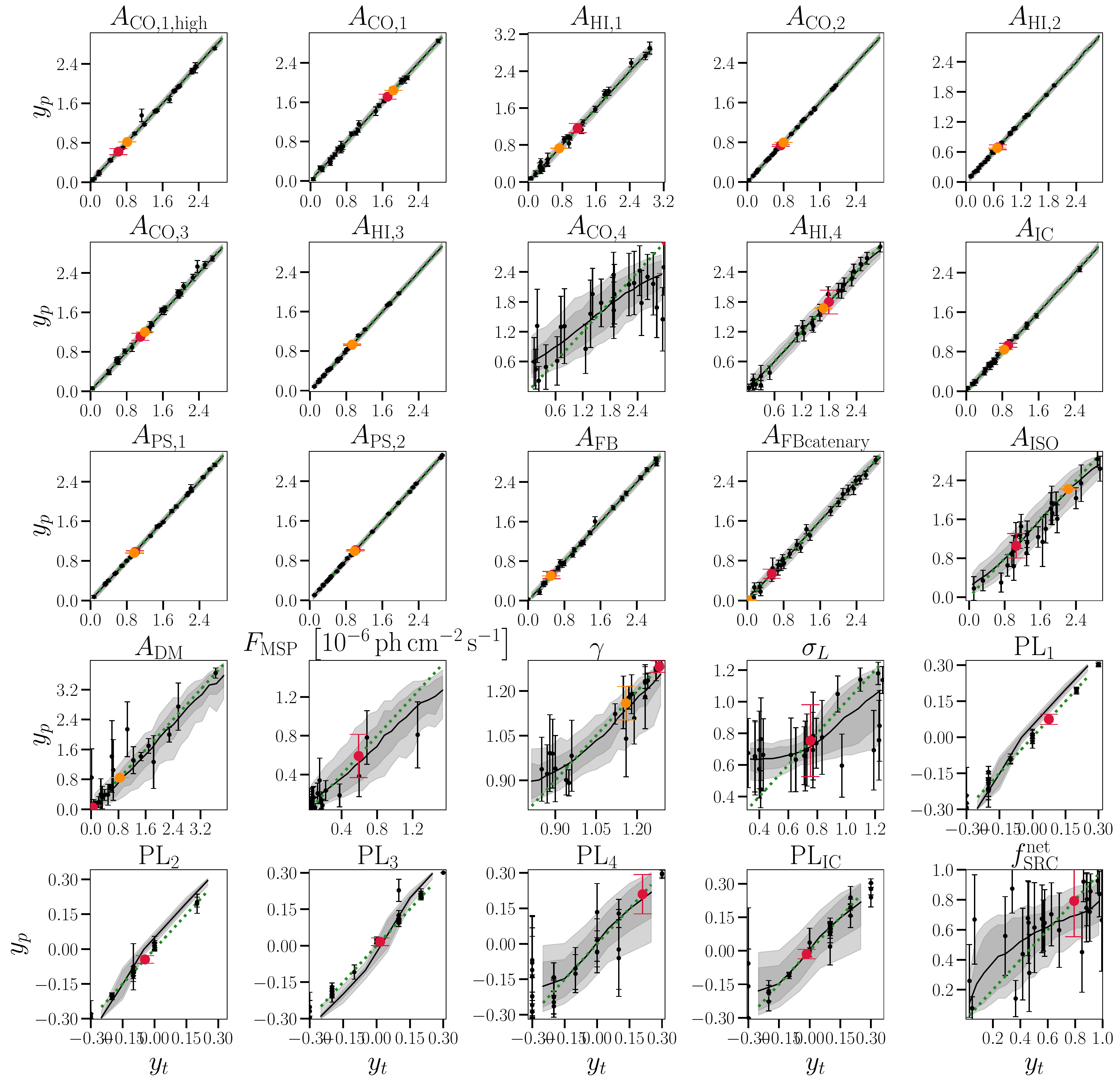}
\end{center}
\caption{Model 3, Scenario B. 
\label{fig:results_modelB3}}
\end{figure}

\section{Spatial and spectral residuals for the examined models and scenarios}
\label{app:residuals}

In this section, we list all remaining plots of spatial and spectral residuals that have not been shown in the main text.

While the structure of the spatial and spectral residuals for Model 1A was a good indicator and motivation for several model extensions, the corresponding figures for other model iterations do not change significantly. Moreover, these residuals are not the result of standard inference methods like a maximum likelihood fit and should therefore not be interpreted as such. The inherent purpose of a maximum likelihood analysis is to find those model parameters that minimize the residual differences between the final best-fit model and the input data. The machine learning approach employed in this work rather aims at deriving a relation between the model parameters and the images it is trained on, i.e.~its task is to establish a parameter reconstruction scheme. The resulting residuals on the image level as shown in, for example, Fig.~\ref{fig:spatial_res_model1_scenA} are not part of the network's loss function in Eq.~\ref{eq:aleatoric_loss}. Therefore, it is by no means guaranteed that even a perfectly unbiased and accurate parameter reconstruction leads to the same result as a maximum likelihood approach. The presence of a pronounced reality gap between the model and true gamma-ray data may enhance this effect. In the explicit example of models and scenarios probed in our work, we find that the spectral residuals in the last two energy bins tend to become worse with increased model complexity whereas the first three energy bins -- those bins that contain the bulk of the GCE emission -- are fit fairly well. Such a behavior seems odd in the context of a maximum likelihood fit but it might be reasonable for the optimization measure created internally by training the neural network. Moreover, the error bars of the spectral residuals are potentially overestimated because we lack information on correlations among the parameters predicted by the network. Retrieving such correlations is a field of active research in the machine-learning community.

\begin{figure}[t!]
\begin{center}
\includegraphics[width=0.45\linewidth]{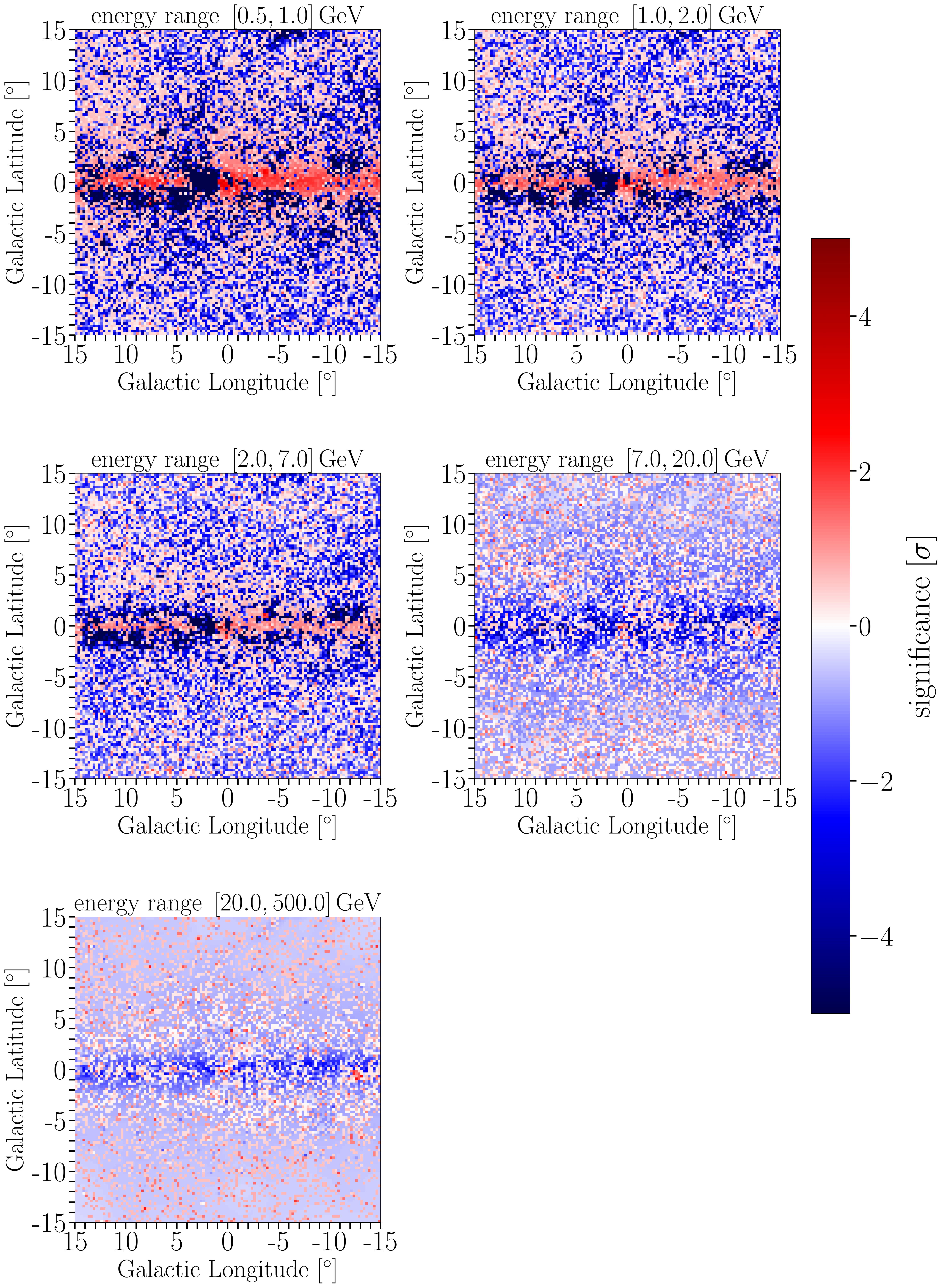}
\includegraphics[width=0.44\linewidth]{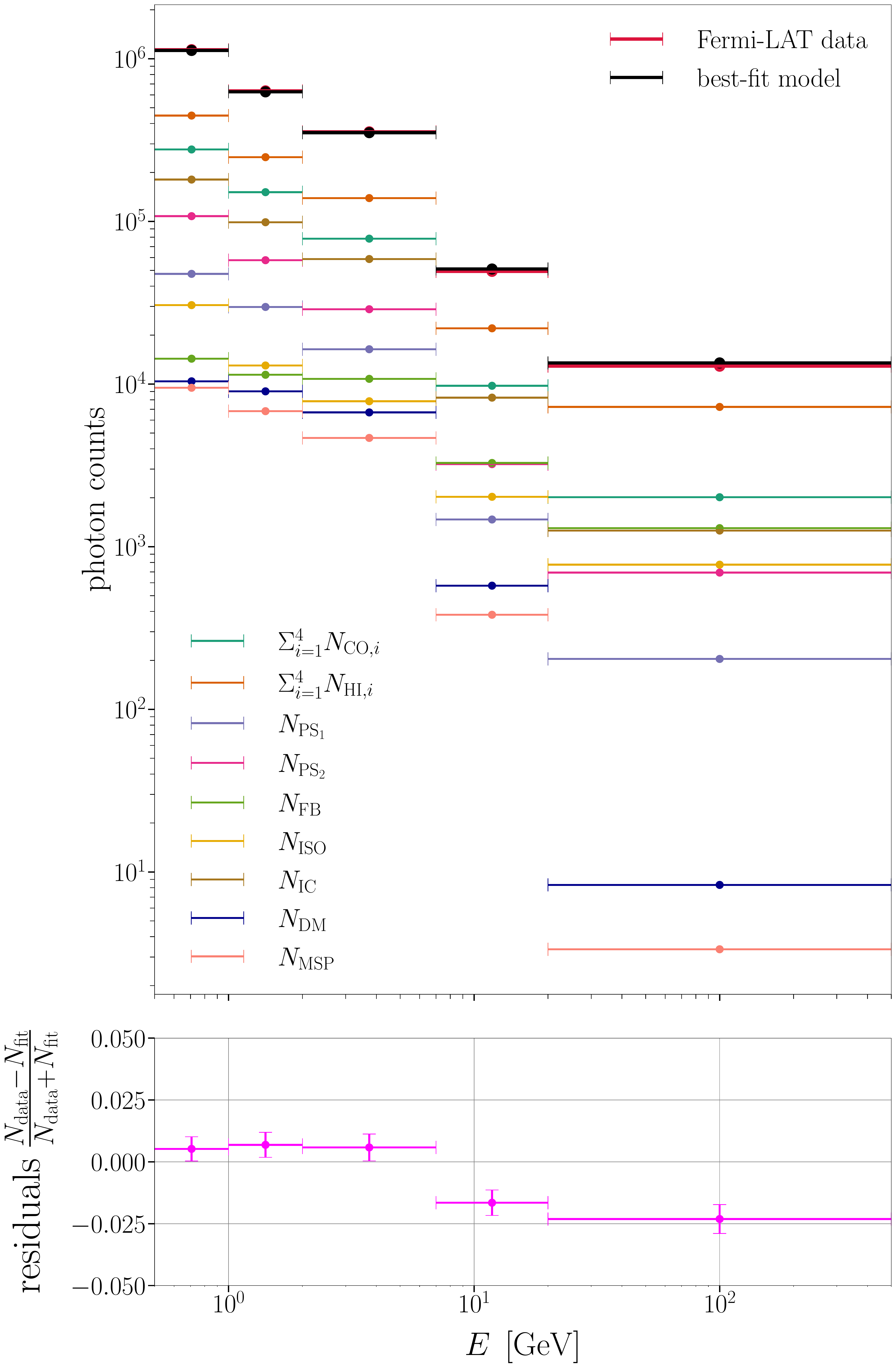}
\end{center}
\caption{The same as Fig.  \ref{fig:spatial_res_model1_scenA} but in the context of Model 2. 
\label{fig:residualsNEWNEW_sc2}}
\end{figure}

\begin{figure}[t!]
\begin{center}
\includegraphics[width=0.45\linewidth]{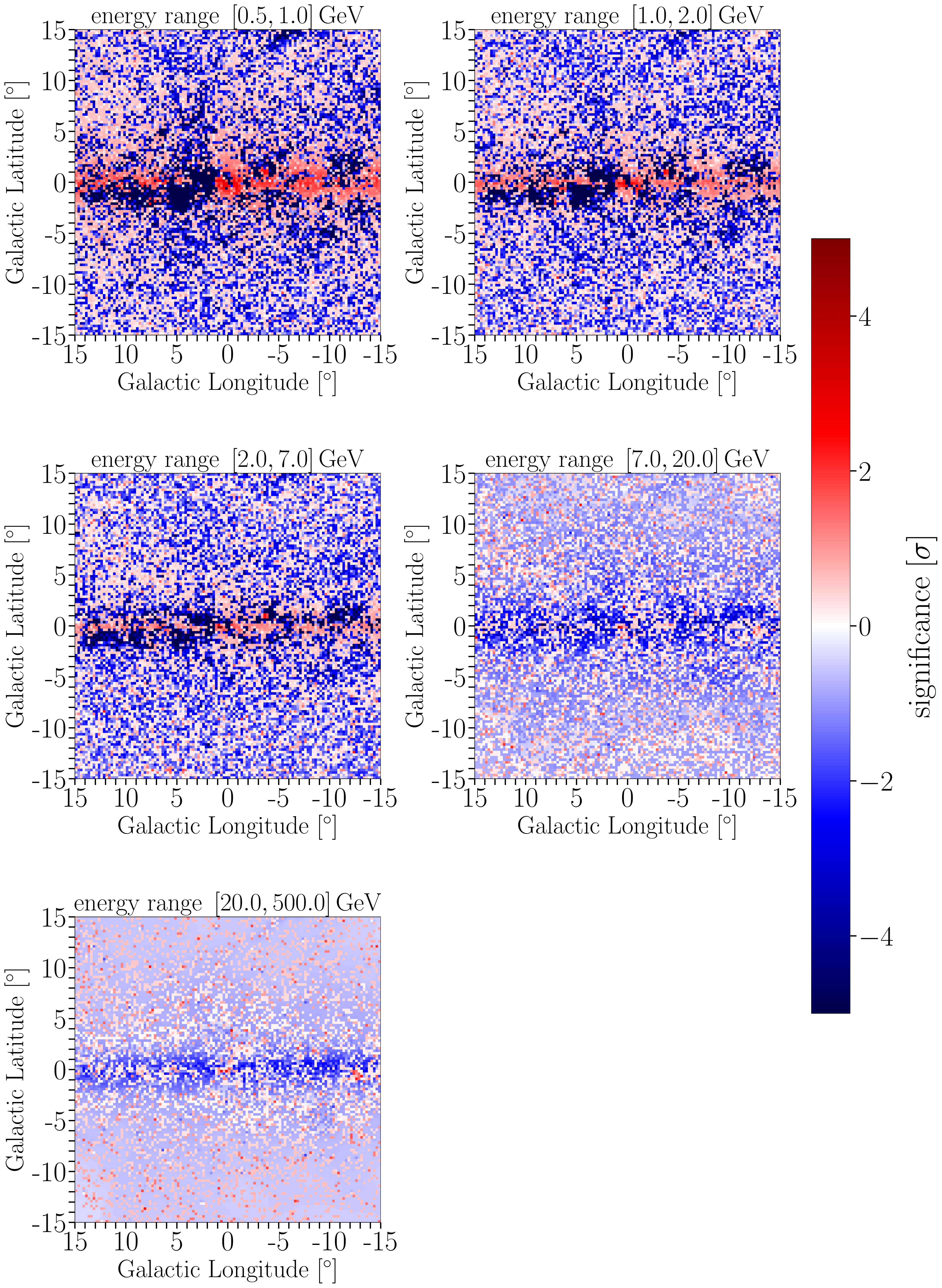}
\includegraphics[width=0.44\linewidth]{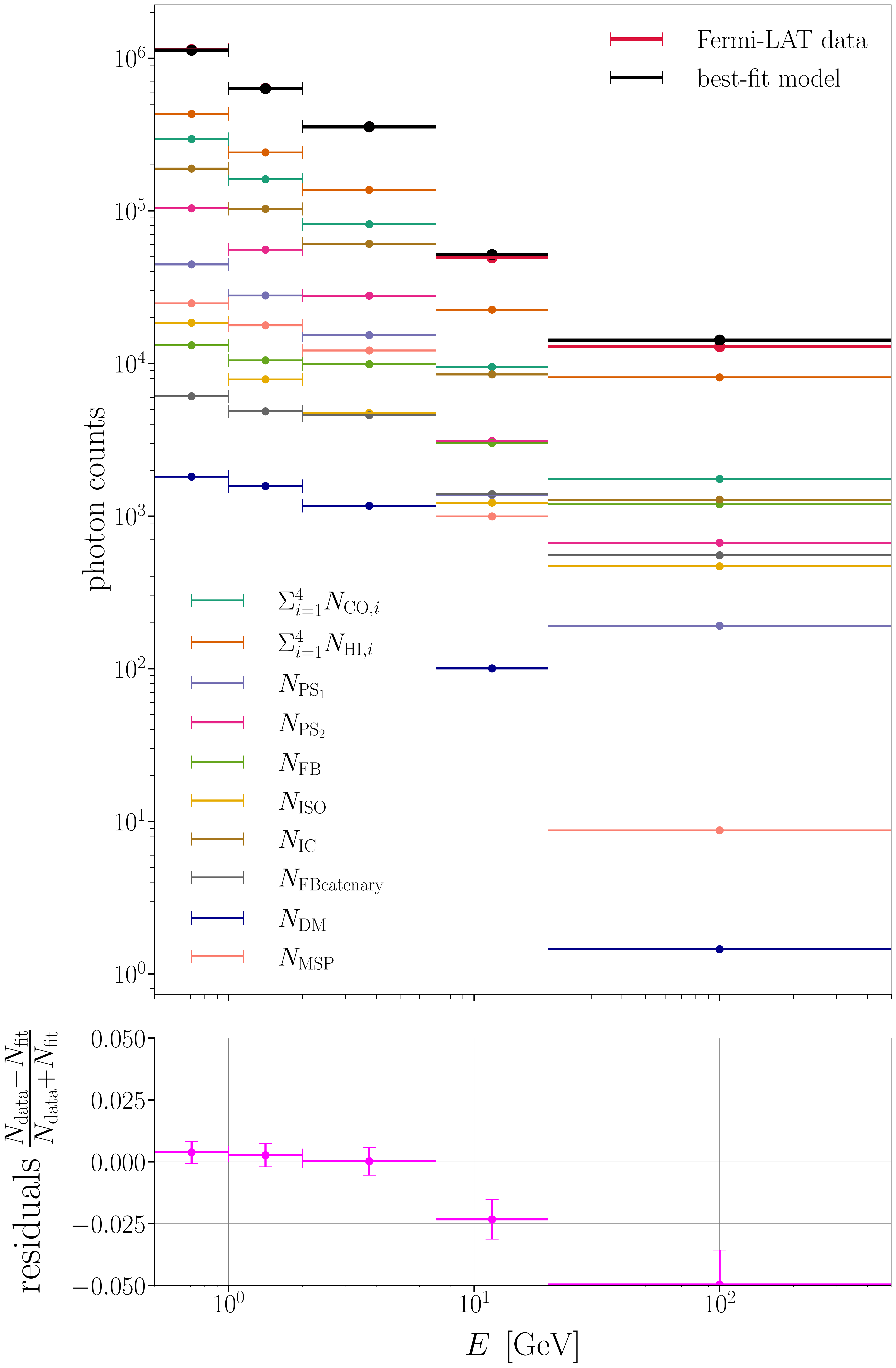}
\end{center}
\caption{The same as Fig.  \ref{fig:spatial_res_model1_scenA} but in the context of Model 3. 
\label{fig:spatial_residuals_extended_2_sc2}}
\end{figure}

\begin{figure}[t!]
\begin{center}
\includegraphics[width=0.45\linewidth]{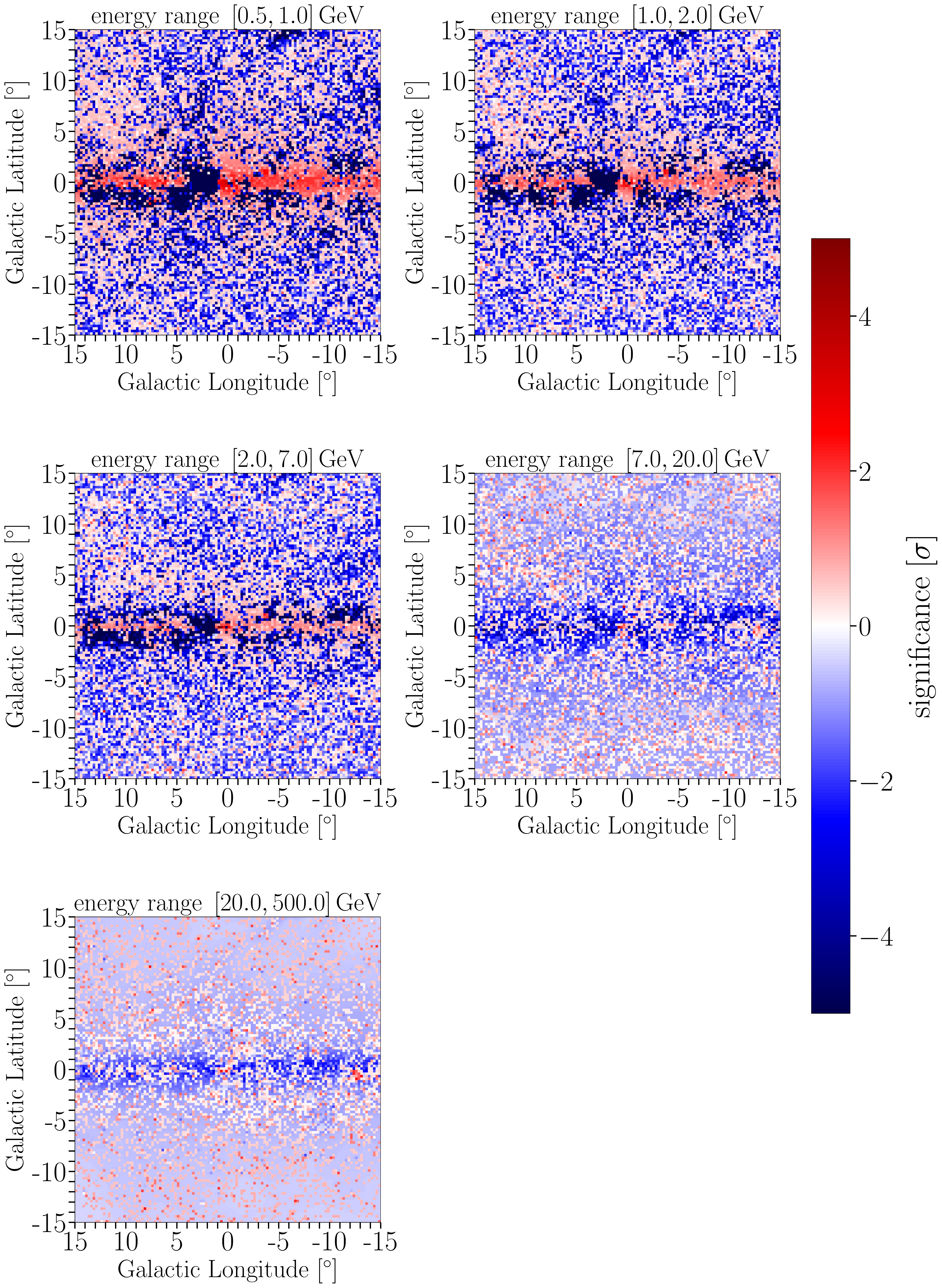}
\includegraphics[width=0.44\linewidth]{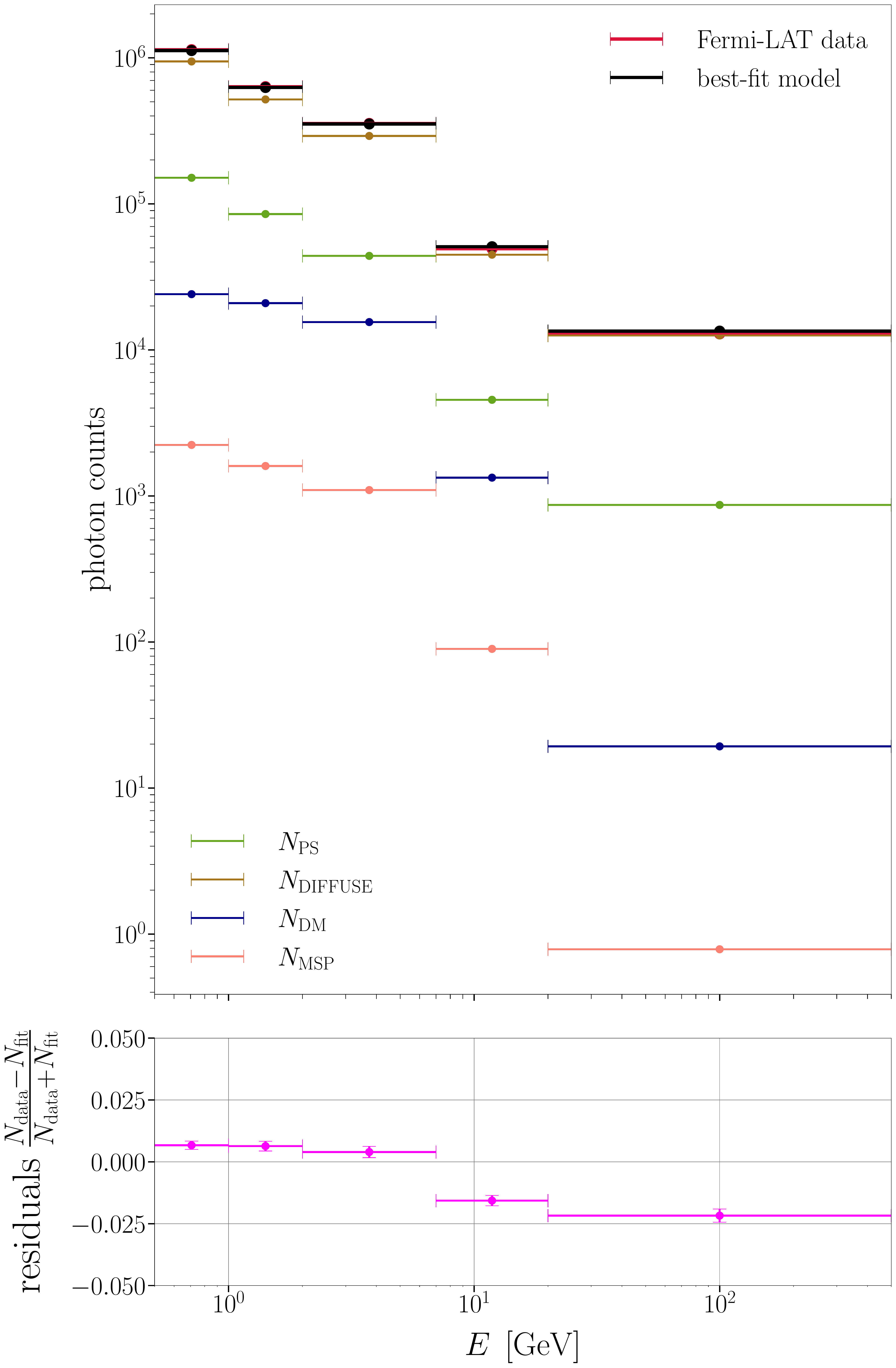}
\end{center}
\caption{The same as  Fig. \ref{fig:spatial_res_model1_scenA} (Model 1), but for Scenario B.  
\label{fig:residuals_modelB1}}
\end{figure}

\begin{figure}[t!]
\begin{center}
\includegraphics[width=0.45\linewidth]{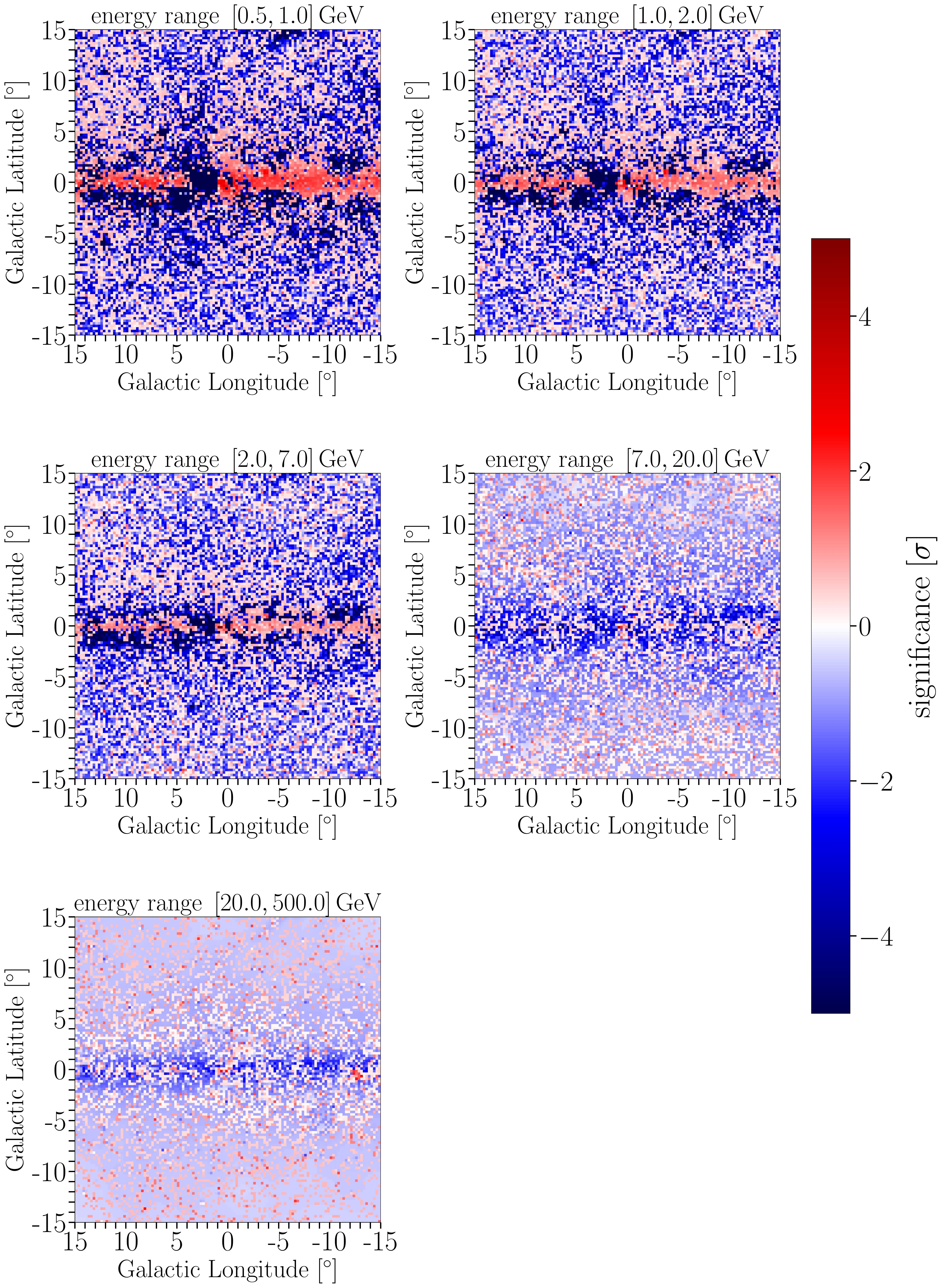}
\includegraphics[width=0.44\linewidth]{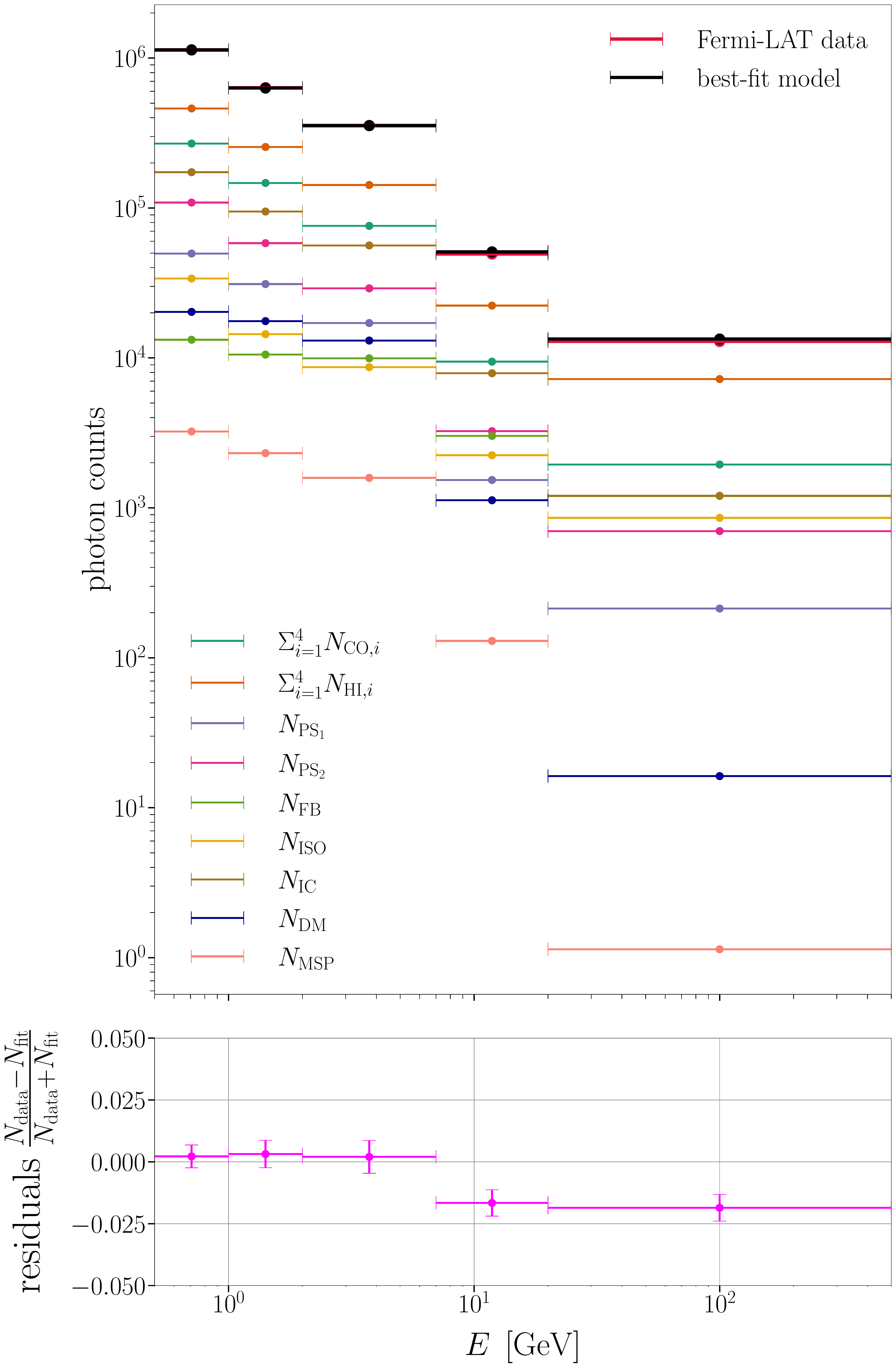}
\end{center}
\caption{Spatial and spectral residuals for Model 2B. 
\label{fig:residuals_modelB2}}
\end{figure}

\begin{figure}[t!]
\begin{center}
\includegraphics[width=0.45\linewidth]{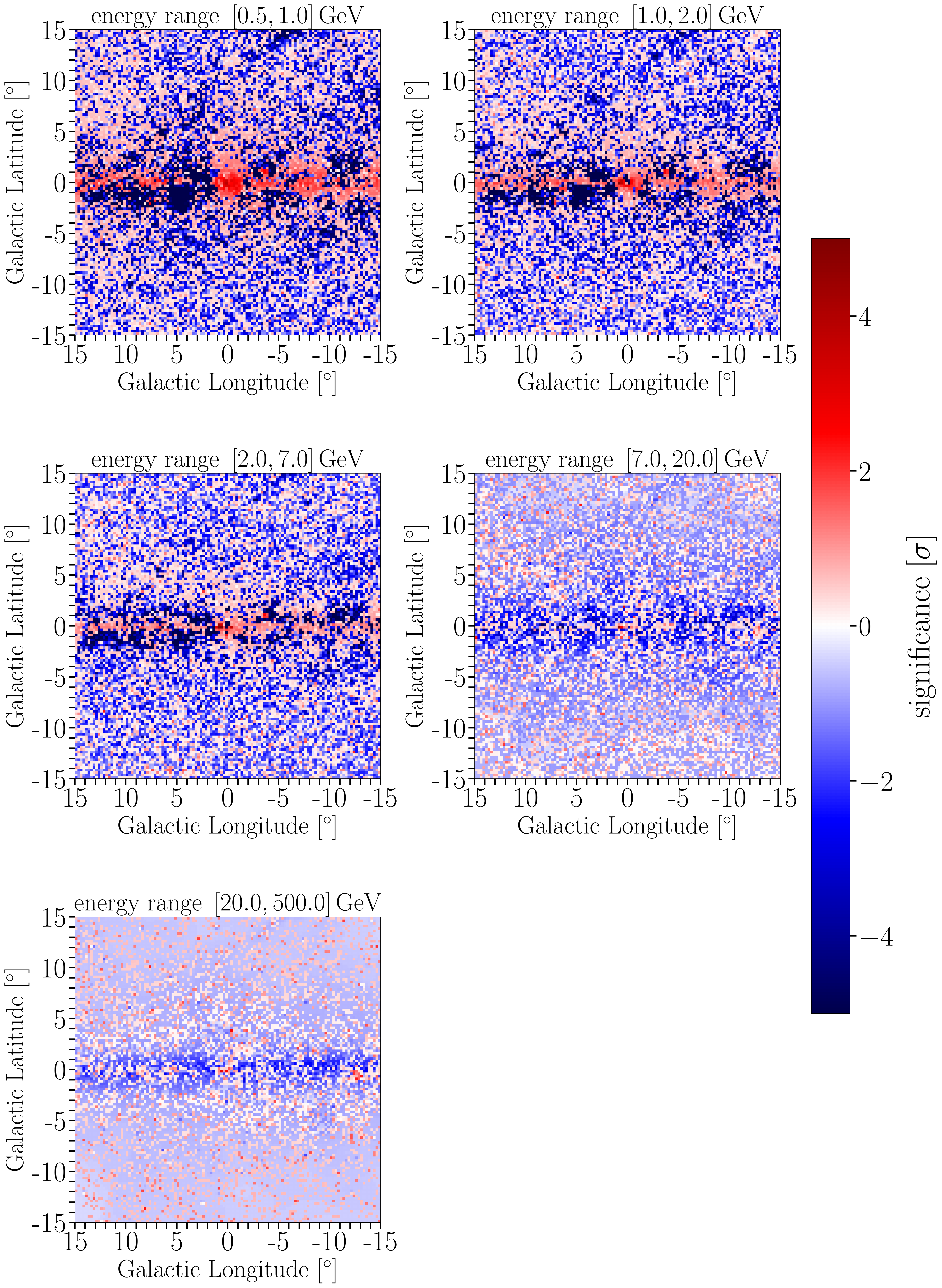}
\includegraphics[width=0.44\linewidth]{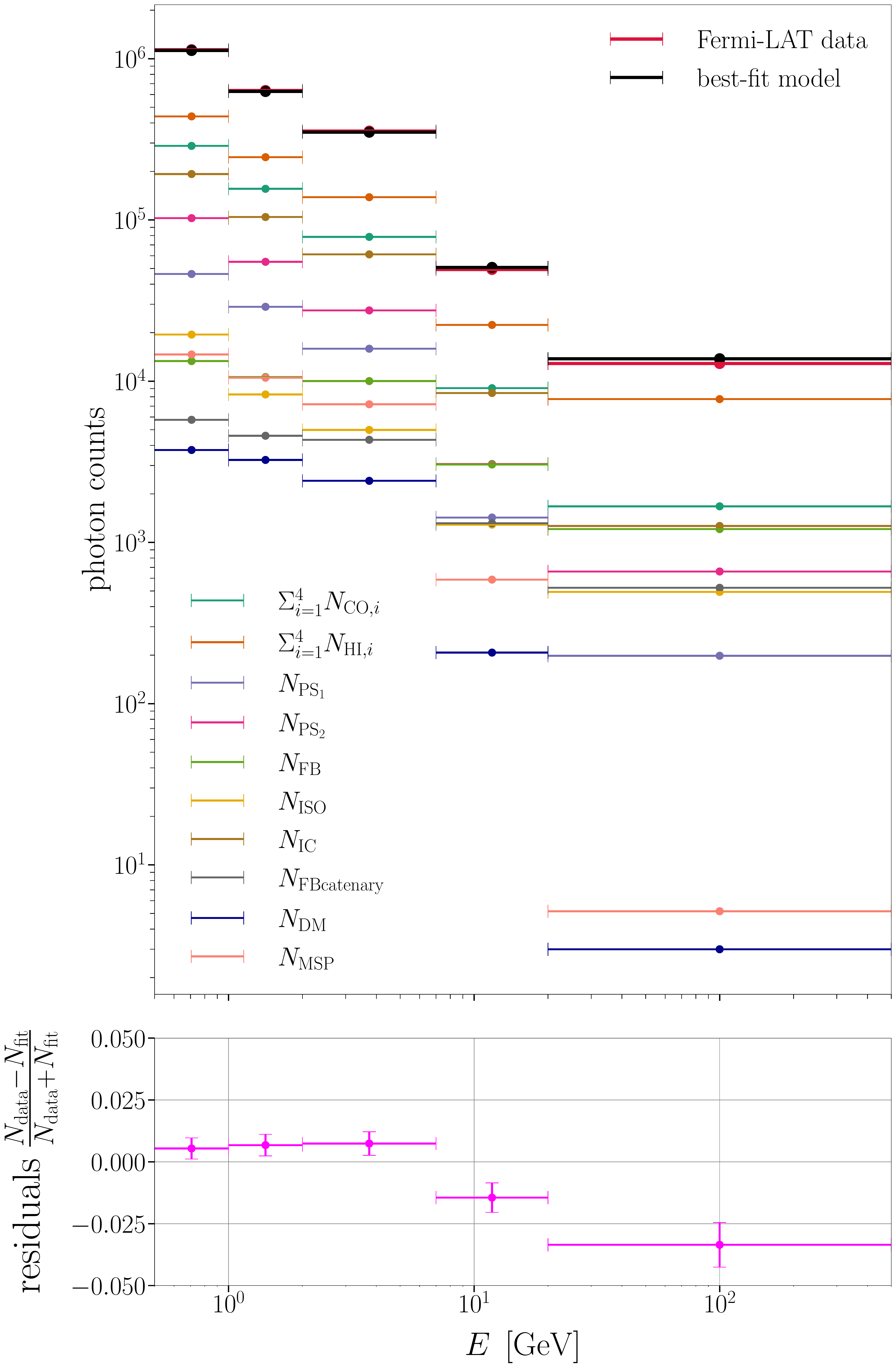}
\end{center}
\caption{Spatial and spectral residuals for Model 3B. 
\label{fig:residuals_modelB3}}
\end{figure}

\section{Consistency check with an alternative diffuse background model}
\label{sec:diffuse_bkg_check}

In this appendix, we list the results of our consistency checks invoking the Pass 8 Fermi diffuse background model used in the context of the 4FGL source catalog supplementing the discussion in the main text (see Sec.~\ref{sec:crossdomain-maintext}).

Figs.~\ref{fig:diffmod_1} to \ref{fig:diffmod_3A} illustrate the bias of the network's predictions induced by applying a validation data set with a diffuse model the network has not been initially trained on. For the comparison, we only select those model parameters that can be reasonably confronted with each other, i.e.~parameters that are shared between the Fermi diffuse validation data set and the original validation data set of the respective model and scenario. 

The qualitative interpretation of these results is identical to the one discussed in the main text using the example of Model 3B. Since the Fermi diffuse model incorporates an additional (smoothed) contribution from the Galactic center region that comprises the GCE and parts of the FBs, we obtain biased network predictions. Each individual network tries to assign the excess gamma-ray emission to several components that may absorb it according to the training data it has seen. What components are affected strongly depends on the respective model and scenario. If the model contains templates for the FBs, the predicted normalization factor is always biased towards higher values in full agreement with the expectations. Another common feature is the bias of the gNFW inner slope parameter $\gamma$ towards larger values since an additional GCE component is present in the validation data. However, the excess emission of the Fermi diffuse model is not consistently attributed to specific templates among models and scenarios. This again illustrates that overcoming the reality gap is essential to \emph{(i)} robustly establish the properties of the GCE and \emph{(ii)} identify the leading process that generates it in the first place.

\begin{figure}[t!]
\begin{center}
\includegraphics[width=0.55\linewidth]{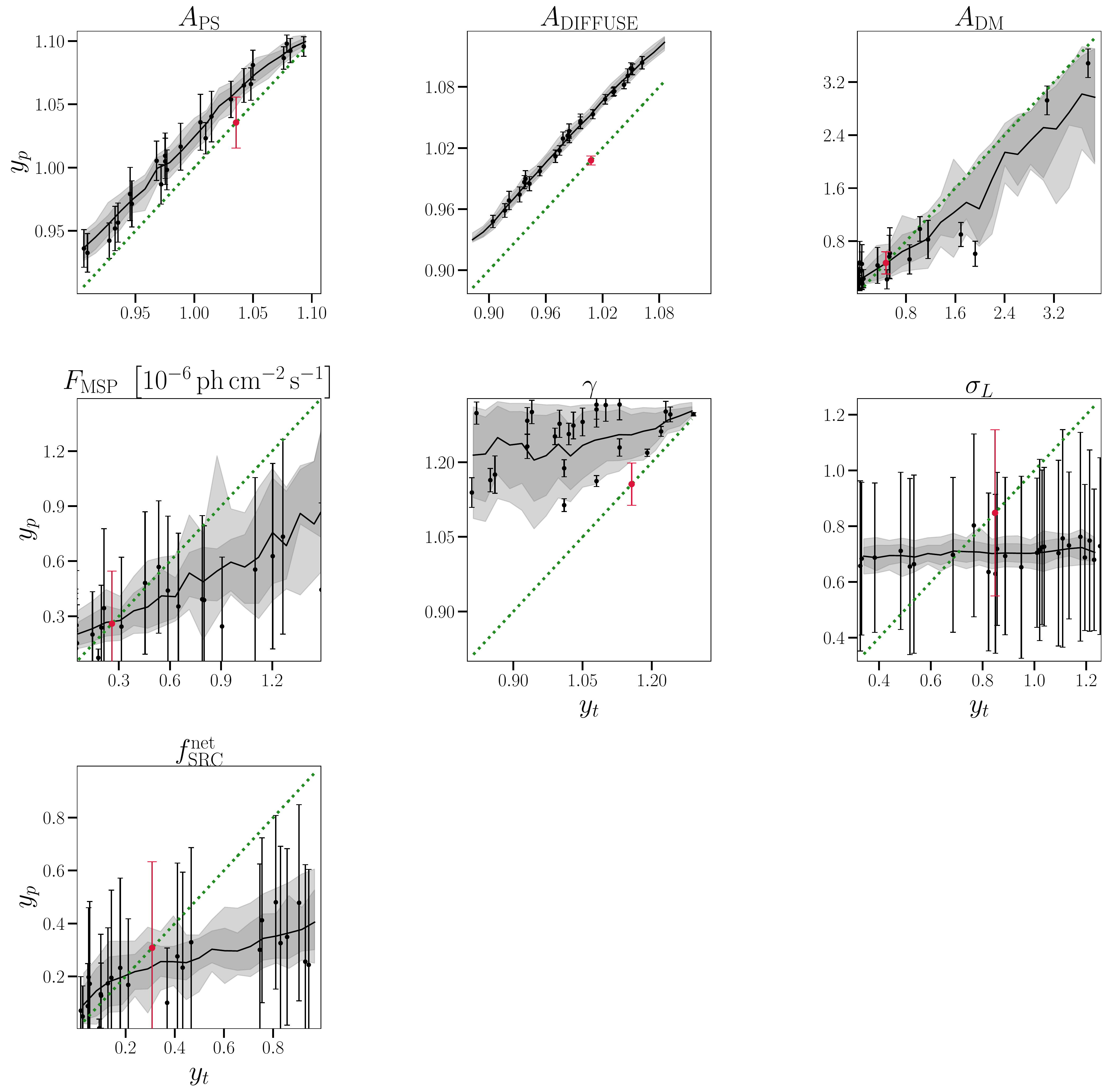}
\includegraphics[width=0.55\linewidth]{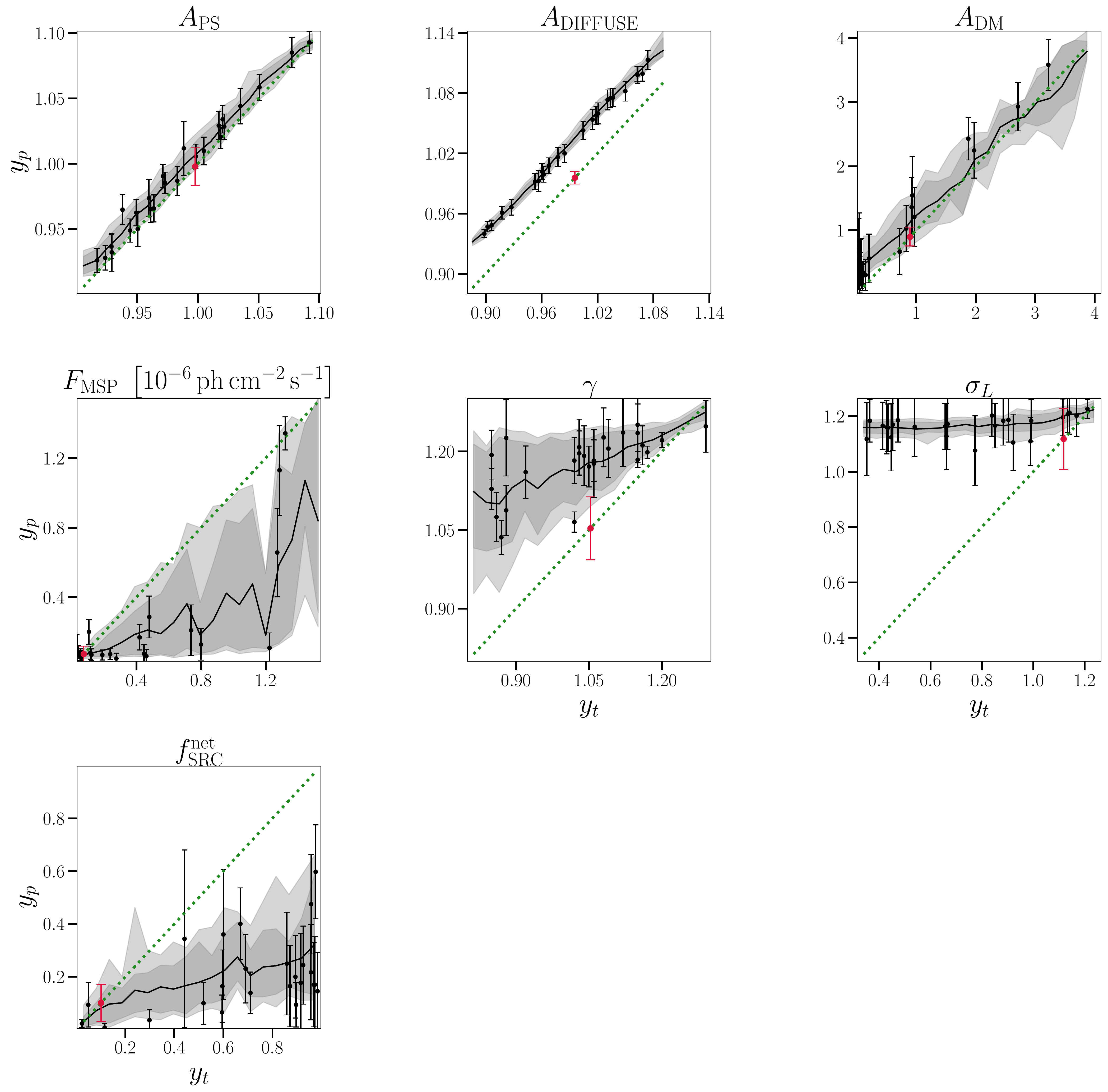}
\end{center}
\caption{Same as Fig.~\ref{fig:Diffmod_3B} for Model 1 ({\it Top: A}); ({\it Bottom: B}). \label{fig:diffmod_1}}
\end{figure}

\begin{figure}[t!]
\begin{center}
\includegraphics[width=0.55\linewidth]{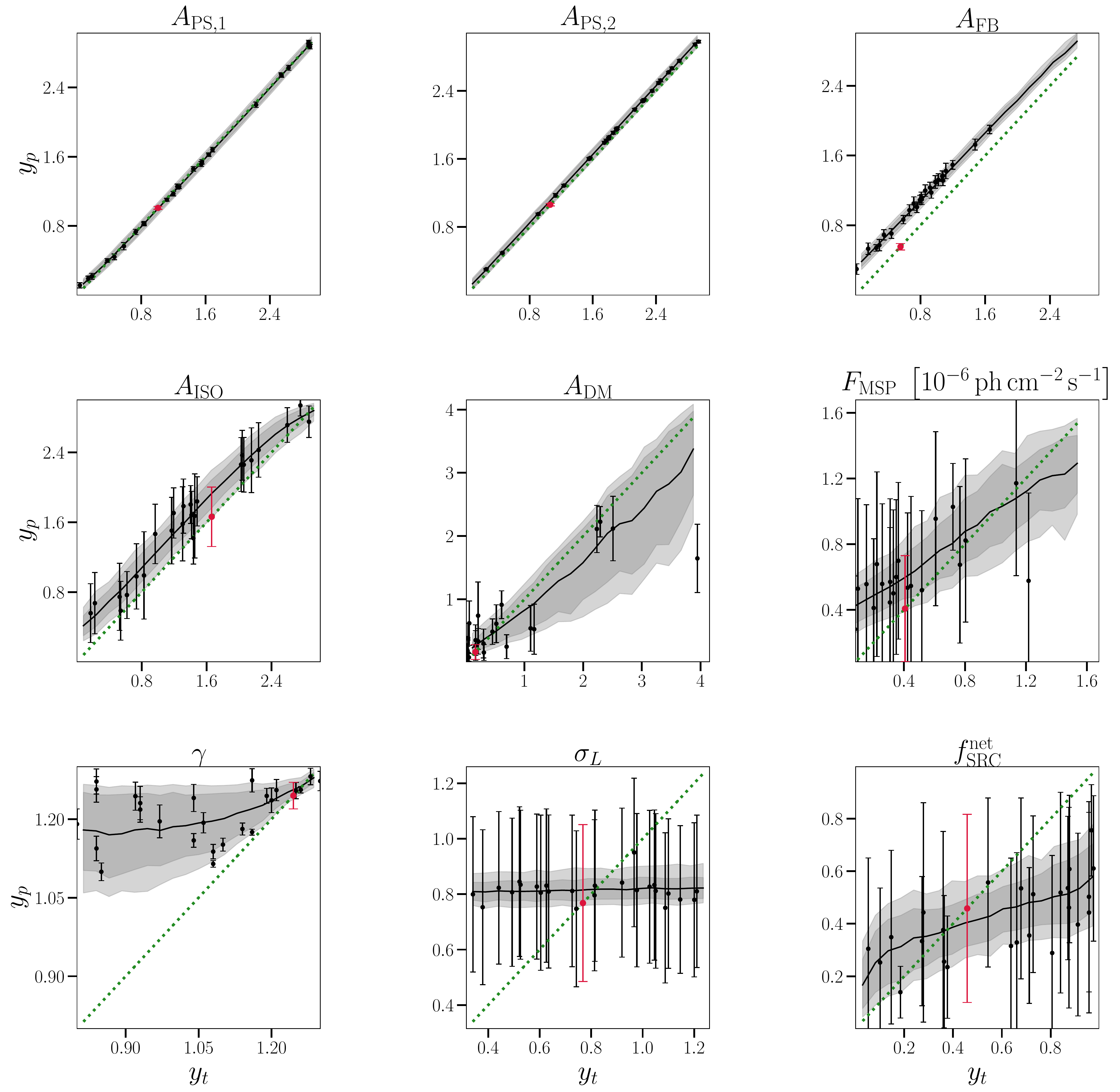}
\includegraphics[width=0.55\linewidth]{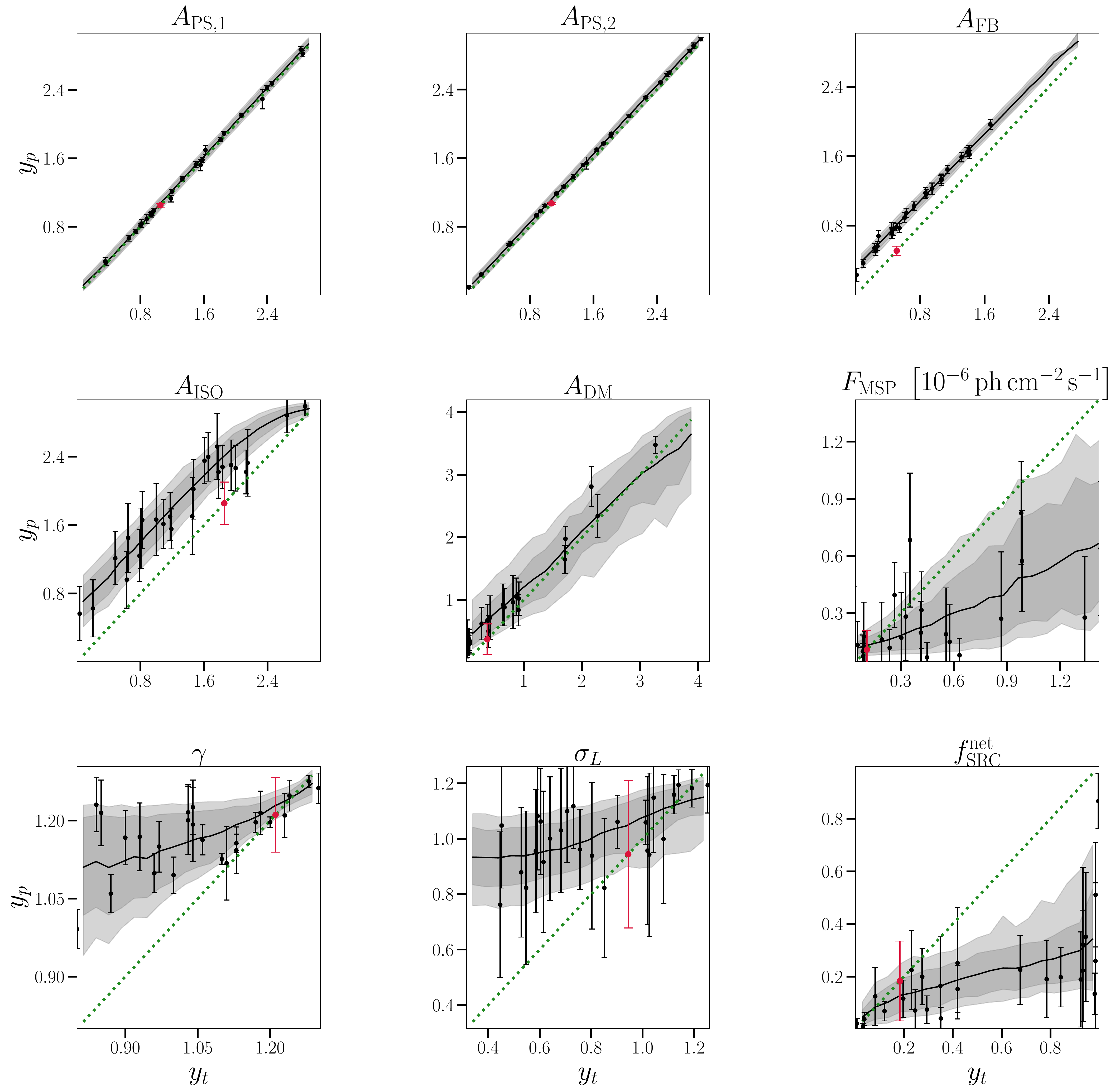}
\end{center}
\caption{Same as Fig.~\ref{fig:Diffmod_3B} for Model 2 ({\it Top: A}); ({\it Bottom: B}). \label{fig:diffmod_2}}
\end{figure}

\begin{figure}[t!]
\begin{center}
\includegraphics[width=0.55\linewidth]{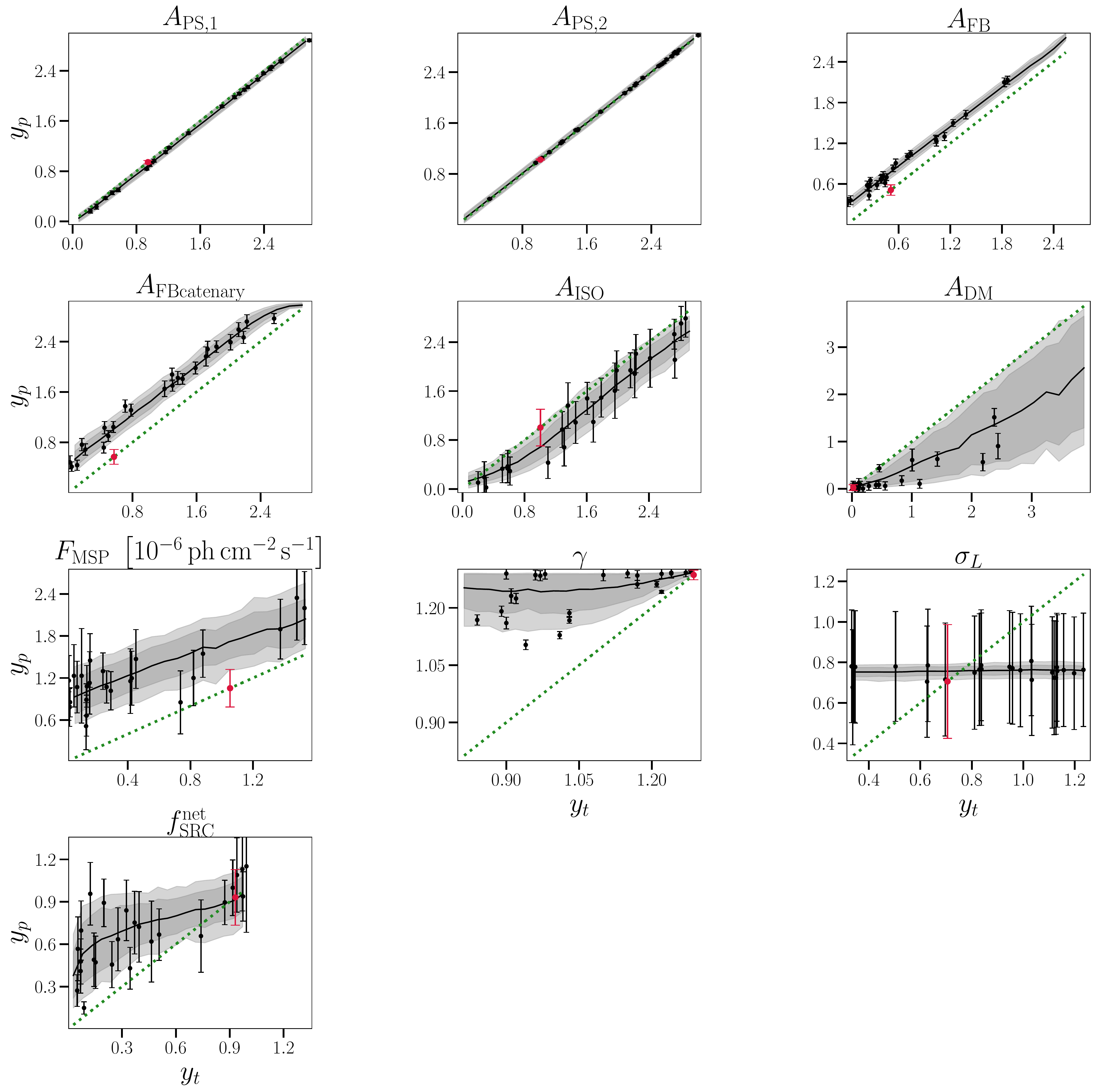}
\end{center}
\caption{Same as Fig.~\ref{fig:Diffmod_3B} for Model 3A. \label{fig:diffmod_3A}}
\end{figure}

\end{appendices}

\end{document}